\documentclass[twocolumn,english,aps,prb,twocolum,superscriptaddress,bibnotes,amsmath,amssymb,floatfix]{revtex4-1}
\usepackage[colorlinks=true,citecolor=blue,linkcolor=magenta]{hyperref}

\usepackage[markup=nocolor, authormarkupposition=left]{changes} 
\usepackage{soul}
\usepackage[utf8]{inputenc}
\usepackage[english]{babel}
\usepackage{amsmath,amsfonts,amssymb}
\usepackage[T1]{fontenc}
\usepackage{url}

\usepackage{amsmath}
\usepackage{amsfonts}
\usepackage{amssymb}

\usepackage{epstopdf}
\usepackage{graphicx}
\graphicspath{{./Figures/}}

\usepackage{changes}

\begin{document}
\title{Space compatibility of emerging, wide-bandgap, ultralow-loss integrated photonics}

\author{Yue Hu}
\affiliation{International Quantum Academy, Shenzhen 518048, China}
\affiliation{Shenzhen Institute for Quantum Science and Engineering, Southern University of Science and Technology, Shenzhen 518055, China}

\author{Xue Bai}
\email[]{baixue@iqasz.cn}
\affiliation{International Quantum Academy, Shenzhen 518048, China}

\author{Baoqi Shi}
\affiliation{International Quantum Academy, Shenzhen 518048, China}

\author{Jiahao Sun}
\affiliation{Shenzhen Institute for Quantum Science and Engineering, Southern University of Science and Technology, Shenzhen 518055, China}
\affiliation{International Quantum Academy, Shenzhen 518048, China}

\author{Yafei Ding} 
\affiliation{International Quantum Academy, Shenzhen 518048, China}
\affiliation{Shenzhen Institute for Quantum Science and Engineering, Southern University of Science and Technology, Shenzhen 518055, China}

\author{Zhenyuan Shang}
\affiliation{International Quantum Academy, Shenzhen 518048, China}
\affiliation{Shenzhen Institute for Quantum Science and Engineering, Southern University of Science and Technology, Shenzhen 518055, China}

 \author{Hanke Feng}
\affiliation{Department of Electrical Engineering \& State Key Laboratory of Terahertz and Millimeter Waves, City University of Hong Kong, Kowloon, China}

\author{Liping Zhou}
\affiliation{State Key Laboratory of Materials for Integrated Circuits, Shanghai Institute of Microsystem and Information Technology, Chinese Academy of Sciences, Shanghai 200050, China}

\author{Bingcheng Yang}
\affiliation{State Key Laboratory of Materials for Integrated Circuits, Shanghai Institute of Microsystem and Information Technology, Chinese Academy of Sciences, Shanghai 200050, China}
\affiliation{The Center of Materials Science and Optoelectronics Engineering, University of Chinese Academy of Sciences, Beijing, China}

\author{Shuting Kang}
\affiliation{MOE Key Laboratory of Weak-Light Nonlinear Photonics, TEDA Applied Physics Institute and School of Physics, Nankai University, Tianjin 300457, China.}

\author{Yuan Chen}
\affiliation{International Quantum Academy, Shenzhen 518048, China}
\affiliation{Shenzhen Institute for Quantum Science and Engineering, Southern University of Science and Technology, Shenzhen 518055, China}

\author{Shuyi Li}
\affiliation{International Quantum Academy, Shenzhen 518048, China}

\author{Jinbao Long}
\affiliation{International Quantum Academy, Shenzhen 518048, China}

\author{Chen Shen}
\affiliation{International Quantum Academy, Shenzhen 518048, China}
\affiliation{Qaleido Photonics, 518048 Shenzhen, China}

\author{Fang Bo}
\email[]{bofang@nankai.edu.cn}
\affiliation{MOE Key Laboratory of Weak-Light Nonlinear Photonics, TEDA Applied Physics Institute and School of Physics, Nankai University, Tianjin 300457, China.}

\author{Xin Ou}
\email[]{ouxin@mail.sim.ac.cn}
\affiliation{State Key Laboratory of Materials for Integrated Circuits, Shanghai Institute of Microsystem and Information Technology, Chinese Academy of Sciences, Shanghai 200050, China}

\author{Cheng Wang}
\email[]{cwang257@cityu.edu.hk}
\affiliation{Department of Electrical Engineering \& State Key Laboratory of Terahertz and Millimeter Waves, City University of Hong Kong, Kowloon, China}

\author{Junqiu Liu}
\email[]{liujq@iqasz.cn}
\affiliation{International Quantum Academy, Shenzhen 518048, China}
\affiliation{Hefei National Laboratory, University of Science and Technology of China, Hefei 230088, China}

\maketitle

\noindent\textbf{
Integrated photonics has revolutionized optical communication, sensing, and computation, offering miniaturized and lightweight solutions for spacecraft with limited size and payload.
Novel chip-scale instruments based on ultralow-loss integrated photonic platforms, including lasers, frequency combs and atomic traps, have been developed for space applications.
Therefore, quantifying the space compatibility of ultralow-loss photonic integrated circuits (PICs), particularly their radiation resistance, is critical.
This study experimentally evaluates the radiation resistance of ultralow-loss Si$_3$N$_4$, 4H-SiC, and LiNbO$_3$ PICs under intense $\gamma$-ray and high-energy proton irradiation.
Results show that proton irradiation with $1.1 \times 10^{10}$ $\mathrm{p/cm^2}$ total flux does not significantly increase optical loss or alter the refractive index of these PICs, while $\gamma$-ray irradiation with 1.2 Mrad accumulated dose only marginally increases their optical loss.
These findings provide preliminary evidence of the excellent space compatibility of ultralow-loss Si$_3$N$_4$, 4H-SiC, and LiNbO$_3$ PICs, highlighting their potential for compact and lightweight space systems.
}

\begin{figure*}[t!]
\centering
\includegraphics{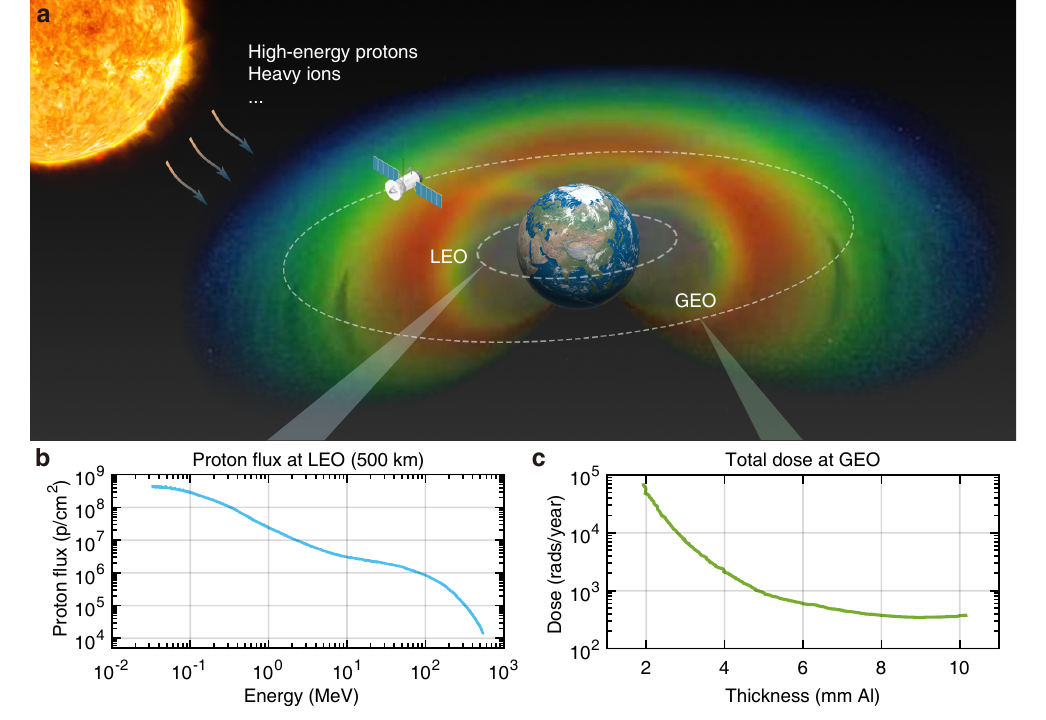}
\caption{
\textbf{Radiation environment of the near-Earth space.}
\textbf{a}. 
Radiation sources comprise not only trapped particles in Van Allen belts, but also high-energy protons, heavy ions and others from solar events and galactic cosmic rays.
The major radiation sources and their doses vary with satellite orbits.
The LEO, operating approximately 500 km above Earth, is mainly influenced by protons from the inner Van Allen belt. 
The GEO, where navigation satellites operate, contains not only electrons from the outer Van Allen belt, but also high-energy particles from solar events, including high-energy protons, heavy ions, and others. 
\textbf{b}.
Energy-flux spectrum of protons in LEO. 
\textbf{c}. 
Total radiation dose in GEO with 2 to 10 mm thick Al protection.
}
\label{Fig:1}
\end{figure*}

\begin{figure*}[t!]
\centering
\includegraphics{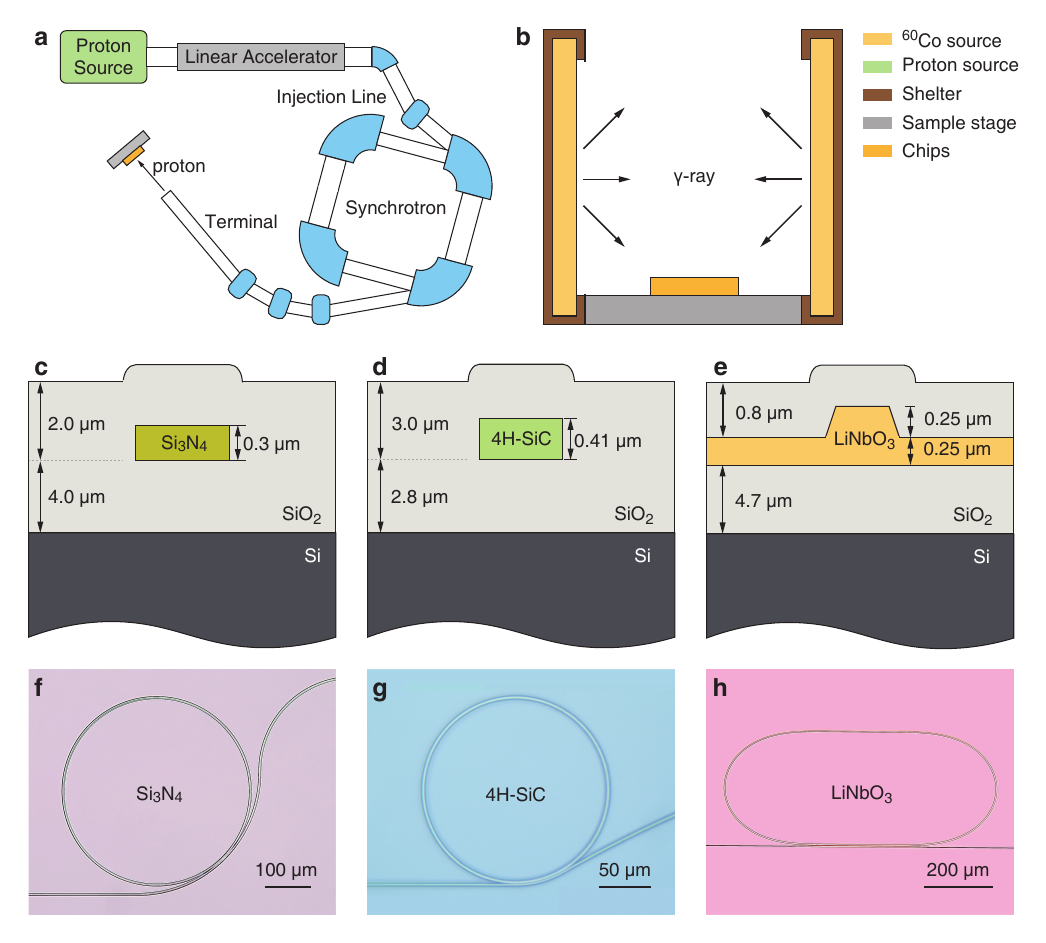}
\caption{
\textbf{Irradiation experimental setups and chip devices under study.} 
\textbf{a}. 
Experimental setup for proton irradiation. 
\textbf{b}.
Experimental setup for $\gamma$-ray irradiation.
\textbf{c--e}.
Chip cross-sections containing the waveguides and the full SiO$_2$ claddings, with marked dimension values.  
\textbf{f--h}. 
Optical microscope images of the microresonators corresponding to c--e.
}
\label{Fig:2}
\end{figure*}

\begin{figure*}[t!]
\centering
\includegraphics{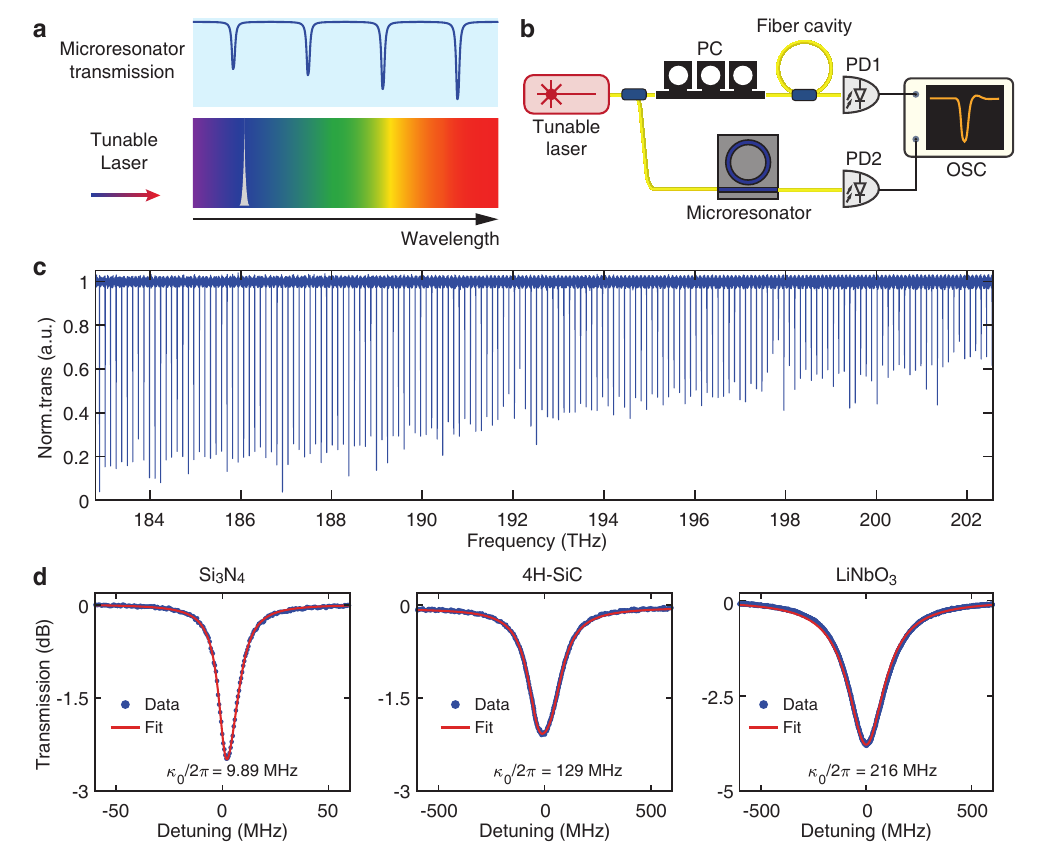}
\caption{
\textbf{Experimental setup to characterize microresonator loss and dispersion in the telecommunication bands.}
\textbf{a,b}. 
Principle and experimental setup. 
A tunable laser chirps from 1480 to 1640 nm (202.7 to 182.9 THz), whose instantaneous frequency is calibrated on-the-fly by a pre-calibrated fiber cavity. 
A portion of the laser output is coupled into optical microresonators under study.
During laser chirping, the frequency-calibrated microresonator transmission spectrum is recorded. 
PC, polarization controller. 
PD, photodetector. 
OSC, oscilloscope 
\textbf{c}. 
A typical microresonator transmission spectrum.
Each resonance is identified and fitted, whose intrinsic loss $\kappa_0/2\pi$ is extracted and analyzed. 
\textbf{d}. 
Three representative resonances with their fits and extracted $\kappa_{0}/2\pi$ values, from the chips corresponding to Figs. 2c--e, respectively. 
}
\label{Fig:3}
\end{figure*}

Integrated photonics~\cite{Thomson:16, Agrell:16} enables synthesis, processing and detection of optical signals via photonic integrated circuits (PICs), and has revolutionized modern optical communication, sensing, and computation.
Particularly, miniaturized, compact, and lightweight integrated devices are favourable for deployment in spacecrafts of limited size and payload. 
Key elements such as microresonators, gratings, interferometers, optical phased arrays, as well as active components such as lasers, detectors and optical amplifiers, are essential for space applications including astronomical instrumentation~\cite{Krainak:19, Roth:23}, spectrographs for astronomy~\cite{Cvetojevic:09, Gatkine:21}, astrocombs~\cite{Obrzud:19, Suh:19}, space optical communication~\cite{Fridlander:18, He:20}, and space sensors~\cite{Ohanian:18, Ziarko:24}.
However, the harsh environment of space~\cite{Barth:04}, especially the cosmic radiation from high-energy particles and cosmic rays, poses challenges to the reliability and stability of integrated photonic devices.
Exposure to cosmic radiation can alter material properties and even degrade optical performance of devices,  by inducing lattice displacement, charge generation, and surface passivation~\cite{Du:23, Bhandaru:15}.
Therefore, investigating and quantifying radiation impact on integrated photonic devices and PICs are mandatory before their deployment in space. 

So far, most studies on radiation effects are on silicon devices and PICs. 
The radiation resistance of silicon devices, such as micro-ring resonators~\cite{Bhandaru:15, Grillanda:16, Du:17, Reghioua:23, Zhou:22a}, Mach-Zehnder interferometers~\cite{Reghioua:23, Zhou:22a, Zhou:22b}, arrayed waveguide gratings~\cite{Zhou:22b}, and modulators~\cite{Sarah:15, Zeiler:17, Hoffman:19, Lalovic:22}, have been studied on the ground. 
Various high-energy radiation sources, such as X-rays, $\gamma$-rays, protons, and neutrons, have been used to simulate the space radiation environment. 
In addition, Ref. ~\cite{Mao:24} has performed experiment in space to further validate the space compatibility of silicon PICs. 

While silicon photonics has made tremendous success, it is well known that silicon PICs suffer from high linear (typically $\alpha>1$ dB/cm) and nonlinear losses. 
The former is due to the high refractive-index contrast to SiO$_2$ cladding, 
and the latter is due to the two-photon absorption (0.25 cm/GW) in the telecommunication wavelength caused by silicon's small bandgap ($\sim1.12$ eV). 
To complement silicon's limitations, numerous material platforms featuring wide bandgap and ultralow optical loss have emerged and quickly matured, such as silicon nitride (Si$_3$N$_4$) \cite{Xuan:16, Ji:17, Liu:21, Spencer:14}, lithium niobate (LiNbO$_3$) \cite{Zhang:17, He:19, GaoR:22} and silicon carbide (SiC) \cite{Guidry:20, Wang:22, Cai:22}. 
These platforms have enabled a new class of chip-scale instruments, such as optical frequency combs~\cite{Kippenberg:18, Gaeta:19, Zhang:19}, narrow-linewidth lasers~\cite{Siddharth:22,Corato-Zanarella:23,Ling:23, Isichenko:24, Nejadriahi:24},  and atomic traps~\cite{Blumenthal:24, Isichenko:23,Lee:22, Ropp:23}. 
Thus they hold great potential for space applications, especially in chip-based optical and atomic clocks~\cite{Kitching:18,Newman:19}. 
These clocks, with reduced size, weight, and power consumption, are critical in fundamental physics experiments~\cite{Will:14,Flechtner:21}, space master clocks~\cite{Lauf:03}, and global navigation satellite systems (GNSS)~\cite{Lechner:00,Grewal:11}.

However, whether these integrated platforms can maintain ultralow optical loss in space environment remain elusive, especially with regard to their radiation resistance.
Limited references~\cite{Brasch:14, Du:17, Du:20, Ma:16} have examined the loss change and refractive index drift in PICs based on Si$_3$N$_4$ and amorphous SiC, whose optical loss values are not as low as the state of the art. 
Moreover, there has not been a report on the radiation resistance of PICs based on single-crystal SiC and LiNbO$_3$.  

Here we experimentally gauge the radiation resistance of ultralow-loss integrated photonics against intense $\gamma$-rays and high-energy protons. 
We select three representative material platforms, i.e. Si$_3$N$_4$, 4H-SiC and LiNbO$_3$, which have emerged as compelling platforms for linear, nonlinear and quantum optics \cite{Moss:13, Xiang:22a, Zhu:21, Boes:23, Lukin:20, Yi:22}.  
Experimentally, we select the irradiation dose of $\gamma$-rays and protons based on the radiation environment of Earth satellite orbits.
Illustrated in Fig. \ref{Fig:1}a, the near-Earth space radiation environment consists mainly of trapped particles in Van Allen belts, and high-energy protons, heavy ions and others from solar events and galactic cosmic rays.  
The major radiation sources and their doses vary with satellite orbits.
For instance, low-Earth-orbit (LEO) satellites, such as the Micius satellite~\cite{LuCY:22}, operate approximately 500 km above Earth.  
Protons from the inner Van Allen belt are the predominant radiation source with energy from 0.01 to 1000 MeV~\cite{Stassinopoulos:88}, while shielding against protons with energy above 10 MeV is challenging. 
Figure \ref{Fig:1}b shows the proton energy spectrum. 
The flux is falling with increasing energy from 10 to 100 MeV, followed by a sharp drop.
Considering this energy-flux relation, we conduct proton-irradiation experiment using a 60-MeV proton beam with $1.1 \times 10^{10}$ p/cm$^2$ total flux. 
This flux well exceeds the cumulative proton exposure expected over 15 years in LEO without shielding~\cite{Stassinopoulos:88}.

Another key application environment is the geostationary orbit (GEO) where navigation satellites operate.
The GEO space environment contains high-energy particles from solar events, including high-energy protons, heavy ions, and others~\cite{Stassinopoulos:80,Stassinopoulos:88}.
Due to the sporadic occurrence and the complex nature of solar flare events~\cite{Curto:20,Leka:18}, we adopt $\gamma$-rays to represent various types of radiation particles, base on the principle of equivalent radiation effect.
Figure \ref{Fig:1}c shows the total dose in GEO with aluminium (Al) shielding of thickness varying from 2 to 10 mm.
Our $\gamma$-ray-irradiation experiment is carried out with 1.2 Mrad accumulated dose, corresponding to the total exposure expected over 15 years in GEO with 2-mm-thick Al shield~\cite{Bhat:05}.

To evaluate the radiation resistance of our ultralow-loss PICs,  we perform two independent irradiation experiments with proton or $\gamma$-ray sources on the ground (not in space). 
Figure \ref{Fig:2}a illustrates the proton irradiation source based on a synchrotron accelerator~\cite{McMillan:45,Wilson:77}, used in our work. 
Initially, low-energy protons are accelerated to 20 keV using a linear accelerator.
Subsequently, these protons are injected into a synchrotron accelerator for acceleration to 10 to 60 MeV. 
Finally, the high-energy protons are extracted to the terminal where they irradiate our chip devices.
Figure \ref{Fig:2}b illustrates the $\gamma$-ray irradiation source employing a dual-grid cobalt source ($^{60}$Co) that is positioned on both sides of the sample stage. 
Our chips are placed at a pre-calculated position on the stage, where they experience the desired dose rate of $\gamma$-ray. 

\noindent \textbf{Chip characterization.}
To characterize the optical loss change of our chips before and after irradiation, we characterize the resonance linewidth values of integrated microresonators made of Si$_3$N$_4$, 4H-SiC and LiNbO$_3$. 
The fabrication processes of these chips are individually described in Supplementary Note 1.
Figures \ref{Fig:2}c--e portray the chip cross-sections containing the waveguides and the full SiO$_2$ claddings, with marked dimension values.    
Figures \ref{Fig:2}f--h display the optical microscope images of the microresonators corresponding to Figs. \ref{Fig:2}c--e.

Conventionally, the optical loss of PICs can be characterized using the cutback method \cite{Bauters:11} or optical-frequency-domain reflectometry (OFDR)~\cite{Soller:05}.
The cutback method estimates optical loss by measuring waveguides of varying lengths, thus the measurement precision is limited. 
In contrast, the OFDR method achieves higher precision by analyzing the interference between a reference laser beam and the light reflected from the waveguide's rear facet. 
Nevertheless, both methods necessitate long spiral or meander waveguides occupying large areas on the chip.  
Here, we instead measure the intrinsic resonance linewidth $\kappa_0/2\pi$ of waveguide-based optical microresonators, which relates to the linear optical loss $\alpha$ (dB/m physical length) as
\begin{equation}
\alpha=27.27\frac{n_g \kappa_0}{2\pi c}
\end{equation}
where $n_g$ is the group index of the waveguide's fundamental optical mode and $c$ is the speed of light in vacuum.
Note that, when using a widely tunable, continuous-wave (CW) laser~\cite{Shi:24}, such microresonator-based measurement provides abundant resonances, enabling statistically improved precision to extract optical loss values. 
In addition, as microresonators are small, many microresonators of identical geometry can be assembled on a single chip, allowing the examination of measurement uniformity and reproducibility. 

We use a vector spectrum analyzer (VSA)~\cite{Luo:24} operating in the telecommunication bands to faithfully characterize the frequency and linewidth of each microresonator's resonance, before and after irradiation. 
The principle and experimental setup of our VSA are illustrated in Figs. \ref{Fig:3}a,b, respectively.
A widely tunable, mode-hop-free, external-cavity diode laser (ECDL, Santec TSL) chirps from 1480 to 1640 nm (202.7 to 182.9 THz), whose instantaneous frequency is calibrated on-the-fly by a pre-calibrated fiber cavity~\cite{Luo:24}. 
A portion of the laser output is coupled into optical microresonators via lensed fibers, inverse tapers and bus waveguides \cite{Liu:18, Pfeiffer:17}.  
During laser chirping, the frequency-calibrated microresonator transmission spectrum is recorded. 
Figure \ref{Fig:3}c presents a typical transmission spectrum.
Each resonance within the measurement range is identified and fitted \cite{Li:13}, whose intrinsic loss $\kappa_0/2\pi$, external coupling strength $\kappa_\text{ex}/2\pi$, and the total (loaded) linewidth $\kappa/2\pi=(\kappa_0+\kappa_\text{ex})/2\pi$ are extracted. 
Details on resonance fit are found in Supplementary Note 2.
Figure \ref{Fig:3}d shows three representative resonances with their fits and extracted $\kappa_{0}/2\pi$ values, from the chips corresponding to Figs. \ref{Fig:2}c--e, respectively. 

In addition, we also characterize the integrated dispersion profile $D_\text{int}/2\pi$ of each microresonator, defined as
\begin{equation}
D_\text{int}(\mu)=\omega_{\mu}-\omega_{0}-D_{1}\mu=\sum\limits_{n=2}^{...} \frac{D_n \mu^n}{n!}
\end{equation}
where $\omega_{\mu}/2\pi$ is the measured frequency value of the $\mu$-th resonance relative to the reference resonance of frequency $\omega_{0}/2\pi$, 
$D_1/2\pi$ is the microresonator's free spectral range (FSR), 
$D_2/2\pi$ describes the group-velocity dispersion (GVD), 
and $D_n/2\pi$ ($n\geqslant3$) terms are high-order dispersion parameters.

\begin{figure*}[t!]
\centering
\includegraphics{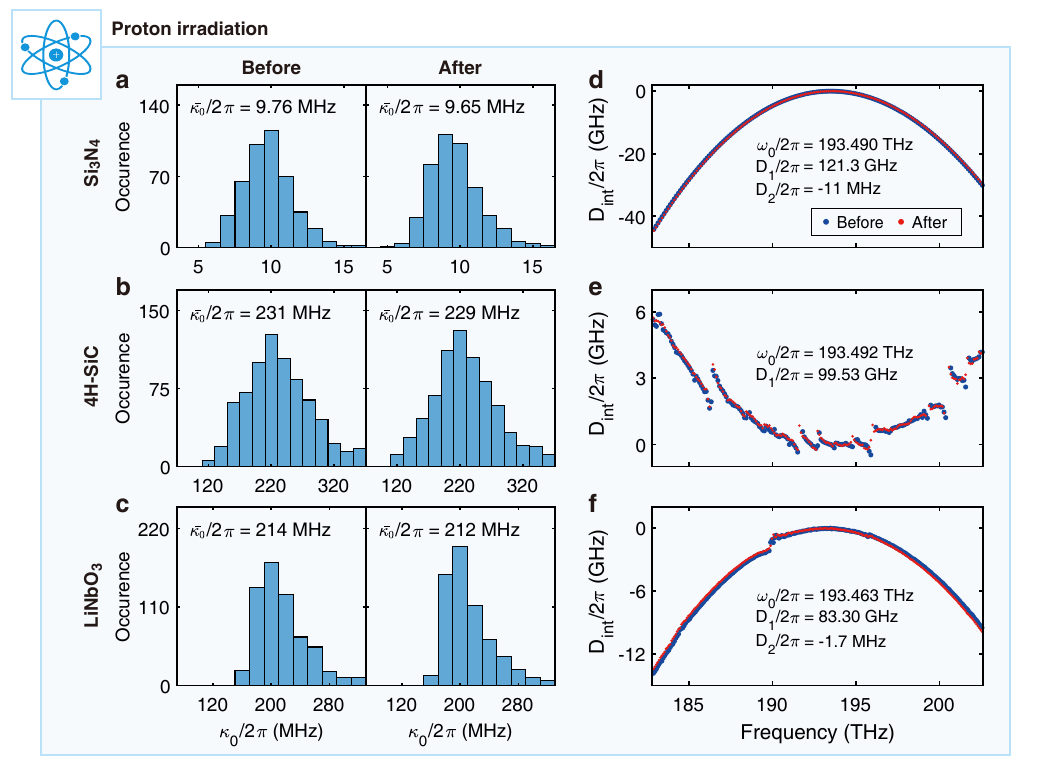}
\caption{
\textbf{Statistical analysis of loss change before and after proton irradiation.}
\textbf{a--c}. 
Histograms of measured $\kappa_0/2\pi$ values for each material corresponding to Figs. 2c--e, respectively. 
Comparisons on the most probable $\kappa_0/2\pi$ values (before and after irradiation) and the average values $\bar{\kappa_0}/2\pi$, evidences that the proton irradiation does not introduce extra optical loss. 
\textbf{d--f}. 
Measured $D_\text{int}/2\pi$ profiles on representative microresonators for each material. 
Comparison on the $D_\text{int}/2\pi$ profiles before (blue dots) and after irradiation (red dots) indicates that the proton irradiation also does not cause refractive index change. 
}
\label{Fig:4}
\end{figure*}

\begin{figure*}[t!]
\centering
\includegraphics{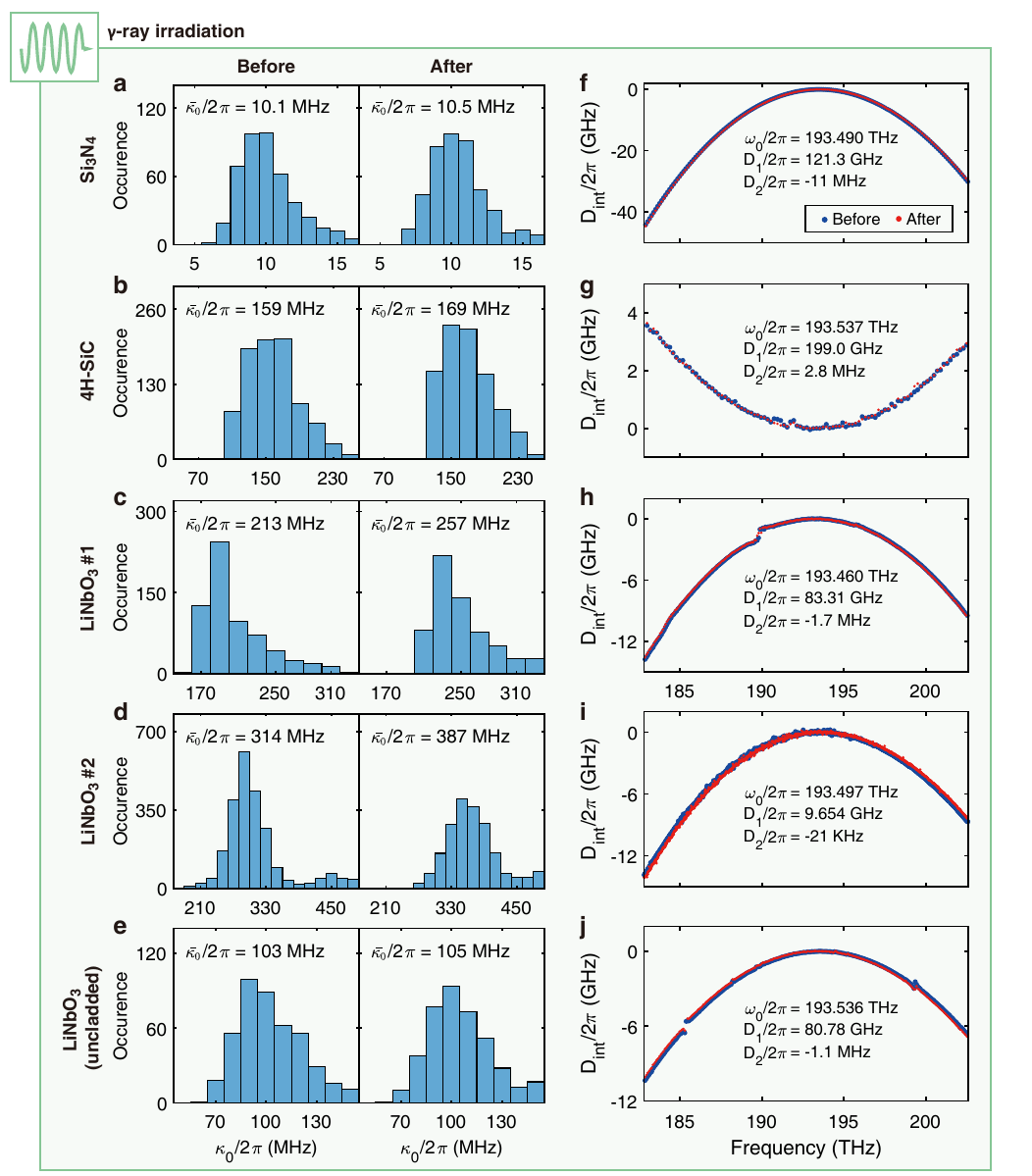}
\caption{
\textbf{Statistical analysis of loss change before and after $\gamma$-ray irradiation.}
\textbf{a--e}. 
Histograms of measured $\kappa_0/2\pi$ values for Si$_3$N$_4$ (as Fig. 2c), 4H-SiC (as Fig. 2d), LiNbO$_3$ with top SiO$_2$ cladding (as Fig. 2e),  and LiNbO$_3$ without top SiO$_2$ cladding. 
Comparison on the average values $\bar{\kappa_0}/2\pi$ (before and after irradiation) reveals marginally increased optical loss, i.e. 4.0\% for Si$_3$N$_4$, 6.3\% for 4H-SiC, and 2.0\% for LiNbO$_3$ without top SiO$_2$ cladding. 
For LiNbO$_3$ with top SiO$_2$ cladding, the $\bar{\kappa_0}/2\pi$ value is increase by 21\% in c and 23\% in d.  
\textbf{f--j}. 
Measured $D_\text{int}/2\pi$ profiles on representative microresonators for each material. 
Comparison on the $D_\text{int}/2\pi$ profiles before (blue dots) and after irradiation (red dots) indicates that the $\gamma$-ray irradiation also does not cause refractive index change. 
}
\label{Fig:5}
\vspace{2cm}
\end{figure*}

For each resonance, we compare the measured $\kappa_0/2\pi$ values before and after irradiation, to reveal the $\kappa_0/2\pi$ change due to irradiation.
In addition, for each microresonator, we compare the fitted $D_\text{int}/2\pi$ profiles before and after irradiation, to reveal the $D_\text{int}/2\pi$ and refractive index change due to irradiation.
Figure \ref{Fig:4} presents the cases with proton irradiation. 
Figures \ref{Fig:4}a--c present the histograms of measured $\kappa_0/2\pi$ values for each material corresponding to Figs. \ref{Fig:2}c--e, respectively. 
Each histogram counts more than 1000 resonances. 
The comparisons (before and after irradiation) on the most probable $\kappa_0/2\pi$ values and the average values $\bar{\kappa_0}/2\pi$, evidence that the proton irradiation does not introduce extra optical loss. 
Meanwhile, for each material, we select a representative microresonator and investigate the $D_\text{int}/2\pi$ change due to proton irradiation. 
Figures \ref{Fig:4}d--f present the comparison on the $D_\text{int}/2\pi$ profiles before and after irradiation, indicating that the proton irradiation also does not cause refractive index change. 

Figure \ref{Fig:5} presents the cases with $\gamma$-ray irradiation. 
Figures \ref{Fig:5}a--c present the histograms of measured $\kappa_0/2\pi$ values for each material corresponding to Figs. \ref{Fig:2}c--e, respectively.
The comparisons (before and after irradiation) on the average values $\bar{\kappa_0}/2\pi$ in Figs. \ref{Fig:5}a,b reveal marginally increased optical loss, i.e. 4.0\% for Si$_3$N$_4$ and 6.3\% for 4H-SiC. 
Interestingly,  this value is elevated to 21\% for LiNbO$_3$ in Fig. \ref{Fig:5}c. 
To examine the reproducibility, we characterize another LiNbO$_3$ chip and again observe $\bar{\kappa_0}/2\pi$ increase by 23\%, as shown in Fig. \ref{Fig:5}d. 

To answer whether the loss increase is due to the radiation impact on the LiNbO$_3$ waveguides, we perform $\gamma$-ray irradiation on another LiNbO$_3$ chip without top SiO$_2$ cladding (see Supplementary Note 3).
Figure \ref{Fig:5}e presents the histograms of measured $\kappa_0/2\pi$ before and after irradiation, where $\bar{\kappa_0}/2\pi$ increase by only 2.0\% is observed. 
Therefore, we infer that the loss increase is due to the radiation impact on the SiO$_2$ top cladding. 
It is very likely that the issue is associated with the specific chemical vapor deposition (CVD) process of SiO$_2$ or the tool used, since the Si$_3$N$_4$ and 4H-SiC chips also have SiO$_2$ top cladding (deposited using different processes in different tools at different places). 
Nevertheless, such loss increase in SiO$_2$-cladded LiNbO$_3$ chip is still moderate. 

Again, for each material, we select a representative microresonator and investigate the $D_\text{int}/2\pi$ change due to $\gamma$-ray irradiation. 
Figure \ref{Fig:5}f--j presents the comparison on the $D_\text{int}/2\pi$ profiles before and after irradiation, indicating that the $\gamma$-ray irradiation also does not cause refractive index change. 

\vspace{0.1cm}
\noindent \textbf{Discussion.}
Our experimental study shows that proton irradiation does not alter optical loss and refractive index of Si$_3$N$_4$, 4H-SiC and LiNbO$_3$ PICs, while $\gamma$-ray irradiation can marginally increase optical loss. 
The latter may be attributed to the total ionizing dose (TID) effect or the displacement damage dose (DDD) effect, which are caused by charge-generation-induced local field modulation and collision-induced displacement, respectively~\cite{Du:23}.
Both effects are long-term and detectable after irradiation. 
The TID effect is caused by trapped charges generated from target atoms upon ion incidence.
These trapped charges modify the target atoms’ polarizability and alter their refractive index~\cite{Du:23}.
In both irradiation experiments, we have not observed refractive index change, suggesting that the TID effect is negligible.

To examine whether the loss increase is caused by the DDD effect, we measure the Raman spectra of Si$_3$N$_4$, SiC and LiNbO$_3$ thin films before and after irradiation. 
Data are found in Supplementary Note 4.
No peak shifts have been detected, indicating that no prominent defect change occurs within the accuracy of our Raman spectrometer.
In addition, we measure the X-ray diffraction (XRD) spectra of SiC and LiNbO$_3$ thin films before and after irradiation. 
Supplementary Note 4 shows that the diffraction patterns remain unchanged, revealing that the lattice structure has not been altered.
From these results, we infer that the loss increase after $\gamma$-ray irradiation is possibly due to a small DDD effect that is below the detection limit of our Raman spectrometer and X-ray diffractometer.
In the proton irradiation experiment, neither the optical nor the material characterization shows loss increase.
This is because that the proton irradiation dose (estimated to be $<8 \times 10^4$ rad based on the measurement in LEO~\cite{Mao:24}) is much smaller than the $\gamma$-ray dose, despite that the higher energy of individual protons compared to $\gamma$-ray photons.

Our study demonstrates that Si$_3$N$_4$, SiC and LiNbO$_3$ PICs exhibit excellent radiation resistance in LEO.
However, for application in GEO, the 2-mm-thick Al shield is insufficient. 
Therefore, a thicker shield is needed in the GEO environment to prevent irradiation damage.
To further explore the space compatibility of these PICs, \textit{in situ} measurements should be considered, as temporary irradiation damage may also affect optical properties.
Additionally, space-based experiments are required to assess the impact of high-energy particles from galactic cosmic rays, which cannot be fully predicted in ground-based tests.

\vspace{0.1cm}
\noindent \textbf{Conclusion.}
In summary, we have performed irradiation experiments using proton and $\gamma$-ray sources with doses of $1.1 \times 10^{10}$ $\mathrm{p/cm^2}$ and 1.2 Mrad, respectively.
We have characterized high-$Q$ Si$_3$N$_4$, 4H-SiC and LiNbO$_3$ microresonators before and after irradiation.
Analysis of intrinsic loss $\kappa_0/2\pi$ and dispersion profile $D_\text{int}/2\pi$ in the telecommunication bands reveals that proton irradiation does not cause observable change of optical loss and refractive index of our devices. 
In contrast, $\gamma$-ray irradiation can marginally increase optical loss, which are 4.0\%, 6.3\% and 21\% for SiO$_2$-cladded Si$_3$N$_4$ , 4H-SiC and LiNbO$_3$, respectively, while refractive indices again remain unchanged.
We further conduct Raman and XRD characterization of these thin films, and have not observed peak shift or pattern change, suggesting that the loss increase is possibly due to the weak DDD effect below the detection limit of our Raman spectrometer and X-ray diffractometer.
Our study shows preliminary evidence of the excellent space compatibility of ultralow-loss integrated photonics based on Si$_3$N$_4$, 4H-SiC and LiNbO$_3$,  and highlights the potential of using these chip devices for space applications which require miniaturized, compact, and lightweight systems. 

\vspace{0.3cm}
\noindent \textbf{Acknowledgments}: 
The authors thank Xinjiang technical institute of physics and chemistry, Chinese academy of sciences, for providing the proton irradiation facility.
The authors thank Shanghai institute of applied physics, Chinese academy of sciences, for providing the $\gamma$-ray irradiation facility.
The authors thank Haifeng Jiang and Jianyu Guan for critical suggestions.

\vspace{0.1cm}
\noindent \textbf{Funding Information}:
Innovation Program for Quantum Science and Technology (2023ZD0301500), 
National Natural Science Foundation of China (Grant No.12261131503 and 62405202), 
National Key R\&D Program of China (No. 2022YFA1404601), 
Shenzhen-Hong Kong Cooperation Zone for Technology and Innovation (HZQB-KCZYB2020050), 
Shenzhen Science and Technology Program (Grant No. RCJC20231211090042078), 
and Guangdong-Hong Kong Technology Cooperation Funding Scheme (Grant No. 2024A0505040008).

\vspace{0.1cm}
\noindent \textbf{Author contributions}:
Y.H. and X.B. investigated and performed the experiment, with assistance from J.S., Y.D., Z.S., Y.C., J. Long and S.L.. 
B.S. and Y.H. built the chip characterization setup. 
Z.S. and C.S. fabricated the Si$_3$N$_4$ devices. 
H.F., C.W., S.K. and F.B. provided the LiNbO$_3$ devices. 
L.Z., B.Y. and X.O. provided the SiC devices. 
Y.H., X.B., B.S. and J.Liu wrote the manuscript with input from others.
J.Liu managed the collaboration and supervised the project.

\vspace{0.1cm}
\noindent \textbf{Conflict of interest}:
Authors declare no conflicts of interest.

\vspace{0.1cm}
\noindent \textbf{Data Availability Statement}: 
The code and data used to produce the plots within this work will be released on the repository \texttt{Zenodo} upon publication of this preprint.

\bibliographystyle{apsrev4-1}
\bibliography{bibliography}
\begin{thebibliography}{78}%
\makeatletter
\providecommand \@ifxundefined [1]{%
 \@ifx{#1\undefined}
}%
\providecommand \@ifnum [1]{%
 \ifnum #1\expandafter \@firstoftwo
 \else \expandafter \@secondoftwo
 \fi
}%
\providecommand \@ifx [1]{%
 \ifx #1\expandafter \@firstoftwo
 \else \expandafter \@secondoftwo
 \fi
}%
\providecommand \natexlab [1]{#1}%
\providecommand \enquote  [1]{``#1''}%
\providecommand \bibnamefont  [1]{#1}%
\providecommand \bibfnamefont [1]{#1}%
\providecommand \citenamefont [1]{#1}%
\providecommand \href@noop [0]{\@secondoftwo}%
\providecommand \href [0]{\begingroup \@sanitize@url \@href}%
\providecommand \@href[1]{\@@startlink{#1}\@@href}%
\providecommand \@@href[1]{\endgroup#1\@@endlink}%
\providecommand \@sanitize@url [0]{\catcode `\\12\catcode `\$12\catcode
  `\&12\catcode `\#12\catcode `\^12\catcode `\_12\catcode `\%12\relax}%
\providecommand \@@startlink[1]{}%
\providecommand \@@endlink[0]{}%
\providecommand \url  [0]{\begingroup\@sanitize@url \@url }%
\providecommand \@url [1]{\endgroup\@href {#1}{\urlprefix }}%
\providecommand \urlprefix  [0]{URL }%
\providecommand \Eprint [0]{\href }%
\providecommand \doibase [0]{http://dx.doi.org/}%
\providecommand \selectlanguage [0]{\@gobble}%
\providecommand \bibinfo  [0]{\@secondoftwo}%
\providecommand \bibfield  [0]{\@secondoftwo}%
\providecommand \translation [1]{[#1]}%
\providecommand \BibitemOpen [0]{}%
\providecommand \bibitemStop [0]{}%
\providecommand \bibitemNoStop [0]{.\EOS\space}%
\providecommand \EOS [0]{\spacefactor3000\relax}%
\providecommand \BibitemShut  [1]{\csname bibitem#1\endcsname}%
\let\auto@bib@innerbib\@empty
\bibitem [{\citenamefont {Thomson}\ \emph {et~al.}(2016)\citenamefont
  {Thomson}, \citenamefont {Zilkie}, \citenamefont {Bowers}, \citenamefont
  {Komljenovic}, \citenamefont {Reed}, \citenamefont {Vivien}, \citenamefont
  {Marris-Morini}, \citenamefont {Cassan}, \citenamefont {Virot}, \citenamefont
  {F{\'{e}}d{\'{e}}li}, \citenamefont {Hartmann}, \citenamefont {Schmid},
  \citenamefont {Xu}, \citenamefont {Boeuf}, \citenamefont {O'Brien},
  \citenamefont {Mashanovich},\ and\ \citenamefont {Nedeljkovic}}]{Thomson:16}%
  \BibitemOpen
  \bibfield  {author} {\bibinfo {author} {\bibfnamefont {D.}~\bibnamefont
  {Thomson}}, \bibinfo {author} {\bibfnamefont {A.}~\bibnamefont {Zilkie}},
  \bibinfo {author} {\bibfnamefont {J.~E.}\ \bibnamefont {Bowers}}, \bibinfo
  {author} {\bibfnamefont {T.}~\bibnamefont {Komljenovic}}, \bibinfo {author}
  {\bibfnamefont {G.~T.}\ \bibnamefont {Reed}}, \bibinfo {author}
  {\bibfnamefont {L.}~\bibnamefont {Vivien}}, \bibinfo {author} {\bibfnamefont
  {D.}~\bibnamefont {Marris-Morini}}, \bibinfo {author} {\bibfnamefont
  {E.}~\bibnamefont {Cassan}}, \bibinfo {author} {\bibfnamefont
  {L.}~\bibnamefont {Virot}}, \bibinfo {author} {\bibfnamefont {J.-M.}\
  \bibnamefont {F{\'{e}}d{\'{e}}li}}, \bibinfo {author} {\bibfnamefont {J.-M.}\
  \bibnamefont {Hartmann}}, \bibinfo {author} {\bibfnamefont {J.~H.}\
  \bibnamefont {Schmid}}, \bibinfo {author} {\bibfnamefont {D.-X.}\
  \bibnamefont {Xu}}, \bibinfo {author} {\bibfnamefont {F.}~\bibnamefont
  {Boeuf}}, \bibinfo {author} {\bibfnamefont {P.}~\bibnamefont {O'Brien}},
  \bibinfo {author} {\bibfnamefont {G.~Z.}\ \bibnamefont {Mashanovich}}, \ and\
  \bibinfo {author} {\bibfnamefont {M.}~\bibnamefont {Nedeljkovic}},\ }\href
  {\doibase 10.1088/2040-8978/18/7/073003} {\bibfield  {journal} {\bibinfo
  {journal} {Journal of Optics}\ }\textbf {\bibinfo {volume} {18}},\ \bibinfo
  {pages} {073003} (\bibinfo {year} {2016})}\BibitemShut {NoStop}%
\bibitem [{\citenamefont {Agrell}\ \emph {et~al.}(2016)\citenamefont {Agrell},
  \citenamefont {Karlsson}, \citenamefont {Chraplyvy}, \citenamefont
  {Richardson}, \citenamefont {Krummrich}, \citenamefont {Winzer},
  \citenamefont {Roberts}, \citenamefont {Fischer}, \citenamefont {Savory},
  \citenamefont {Eggleton}, \citenamefont {Secondini}, \citenamefont
  {Kschischang}, \citenamefont {Lord}, \citenamefont {Prat}, \citenamefont
  {Tomkos}, \citenamefont {Bowers}, \citenamefont {Srinivasan}, \citenamefont
  {Brandt-Pearce},\ and\ \citenamefont {Gisin}}]{Agrell:16}%
  \BibitemOpen
  \bibfield  {author} {\bibinfo {author} {\bibfnamefont {E.}~\bibnamefont
  {Agrell}}, \bibinfo {author} {\bibfnamefont {M.}~\bibnamefont {Karlsson}},
  \bibinfo {author} {\bibfnamefont {A.~R.}\ \bibnamefont {Chraplyvy}}, \bibinfo
  {author} {\bibfnamefont {D.~J.}\ \bibnamefont {Richardson}}, \bibinfo
  {author} {\bibfnamefont {P.~M.}\ \bibnamefont {Krummrich}}, \bibinfo {author}
  {\bibfnamefont {P.}~\bibnamefont {Winzer}}, \bibinfo {author} {\bibfnamefont
  {K.}~\bibnamefont {Roberts}}, \bibinfo {author} {\bibfnamefont {J.~K.}\
  \bibnamefont {Fischer}}, \bibinfo {author} {\bibfnamefont {S.~J.}\
  \bibnamefont {Savory}}, \bibinfo {author} {\bibfnamefont {B.~J.}\
  \bibnamefont {Eggleton}}, \bibinfo {author} {\bibfnamefont {M.}~\bibnamefont
  {Secondini}}, \bibinfo {author} {\bibfnamefont {F.~R.}\ \bibnamefont
  {Kschischang}}, \bibinfo {author} {\bibfnamefont {A.}~\bibnamefont {Lord}},
  \bibinfo {author} {\bibfnamefont {J.}~\bibnamefont {Prat}}, \bibinfo {author}
  {\bibfnamefont {I.}~\bibnamefont {Tomkos}}, \bibinfo {author} {\bibfnamefont
  {J.~E.}\ \bibnamefont {Bowers}}, \bibinfo {author} {\bibfnamefont
  {S.}~\bibnamefont {Srinivasan}}, \bibinfo {author} {\bibfnamefont
  {M.}~\bibnamefont {Brandt-Pearce}}, \ and\ \bibinfo {author} {\bibfnamefont
  {N.}~\bibnamefont {Gisin}},\ }\href {\doibase 10.1088/2040-8978/18/6/063002}
  {\bibfield  {journal} {\bibinfo  {journal} {Journal of Optics}\ }\textbf
  {\bibinfo {volume} {18}},\ \bibinfo {pages} {063002} (\bibinfo {year}
  {2016})}\BibitemShut {NoStop}%
\bibitem [{\citenamefont {Krainak}\ \emph {et~al.}(2019)\citenamefont
  {Krainak}, \citenamefont {Stephen}, \citenamefont {Troupaki}, \citenamefont
  {Tedder}, \citenamefont {Reyna}, \citenamefont {Klamkin}, \citenamefont
  {Zhao}, \citenamefont {Song}, \citenamefont {Fridlander}, \citenamefont
  {Tran}, \citenamefont {Bowers}, \citenamefont {Bergman}, \citenamefont
  {Lipson}, \citenamefont {Rizzo}, \citenamefont {Datta}, \citenamefont
  {Abrams}, \citenamefont {Mookherjea}, \citenamefont {Ho}, \citenamefont
  {Bei}, \citenamefont {Huang}, \citenamefont {Tu}, \citenamefont {Moslehi},
  \citenamefont {Harris}, \citenamefont {Matsko}, \citenamefont {Savchenkov},
  \citenamefont {Liu}, \citenamefont {Proietti}, \citenamefont {Yoo},
  \citenamefont {Johansson}, \citenamefont {Dorrer}, \citenamefont
  {Arteaga-Sierra}, \citenamefont {Qiao}, \citenamefont {Gong}, \citenamefont
  {Gu}, \citenamefont {III}, \citenamefont {Ni}, \citenamefont {Ding},
  \citenamefont {Duan}, \citenamefont {Dalir}, \citenamefont {Chen},
  \citenamefont {Sorger},\ and\ \citenamefont {Komljenovic}}]{Krainak:19}%
  \BibitemOpen
  \bibfield  {author} {\bibinfo {author} {\bibfnamefont {M.}~\bibnamefont
  {Krainak}}, \bibinfo {author} {\bibfnamefont {M.}~\bibnamefont {Stephen}},
  \bibinfo {author} {\bibfnamefont {E.}~\bibnamefont {Troupaki}}, \bibinfo
  {author} {\bibfnamefont {S.}~\bibnamefont {Tedder}}, \bibinfo {author}
  {\bibfnamefont {B.}~\bibnamefont {Reyna}}, \bibinfo {author} {\bibfnamefont
  {J.}~\bibnamefont {Klamkin}}, \bibinfo {author} {\bibfnamefont
  {H.}~\bibnamefont {Zhao}}, \bibinfo {author} {\bibfnamefont {B.}~\bibnamefont
  {Song}}, \bibinfo {author} {\bibfnamefont {J.}~\bibnamefont {Fridlander}},
  \bibinfo {author} {\bibfnamefont {M.}~\bibnamefont {Tran}}, \bibinfo {author}
  {\bibfnamefont {J.~E.}\ \bibnamefont {Bowers}}, \bibinfo {author}
  {\bibfnamefont {K.}~\bibnamefont {Bergman}}, \bibinfo {author} {\bibfnamefont
  {M.}~\bibnamefont {Lipson}}, \bibinfo {author} {\bibfnamefont
  {A.}~\bibnamefont {Rizzo}}, \bibinfo {author} {\bibfnamefont
  {I.}~\bibnamefont {Datta}}, \bibinfo {author} {\bibfnamefont
  {N.}~\bibnamefont {Abrams}}, \bibinfo {author} {\bibfnamefont
  {S.}~\bibnamefont {Mookherjea}}, \bibinfo {author} {\bibfnamefont {S.-T.}\
  \bibnamefont {Ho}}, \bibinfo {author} {\bibfnamefont {Q.}~\bibnamefont
  {Bei}}, \bibinfo {author} {\bibfnamefont {Y.}~\bibnamefont {Huang}}, \bibinfo
  {author} {\bibfnamefont {Y.}~\bibnamefont {Tu}}, \bibinfo {author}
  {\bibfnamefont {B.}~\bibnamefont {Moslehi}}, \bibinfo {author} {\bibfnamefont
  {J.}~\bibnamefont {Harris}}, \bibinfo {author} {\bibfnamefont
  {A.}~\bibnamefont {Matsko}}, \bibinfo {author} {\bibfnamefont
  {A.}~\bibnamefont {Savchenkov}}, \bibinfo {author} {\bibfnamefont
  {G.}~\bibnamefont {Liu}}, \bibinfo {author} {\bibfnamefont {R.}~\bibnamefont
  {Proietti}}, \bibinfo {author} {\bibfnamefont {S.~J.~B.}\ \bibnamefont
  {Yoo}}, \bibinfo {author} {\bibfnamefont {L.}~\bibnamefont {Johansson}},
  \bibinfo {author} {\bibfnamefont {C.}~\bibnamefont {Dorrer}}, \bibinfo
  {author} {\bibfnamefont {F.~R.}\ \bibnamefont {Arteaga-Sierra}}, \bibinfo
  {author} {\bibfnamefont {J.}~\bibnamefont {Qiao}}, \bibinfo {author}
  {\bibfnamefont {S.}~\bibnamefont {Gong}}, \bibinfo {author} {\bibfnamefont
  {T.}~\bibnamefont {Gu}}, \bibinfo {author} {\bibfnamefont {O.~J.~O.}\
  \bibnamefont {III}}, \bibinfo {author} {\bibfnamefont {X.}~\bibnamefont
  {Ni}}, \bibinfo {author} {\bibfnamefont {Y.}~\bibnamefont {Ding}}, \bibinfo
  {author} {\bibfnamefont {Y.}~\bibnamefont {Duan}}, \bibinfo {author}
  {\bibfnamefont {H.}~\bibnamefont {Dalir}}, \bibinfo {author} {\bibfnamefont
  {R.~T.}\ \bibnamefont {Chen}}, \bibinfo {author} {\bibfnamefont {V.~J.}\
  \bibnamefont {Sorger}}, \ and\ \bibinfo {author} {\bibfnamefont
  {T.}~\bibnamefont {Komljenovic}},\ }in\ \href {\doibase 10.1117/12.2509808}
  {\emph {\bibinfo {booktitle} {Components and Packaging for Laser Systems
  V}}},\ Vol.\ \bibinfo {volume} {10899},\ \bibinfo {editor} {edited by\
  \bibinfo {editor} {\bibfnamefont {A.~L.}\ \bibnamefont {Glebov}}\ and\
  \bibinfo {editor} {\bibfnamefont {P.~O.}\ \bibnamefont {Leisher}}},\ \bibinfo
  {organization} {International Society for Optics and Photonics}\ (\bibinfo
  {publisher} {SPIE},\ \bibinfo {year} {2019})\ p.\ \bibinfo {pages}
  {108990F}\BibitemShut {NoStop}%
\bibitem [{\citenamefont {Roth}\ \emph {et~al.}(2023)\citenamefont {Roth},
  \citenamefont {Madhav}, \citenamefont {Stoll}, \citenamefont
  {Bodenm{\"u}ller}, \citenamefont {Dinkelaker}, \citenamefont {Rahman},
  \citenamefont {Hernandez}, \citenamefont {G{\"u}nther},\ and\ \citenamefont
  {Vjesnica}}]{Roth:23}%
  \BibitemOpen
  \bibfield  {author} {\bibinfo {author} {\bibfnamefont {M.~M.}\ \bibnamefont
  {Roth}}, \bibinfo {author} {\bibfnamefont {K.}~\bibnamefont {Madhav}},
  \bibinfo {author} {\bibfnamefont {A.}~\bibnamefont {Stoll}}, \bibinfo
  {author} {\bibfnamefont {D.}~\bibnamefont {Bodenm{\"u}ller}}, \bibinfo
  {author} {\bibfnamefont {A.~N.}\ \bibnamefont {Dinkelaker}}, \bibinfo
  {author} {\bibfnamefont {A.}~\bibnamefont {Rahman}}, \bibinfo {author}
  {\bibfnamefont {E.}~\bibnamefont {Hernandez}}, \bibinfo {author}
  {\bibfnamefont {A.}~\bibnamefont {G{\"u}nther}}, \ and\ \bibinfo {author}
  {\bibfnamefont {S.}~\bibnamefont {Vjesnica}},\ }in\ \href {\doibase
  10.1117/12.2655630} {\emph {\bibinfo {booktitle} {Integrated Optics: Devices,
  Materials, and Technologies XXVII}}},\ Vol.\ \bibinfo {volume} {12424},\
  \bibinfo {editor} {edited by\ \bibinfo {editor} {\bibfnamefont {S.~M.}\
  \bibnamefont {Garc{\'\i}a-Blanco}}\ and\ \bibinfo {editor} {\bibfnamefont
  {P.}~\bibnamefont {Cheben}}},\ \bibinfo {organization} {International Society
  for Optics and Photonics}\ (\bibinfo  {publisher} {SPIE},\ \bibinfo {year}
  {2023})\ p.\ \bibinfo {pages} {124240B}\BibitemShut {NoStop}%
\bibitem [{\citenamefont {Cvetojevic}\ \emph {et~al.}(2009)\citenamefont
  {Cvetojevic}, \citenamefont {Lawrence}, \citenamefont {Ellis}, \citenamefont
  {Bland-Hawthorn}, \citenamefont {Haynes},\ and\ \citenamefont
  {Horton}}]{Cvetojevic:09}%
  \BibitemOpen
  \bibfield  {author} {\bibinfo {author} {\bibfnamefont {N.}~\bibnamefont
  {Cvetojevic}}, \bibinfo {author} {\bibfnamefont {J.~S.}\ \bibnamefont
  {Lawrence}}, \bibinfo {author} {\bibfnamefont {S.~C.}\ \bibnamefont {Ellis}},
  \bibinfo {author} {\bibfnamefont {J.}~\bibnamefont {Bland-Hawthorn}},
  \bibinfo {author} {\bibfnamefont {R.}~\bibnamefont {Haynes}}, \ and\ \bibinfo
  {author} {\bibfnamefont {A.}~\bibnamefont {Horton}},\ }\href {\doibase
  10.1364/OE.17.018643} {\bibfield  {journal} {\bibinfo  {journal} {Opt.
  Express}\ }\textbf {\bibinfo {volume} {17}},\ \bibinfo {pages} {18643}
  (\bibinfo {year} {2009})}\BibitemShut {NoStop}%
\bibitem [{\citenamefont {Gatkine}\ \emph {et~al.}(2021)\citenamefont
  {Gatkine}, \citenamefont {Jovanovic}, \citenamefont {Hopgood}, \citenamefont
  {Ellis}, \citenamefont {Broeke}, \citenamefont {{\L}awniczuk}, \citenamefont
  {Jewell}, \citenamefont {Wallace},\ and\ \citenamefont {Mawet}}]{Gatkine:21}%
  \BibitemOpen
  \bibfield  {author} {\bibinfo {author} {\bibfnamefont {P.}~\bibnamefont
  {Gatkine}}, \bibinfo {author} {\bibfnamefont {N.}~\bibnamefont {Jovanovic}},
  \bibinfo {author} {\bibfnamefont {C.}~\bibnamefont {Hopgood}}, \bibinfo
  {author} {\bibfnamefont {S.}~\bibnamefont {Ellis}}, \bibinfo {author}
  {\bibfnamefont {R.}~\bibnamefont {Broeke}}, \bibinfo {author} {\bibfnamefont
  {K.}~\bibnamefont {{\L}awniczuk}}, \bibinfo {author} {\bibfnamefont
  {J.}~\bibnamefont {Jewell}}, \bibinfo {author} {\bibfnamefont {J.~K.}\
  \bibnamefont {Wallace}}, \ and\ \bibinfo {author} {\bibfnamefont
  {D.}~\bibnamefont {Mawet}},\ }\href {\doibase 10.1364/AO.423439} {\bibfield
  {journal} {\bibinfo  {journal} {Appl. Opt.}\ }\textbf {\bibinfo {volume}
  {60}},\ \bibinfo {pages} {D15} (\bibinfo {year} {2021})}\BibitemShut
  {NoStop}%
\bibitem [{\citenamefont {Obrzud}\ \emph {et~al.}(2019)\citenamefont {Obrzud},
  \citenamefont {Rainer}, \citenamefont {Harutyunyan}, \citenamefont
  {Anderson}, \citenamefont {Liu}, \citenamefont {Geiselmann}, \citenamefont
  {Chazelas}, \citenamefont {Kundermann}, \citenamefont {Lecomte},
  \citenamefont {Cecconi}, \citenamefont {Ghedina}, \citenamefont {Molinari},
  \citenamefont {Pepe}, \citenamefont {Wildi}, \citenamefont {Bouchy},
  \citenamefont {Kippenberg},\ and\ \citenamefont {Herr}}]{Obrzud:19}%
  \BibitemOpen
  \bibfield  {author} {\bibinfo {author} {\bibfnamefont {E.}~\bibnamefont
  {Obrzud}}, \bibinfo {author} {\bibfnamefont {M.}~\bibnamefont {Rainer}},
  \bibinfo {author} {\bibfnamefont {A.}~\bibnamefont {Harutyunyan}}, \bibinfo
  {author} {\bibfnamefont {M.~H.}\ \bibnamefont {Anderson}}, \bibinfo {author}
  {\bibfnamefont {J.}~\bibnamefont {Liu}}, \bibinfo {author} {\bibfnamefont
  {M.}~\bibnamefont {Geiselmann}}, \bibinfo {author} {\bibfnamefont
  {B.}~\bibnamefont {Chazelas}}, \bibinfo {author} {\bibfnamefont
  {S.}~\bibnamefont {Kundermann}}, \bibinfo {author} {\bibfnamefont
  {S.}~\bibnamefont {Lecomte}}, \bibinfo {author} {\bibfnamefont
  {M.}~\bibnamefont {Cecconi}}, \bibinfo {author} {\bibfnamefont
  {A.}~\bibnamefont {Ghedina}}, \bibinfo {author} {\bibfnamefont
  {E.}~\bibnamefont {Molinari}}, \bibinfo {author} {\bibfnamefont
  {F.}~\bibnamefont {Pepe}}, \bibinfo {author} {\bibfnamefont {F.}~\bibnamefont
  {Wildi}}, \bibinfo {author} {\bibfnamefont {F.}~\bibnamefont {Bouchy}},
  \bibinfo {author} {\bibfnamefont {T.~J.}\ \bibnamefont {Kippenberg}}, \ and\
  \bibinfo {author} {\bibfnamefont {T.}~\bibnamefont {Herr}},\ }\href {\doibase
  10.1038/s41566-018-0309-y} {\bibfield  {journal} {\bibinfo  {journal} {Nature
  Photonics}\ }\textbf {\bibinfo {volume} {13}},\ \bibinfo {pages} {31}
  (\bibinfo {year} {2019})}\BibitemShut {NoStop}%
\bibitem [{\citenamefont {Suh}\ \emph {et~al.}(2019)\citenamefont {Suh},
  \citenamefont {Yi}, \citenamefont {Lai}, \citenamefont {Leifer},
  \citenamefont {Grudinin}, \citenamefont {Vasisht}, \citenamefont {Martin},
  \citenamefont {Fitzgerald}, \citenamefont {Doppmann}, \citenamefont {Wang},
  \citenamefont {Mawet}, \citenamefont {Papp}, \citenamefont {Diddams},
  \citenamefont {Beichman},\ and\ \citenamefont {Vahala}}]{Suh:19}%
  \BibitemOpen
  \bibfield  {author} {\bibinfo {author} {\bibfnamefont {M.-G.}\ \bibnamefont
  {Suh}}, \bibinfo {author} {\bibfnamefont {X.}~\bibnamefont {Yi}}, \bibinfo
  {author} {\bibfnamefont {Y.-H.}\ \bibnamefont {Lai}}, \bibinfo {author}
  {\bibfnamefont {S.}~\bibnamefont {Leifer}}, \bibinfo {author} {\bibfnamefont
  {I.~S.}\ \bibnamefont {Grudinin}}, \bibinfo {author} {\bibfnamefont
  {G.}~\bibnamefont {Vasisht}}, \bibinfo {author} {\bibfnamefont {E.~C.}\
  \bibnamefont {Martin}}, \bibinfo {author} {\bibfnamefont {M.~P.}\
  \bibnamefont {Fitzgerald}}, \bibinfo {author} {\bibfnamefont
  {G.}~\bibnamefont {Doppmann}}, \bibinfo {author} {\bibfnamefont
  {J.}~\bibnamefont {Wang}}, \bibinfo {author} {\bibfnamefont {D.}~\bibnamefont
  {Mawet}}, \bibinfo {author} {\bibfnamefont {S.~B.}\ \bibnamefont {Papp}},
  \bibinfo {author} {\bibfnamefont {S.~A.}\ \bibnamefont {Diddams}}, \bibinfo
  {author} {\bibfnamefont {C.}~\bibnamefont {Beichman}}, \ and\ \bibinfo
  {author} {\bibfnamefont {K.}~\bibnamefont {Vahala}},\ }\href {\doibase
  10.1038/s41566-018-0312-3} {\bibfield  {journal} {\bibinfo  {journal} {Nature
  Photonics}\ }\textbf {\bibinfo {volume} {13}},\ \bibinfo {pages} {25}
  (\bibinfo {year} {2019})}\BibitemShut {NoStop}%
\bibitem [{\citenamefont {Fridlander}\ \emph {et~al.}(2018)\citenamefont
  {Fridlander}, \citenamefont {Pinna}, \citenamefont {Rosborough},
  \citenamefont {Estrella}, \citenamefont {Johansson},\ and\ \citenamefont
  {Klamkin}}]{Fridlander:18}%
  \BibitemOpen
  \bibfield  {author} {\bibinfo {author} {\bibfnamefont {J.}~\bibnamefont
  {Fridlander}}, \bibinfo {author} {\bibfnamefont {S.}~\bibnamefont {Pinna}},
  \bibinfo {author} {\bibfnamefont {V.}~\bibnamefont {Rosborough}}, \bibinfo
  {author} {\bibfnamefont {S.}~\bibnamefont {Estrella}}, \bibinfo {author}
  {\bibfnamefont {L.}~\bibnamefont {Johansson}}, \ and\ \bibinfo {author}
  {\bibfnamefont {J.}~\bibnamefont {Klamkin}},\ }in\ \href {\doibase
  10.1117/12.2287557} {\emph {\bibinfo {booktitle} {Free-Space Laser
  Communication and Atmospheric Propagation XXX}}},\ Vol.\ \bibinfo {volume}
  {10524},\ \bibinfo {editor} {edited by\ \bibinfo {editor} {\bibfnamefont
  {H.}~\bibnamefont {Hemmati}}\ and\ \bibinfo {editor} {\bibfnamefont {D.~M.}\
  \bibnamefont {Boroson}}},\ \bibinfo {organization} {International Society for
  Optics and Photonics}\ (\bibinfo  {publisher} {SPIE},\ \bibinfo {year}
  {2018})\ p.\ \bibinfo {pages} {105240Y}\BibitemShut {NoStop}%
\bibitem [{\citenamefont {He}\ \emph {et~al.}(2020)\citenamefont {He},
  \citenamefont {Dong},\ and\ \citenamefont {Xu}}]{He:20}%
  \BibitemOpen
  \bibfield  {author} {\bibinfo {author} {\bibfnamefont {J.}~\bibnamefont
  {He}}, \bibinfo {author} {\bibfnamefont {T.}~\bibnamefont {Dong}}, \ and\
  \bibinfo {author} {\bibfnamefont {Y.}~\bibnamefont {Xu}},\ }\href {\doibase
  10.1109/ACCESS.2020.3030627} {\bibfield  {journal} {\bibinfo  {journal} {IEEE
  Access}\ }\textbf {\bibinfo {volume} {8}},\ \bibinfo {pages} {188284}
  (\bibinfo {year} {2020})}\BibitemShut {NoStop}%
\bibitem [{\citenamefont {Ohanian}\ \emph {et~al.}(2018)\citenamefont
  {Ohanian}, \citenamefont {Yakusheva}, \citenamefont {Kreger}, \citenamefont
  {Kominsky}, \citenamefont {Soller}, \citenamefont {Tran}, \citenamefont
  {Komljenovic},\ and\ \citenamefont {Bowers}}]{Ohanian:18}%
  \BibitemOpen
  \bibfield  {author} {\bibinfo {author} {\bibfnamefont {O.~J.}\ \bibnamefont
  {Ohanian}}, \bibinfo {author} {\bibfnamefont {A.~A.}\ \bibnamefont
  {Yakusheva}}, \bibinfo {author} {\bibfnamefont {S.~T.}\ \bibnamefont
  {Kreger}}, \bibinfo {author} {\bibfnamefont {D.}~\bibnamefont {Kominsky}},
  \bibinfo {author} {\bibfnamefont {B.~J.}\ \bibnamefont {Soller}}, \bibinfo
  {author} {\bibfnamefont {M.}~\bibnamefont {Tran}}, \bibinfo {author}
  {\bibfnamefont {T.}~\bibnamefont {Komljenovic}}, \ and\ \bibinfo {author}
  {\bibfnamefont {J.~E.}\ \bibnamefont {Bowers}},\ }in\ \href {\doibase
  10.1364/OFS.2018.WB1} {\emph {\bibinfo {booktitle} {26th International
  Conference on Optical Fiber Sensors}}}\ (\bibinfo  {publisher} {Optica
  Publishing Group},\ \bibinfo {year} {2018})\ p.\ \bibinfo {pages}
  {WB1}\BibitemShut {NoStop}%
\bibitem [{\citenamefont {Ziarko}\ \emph {et~al.}(2024)\citenamefont {Ziarko},
  \citenamefont {Terrasanta}, \citenamefont {Bergamasco},\ and\ \citenamefont
  {Poliak}}]{Ziarko:24}%
  \BibitemOpen
  \bibfield  {author} {\bibinfo {author} {\bibfnamefont {M.}~\bibnamefont
  {Ziarko}}, \bibinfo {author} {\bibfnamefont {G.}~\bibnamefont {Terrasanta}},
  \bibinfo {author} {\bibfnamefont {N.}~\bibnamefont {Bergamasco}}, \ and\
  \bibinfo {author} {\bibfnamefont {J.}~\bibnamefont {Poliak}},\ }in\ \href
  {\doibase 10.1117/12.3001792} {\emph {\bibinfo {booktitle} {Free-Space Laser
  Communications XXXVI}}},\ Vol.\ \bibinfo {volume} {12877},\ \bibinfo {editor}
  {edited by\ \bibinfo {editor} {\bibfnamefont {H.}~\bibnamefont {Hemmati}}\
  and\ \bibinfo {editor} {\bibfnamefont {B.~S.}\ \bibnamefont {Robinson}}},\
  \bibinfo {organization} {International Society for Optics and Photonics}\
  (\bibinfo  {publisher} {SPIE},\ \bibinfo {year} {2024})\ p.\ \bibinfo {pages}
  {128770G}\BibitemShut {NoStop}%
\bibitem [{\citenamefont {Barth}(2003)}]{Barth:04}%
  \BibitemOpen
  \bibfield  {author} {\bibinfo {author} {\bibfnamefont {J.~L.}\ \bibnamefont
  {Barth}},\ }in\ \href {\doibase 10.1007/1-4020-2595-5_2} {\emph {\bibinfo
  {booktitle} {Protection of Materials and Structures from Space
  Environment}}},\ \bibinfo {editor} {edited by\ \bibinfo {editor}
  {\bibfnamefont {J.~I.}\ \bibnamefont {Kleiman}}\ and\ \bibinfo {editor}
  {\bibfnamefont {Z.}~\bibnamefont {Iskanderova}}}\ (\bibinfo  {publisher}
  {Springer Netherlands},\ \bibinfo {address} {Dordrecht},\ \bibinfo {year}
  {2003})\ pp.\ \bibinfo {pages} {7--29}\BibitemShut {NoStop}%
\bibitem [{\citenamefont {Du}(2023)}]{Du:23}%
  \BibitemOpen
  \bibfield  {author} {\bibinfo {author} {\bibfnamefont {Q.}~\bibnamefont
  {Du}},\ }\href {\doibase 10.1364/OME.476935} {\bibfield  {journal} {\bibinfo
  {journal} {Opt. Mater. Express}\ }\textbf {\bibinfo {volume} {13}},\ \bibinfo
  {pages} {403} (\bibinfo {year} {2023})}\BibitemShut {NoStop}%
\bibitem [{\citenamefont {Bhandaru}\ \emph {et~al.}(2015)\citenamefont
  {Bhandaru}, \citenamefont {Hu}, \citenamefont {Fleetwood},\ and\
  \citenamefont {Weiss}}]{Bhandaru:15}%
  \BibitemOpen
  \bibfield  {author} {\bibinfo {author} {\bibfnamefont {S.}~\bibnamefont
  {Bhandaru}}, \bibinfo {author} {\bibfnamefont {S.}~\bibnamefont {Hu}},
  \bibinfo {author} {\bibfnamefont {D.~M.}\ \bibnamefont {Fleetwood}}, \ and\
  \bibinfo {author} {\bibfnamefont {S.~M.}\ \bibnamefont {Weiss}},\ }\href
  {\doibase 10.1109/TNS.2014.2387772} {\bibfield  {journal} {\bibinfo
  {journal} {IEEE Transactions on Nuclear Science}\ }\textbf {\bibinfo {volume}
  {62}},\ \bibinfo {pages} {323} (\bibinfo {year} {2015})}\BibitemShut
  {NoStop}%
\bibitem [{\citenamefont {Grillanda}\ \emph {et~al.}(2016)\citenamefont
  {Grillanda}, \citenamefont {Singh}, \citenamefont {Raghunathan},
  \citenamefont {Morichetti}, \citenamefont {Melloni}, \citenamefont
  {Kimerling},\ and\ \citenamefont {Agarwal}}]{Grillanda:16}%
  \BibitemOpen
  \bibfield  {author} {\bibinfo {author} {\bibfnamefont {S.}~\bibnamefont
  {Grillanda}}, \bibinfo {author} {\bibfnamefont {V.}~\bibnamefont {Singh}},
  \bibinfo {author} {\bibfnamefont {V.}~\bibnamefont {Raghunathan}}, \bibinfo
  {author} {\bibfnamefont {F.}~\bibnamefont {Morichetti}}, \bibinfo {author}
  {\bibfnamefont {A.}~\bibnamefont {Melloni}}, \bibinfo {author} {\bibfnamefont
  {L.}~\bibnamefont {Kimerling}}, \ and\ \bibinfo {author} {\bibfnamefont
  {A.~M.}\ \bibnamefont {Agarwal}},\ }\href {\doibase 10.1364/OL.41.003053}
  {\bibfield  {journal} {\bibinfo  {journal} {Opt. Lett.}\ }\textbf {\bibinfo
  {volume} {41}},\ \bibinfo {pages} {3053} (\bibinfo {year}
  {2016})}\BibitemShut {NoStop}%
\bibitem [{\citenamefont {Du}\ \emph {et~al.}(2017)\citenamefont {Du},
  \citenamefont {Huang}, \citenamefont {Ogbuu}, \citenamefont {Zhang},
  \citenamefont {Li}, \citenamefont {Singh}, \citenamefont {Agarwal},\ and\
  \citenamefont {Hu}}]{Du:17}%
  \BibitemOpen
  \bibfield  {author} {\bibinfo {author} {\bibfnamefont {Q.}~\bibnamefont
  {Du}}, \bibinfo {author} {\bibfnamefont {Y.}~\bibnamefont {Huang}}, \bibinfo
  {author} {\bibfnamefont {O.}~\bibnamefont {Ogbuu}}, \bibinfo {author}
  {\bibfnamefont {W.}~\bibnamefont {Zhang}}, \bibinfo {author} {\bibfnamefont
  {J.}~\bibnamefont {Li}}, \bibinfo {author} {\bibfnamefont {V.}~\bibnamefont
  {Singh}}, \bibinfo {author} {\bibfnamefont {A.~M.}\ \bibnamefont {Agarwal}},
  \ and\ \bibinfo {author} {\bibfnamefont {J.}~\bibnamefont {Hu}},\ }\href
  {\doibase 10.1364/OL.42.000587} {\bibfield  {journal} {\bibinfo  {journal}
  {Opt. Lett.}\ }\textbf {\bibinfo {volume} {42}},\ \bibinfo {pages} {587}
  (\bibinfo {year} {2017})}\BibitemShut {NoStop}%
\bibitem [{\citenamefont {Reghioua}\ \emph {et~al.}(2023)\citenamefont
  {Reghioua}, \citenamefont {Girard}, \citenamefont {Morana}, \citenamefont
  {Lambert}, \citenamefont {Faugier-Tovar}, \citenamefont {Grosse},
  \citenamefont {Garcia}, \citenamefont {Kazar-Mendes}, \citenamefont
  {Ferrari}, \citenamefont {Paillet},\ and\ \citenamefont
  {Szelag}}]{Reghioua:23}%
  \BibitemOpen
  \bibfield  {author} {\bibinfo {author} {\bibfnamefont {I.}~\bibnamefont
  {Reghioua}}, \bibinfo {author} {\bibfnamefont {S.}~\bibnamefont {Girard}},
  \bibinfo {author} {\bibfnamefont {A.}~\bibnamefont {Morana}}, \bibinfo
  {author} {\bibfnamefont {D.}~\bibnamefont {Lambert}}, \bibinfo {author}
  {\bibfnamefont {J.}~\bibnamefont {Faugier-Tovar}}, \bibinfo {author}
  {\bibfnamefont {P.}~\bibnamefont {Grosse}}, \bibinfo {author} {\bibfnamefont
  {S.}~\bibnamefont {Garcia}}, \bibinfo {author} {\bibfnamefont
  {M.}~\bibnamefont {Kazar-Mendes}}, \bibinfo {author} {\bibfnamefont
  {M.}~\bibnamefont {Ferrari}}, \bibinfo {author} {\bibfnamefont
  {P.}~\bibnamefont {Paillet}}, \ and\ \bibinfo {author} {\bibfnamefont
  {B.}~\bibnamefont {Szelag}},\ }\href {\doibase 10.1109/TNS.2023.3296985}
  {\bibfield  {journal} {\bibinfo  {journal} {IEEE Transactions on Nuclear
  Science}\ }\textbf {\bibinfo {volume} {70}},\ \bibinfo {pages} {1973}
  (\bibinfo {year} {2023})}\BibitemShut {NoStop}%
\bibitem [{\citenamefont {Zhou}\ \emph
  {et~al.}(2022{\natexlab{a}})\citenamefont {Zhou}, \citenamefont {Bi},
  \citenamefont {Wang}, \citenamefont {Wu}, \citenamefont {Huang},
  \citenamefont {Zhang}, \citenamefont {Fleetwood},\ and\ \citenamefont
  {Wu}}]{Zhou:22a}%
  \BibitemOpen
  \bibfield  {author} {\bibinfo {author} {\bibfnamefont {Y.}~\bibnamefont
  {Zhou}}, \bibinfo {author} {\bibfnamefont {D.}~\bibnamefont {Bi}}, \bibinfo
  {author} {\bibfnamefont {S.}~\bibnamefont {Wang}}, \bibinfo {author}
  {\bibfnamefont {L.}~\bibnamefont {Wu}}, \bibinfo {author} {\bibfnamefont
  {Y.}~\bibnamefont {Huang}}, \bibinfo {author} {\bibfnamefont
  {E.}~\bibnamefont {Zhang}}, \bibinfo {author} {\bibfnamefont {D.~M.}\
  \bibnamefont {Fleetwood}}, \ and\ \bibinfo {author} {\bibfnamefont
  {A.}~\bibnamefont {Wu}},\ }\href {\doibase 10.1364/OE.447160} {\bibfield
  {journal} {\bibinfo  {journal} {Opt. Express}\ }\textbf {\bibinfo {volume}
  {30}},\ \bibinfo {pages} {4017} (\bibinfo {year}
  {2022}{\natexlab{a}})}\BibitemShut {NoStop}%
\bibitem [{\citenamefont {Zhou}\ \emph
  {et~al.}(2022{\natexlab{b}})\citenamefont {Zhou}, \citenamefont {Lv},
  \citenamefont {Bi}, \citenamefont {Wu}, \citenamefont {Wang}, \citenamefont
  {Ma}, \citenamefont {Zhang}, \citenamefont {Fleetwood},\ and\ \citenamefont
  {Wu}}]{Zhou:22b}%
  \BibitemOpen
  \bibfield  {author} {\bibinfo {author} {\bibfnamefont {Y.}~\bibnamefont
  {Zhou}}, \bibinfo {author} {\bibfnamefont {D.}~\bibnamefont {Lv}}, \bibinfo
  {author} {\bibfnamefont {D.}~\bibnamefont {Bi}}, \bibinfo {author}
  {\bibfnamefont {L.}~\bibnamefont {Wu}}, \bibinfo {author} {\bibfnamefont
  {R.}~\bibnamefont {Wang}}, \bibinfo {author} {\bibfnamefont {S.}~\bibnamefont
  {Ma}}, \bibinfo {author} {\bibfnamefont {E.~X.}\ \bibnamefont {Zhang}},
  \bibinfo {author} {\bibfnamefont {D.~M.}\ \bibnamefont {Fleetwood}}, \ and\
  \bibinfo {author} {\bibfnamefont {A.}~\bibnamefont {Wu}},\ }\href {\doibase
  10.1364/OE.453903} {\bibfield  {journal} {\bibinfo  {journal} {Opt. Express}\
  }\textbf {\bibinfo {volume} {30}},\ \bibinfo {pages} {16921} (\bibinfo {year}
  {2022}{\natexlab{b}})}\BibitemShut {NoStop}%
\bibitem [{\citenamefont {Seif El Nasr-Storey}\ \emph
  {et~al.}(2015)\citenamefont {Seif El Nasr-Storey}, \citenamefont {Boeuf},
  \citenamefont {Baudot}, \citenamefont {Detraz}, \citenamefont {Fedeli},
  \citenamefont {Marris-Morini}, \citenamefont {Olantera}, \citenamefont
  {Pezzullo}, \citenamefont {Sigaud}, \citenamefont {Soos}, \citenamefont
  {Troska}, \citenamefont {Vasey}, \citenamefont {Vivien}, \citenamefont
  {Zeiler},\ and\ \citenamefont {Ziebell}}]{Sarah:15}%
  \BibitemOpen
  \bibfield  {author} {\bibinfo {author} {\bibfnamefont {S.}~\bibnamefont {Seif
  El Nasr-Storey}}, \bibinfo {author} {\bibfnamefont {F.}~\bibnamefont
  {Boeuf}}, \bibinfo {author} {\bibfnamefont {C.}~\bibnamefont {Baudot}},
  \bibinfo {author} {\bibfnamefont {S.}~\bibnamefont {Detraz}}, \bibinfo
  {author} {\bibfnamefont {J.~M.}\ \bibnamefont {Fedeli}}, \bibinfo {author}
  {\bibfnamefont {D.}~\bibnamefont {Marris-Morini}}, \bibinfo {author}
  {\bibfnamefont {L.}~\bibnamefont {Olantera}}, \bibinfo {author}
  {\bibfnamefont {G.}~\bibnamefont {Pezzullo}}, \bibinfo {author}
  {\bibfnamefont {C.}~\bibnamefont {Sigaud}}, \bibinfo {author} {\bibfnamefont
  {C.}~\bibnamefont {Soos}}, \bibinfo {author} {\bibfnamefont {J.}~\bibnamefont
  {Troska}}, \bibinfo {author} {\bibfnamefont {F.}~\bibnamefont {Vasey}},
  \bibinfo {author} {\bibfnamefont {L.}~\bibnamefont {Vivien}}, \bibinfo
  {author} {\bibfnamefont {M.}~\bibnamefont {Zeiler}}, \ and\ \bibinfo {author}
  {\bibfnamefont {M.}~\bibnamefont {Ziebell}},\ }\href {\doibase
  10.1109/TNS.2015.2388546} {\bibfield  {journal} {\bibinfo  {journal} {IEEE
  Transactions on Nuclear Science}\ }\textbf {\bibinfo {volume} {62}},\
  \bibinfo {pages} {329} (\bibinfo {year} {2015})}\BibitemShut {NoStop}%
\bibitem [{\citenamefont {Zeiler}\ \emph {et~al.}(2017)\citenamefont {Zeiler},
  \citenamefont {El~Nasr-Storey}, \citenamefont {Detraz}, \citenamefont
  {Kraxner}, \citenamefont {Olantera}, \citenamefont {Scarcella}, \citenamefont
  {Sigaud}, \citenamefont {Soos}, \citenamefont {Troska},\ and\ \citenamefont
  {Vasey}}]{Zeiler:17}%
  \BibitemOpen
  \bibfield  {author} {\bibinfo {author} {\bibfnamefont {M.}~\bibnamefont
  {Zeiler}}, \bibinfo {author} {\bibfnamefont {S.~S.}\ \bibnamefont
  {El~Nasr-Storey}}, \bibinfo {author} {\bibfnamefont {S.}~\bibnamefont
  {Detraz}}, \bibinfo {author} {\bibfnamefont {A.}~\bibnamefont {Kraxner}},
  \bibinfo {author} {\bibfnamefont {L.}~\bibnamefont {Olantera}}, \bibinfo
  {author} {\bibfnamefont {C.}~\bibnamefont {Scarcella}}, \bibinfo {author}
  {\bibfnamefont {C.}~\bibnamefont {Sigaud}}, \bibinfo {author} {\bibfnamefont
  {C.}~\bibnamefont {Soos}}, \bibinfo {author} {\bibfnamefont {J.}~\bibnamefont
  {Troska}}, \ and\ \bibinfo {author} {\bibfnamefont {F.}~\bibnamefont
  {Vasey}},\ }\href {\doibase 10.1109/TNS.2017.2754948} {\bibfield  {journal}
  {\bibinfo  {journal} {IEEE Transactions on Nuclear Science}\ }\textbf
  {\bibinfo {volume} {64}},\ \bibinfo {pages} {2794} (\bibinfo {year}
  {2017})}\BibitemShut {NoStop}%
\bibitem [{\citenamefont {Hoffman}\ \emph {et~al.}(2019)\citenamefont
  {Hoffman}, \citenamefont {Gehl}, \citenamefont {Martinez}, \citenamefont
  {Trotter}, \citenamefont {Starbuck}, \citenamefont {Pomerene}, \citenamefont
  {Dallo}, \citenamefont {Hood}, \citenamefont {Dodd}, \citenamefont {Swanson},
  \citenamefont {Long}, \citenamefont {Derose},\ and\ \citenamefont
  {Lentine}}]{Hoffman:19}%
  \BibitemOpen
  \bibfield  {author} {\bibinfo {author} {\bibfnamefont {G.~B.}\ \bibnamefont
  {Hoffman}}, \bibinfo {author} {\bibfnamefont {M.}~\bibnamefont {Gehl}},
  \bibinfo {author} {\bibfnamefont {N.~J.}\ \bibnamefont {Martinez}}, \bibinfo
  {author} {\bibfnamefont {D.~C.}\ \bibnamefont {Trotter}}, \bibinfo {author}
  {\bibfnamefont {A.~L.}\ \bibnamefont {Starbuck}}, \bibinfo {author}
  {\bibfnamefont {A.}~\bibnamefont {Pomerene}}, \bibinfo {author}
  {\bibfnamefont {C.~M.}\ \bibnamefont {Dallo}}, \bibinfo {author}
  {\bibfnamefont {D.}~\bibnamefont {Hood}}, \bibinfo {author} {\bibfnamefont
  {P.~E.}\ \bibnamefont {Dodd}}, \bibinfo {author} {\bibfnamefont {S.~E.}\
  \bibnamefont {Swanson}}, \bibinfo {author} {\bibfnamefont {C.~M.}\
  \bibnamefont {Long}}, \bibinfo {author} {\bibfnamefont {C.~T.}\ \bibnamefont
  {Derose}}, \ and\ \bibinfo {author} {\bibfnamefont {A.~L.}\ \bibnamefont
  {Lentine}},\ }\href {\doibase 10.1109/TNS.2019.2907582} {\bibfield  {journal}
  {\bibinfo  {journal} {IEEE Transactions on Nuclear Science}\ }\textbf
  {\bibinfo {volume} {66}},\ \bibinfo {pages} {801} (\bibinfo {year}
  {2019})}\BibitemShut {NoStop}%
\bibitem [{\citenamefont {Lalovi{\'c}}\ \emph {et~al.}(2022)\citenamefont
  {Lalovi{\'c}}, \citenamefont {Scarcella}, \citenamefont {Bulling},
  \citenamefont {Detraz}, \citenamefont {Marcon}, \citenamefont {Olanter{\"a}},
  \citenamefont {Prousalidi}, \citenamefont {Sandven}, \citenamefont {Sigaud},
  \citenamefont {Soos},\ and\ \citenamefont {Troska}}]{Lalovic:22}%
  \BibitemOpen
  \bibfield  {author} {\bibinfo {author} {\bibfnamefont {M.}~\bibnamefont
  {Lalovi{\'c}}}, \bibinfo {author} {\bibfnamefont {C.}~\bibnamefont
  {Scarcella}}, \bibinfo {author} {\bibfnamefont {A.}~\bibnamefont {Bulling}},
  \bibinfo {author} {\bibfnamefont {S.}~\bibnamefont {Detraz}}, \bibinfo
  {author} {\bibfnamefont {L.}~\bibnamefont {Marcon}}, \bibinfo {author}
  {\bibfnamefont {L.}~\bibnamefont {Olanter{\"a}}}, \bibinfo {author}
  {\bibfnamefont {T.}~\bibnamefont {Prousalidi}}, \bibinfo {author}
  {\bibfnamefont {U.}~\bibnamefont {Sandven}}, \bibinfo {author} {\bibfnamefont
  {C.}~\bibnamefont {Sigaud}}, \bibinfo {author} {\bibfnamefont
  {C.}~\bibnamefont {Soos}}, \ and\ \bibinfo {author} {\bibfnamefont
  {J.}~\bibnamefont {Troska}},\ }\href {\doibase 10.1109/TNS.2022.3148579}
  {\bibfield  {journal} {\bibinfo  {journal} {IEEE Transactions on Nuclear
  Science}\ }\textbf {\bibinfo {volume} {69}},\ \bibinfo {pages} {1521}
  (\bibinfo {year} {2022})}\BibitemShut {NoStop}%
\bibitem [{\citenamefont {Mao}\ \emph {et~al.}(2024)\citenamefont {Mao},
  \citenamefont {Chang}, \citenamefont {Lee}, \citenamefont {Yu}, \citenamefont
  {Maruca}, \citenamefont {Ullah}, \citenamefont {Matthaeus}, \citenamefont
  {Krainak}, \citenamefont {Dong},\ and\ \citenamefont {Gu}}]{Mao:24}%
  \BibitemOpen
  \bibfield  {author} {\bibinfo {author} {\bibfnamefont {D.}~\bibnamefont
  {Mao}}, \bibinfo {author} {\bibfnamefont {L.}~\bibnamefont {Chang}}, \bibinfo
  {author} {\bibfnamefont {H.}~\bibnamefont {Lee}}, \bibinfo {author}
  {\bibfnamefont {A.~W.}\ \bibnamefont {Yu}}, \bibinfo {author} {\bibfnamefont
  {B.~A.}\ \bibnamefont {Maruca}}, \bibinfo {author} {\bibfnamefont
  {K.}~\bibnamefont {Ullah}}, \bibinfo {author} {\bibfnamefont {W.~H.}\
  \bibnamefont {Matthaeus}}, \bibinfo {author} {\bibfnamefont {M.~A.}\
  \bibnamefont {Krainak}}, \bibinfo {author} {\bibfnamefont {P.}~\bibnamefont
  {Dong}}, \ and\ \bibinfo {author} {\bibfnamefont {T.}~\bibnamefont {Gu}},\
  }\href {\doibase 10.1126/sciadv.adi9171} {\bibfield  {journal} {\bibinfo
  {journal} {Science Advances}\ }\textbf {\bibinfo {volume} {10}},\ \bibinfo
  {pages} {eadi9171} (\bibinfo {year} {2024})}\BibitemShut {NoStop}%
\bibitem [{\citenamefont {Xuan}\ \emph {et~al.}(2016)\citenamefont {Xuan},
  \citenamefont {Liu}, \citenamefont {Varghese}, \citenamefont {Metcalf},
  \citenamefont {Xue}, \citenamefont {Wang}, \citenamefont {Han}, \citenamefont
  {Jaramillo-Villegas}, \citenamefont {Noman}, \citenamefont {Wang},
  \citenamefont {Kim}, \citenamefont {Teng}, \citenamefont {Lee}, \citenamefont
  {Niu}, \citenamefont {Fan}, \citenamefont {Wang}, \citenamefont {Leaird},
  \citenamefont {Weiner},\ and\ \citenamefont {Qi}}]{Xuan:16}%
  \BibitemOpen
  \bibfield  {author} {\bibinfo {author} {\bibfnamefont {Y.}~\bibnamefont
  {Xuan}}, \bibinfo {author} {\bibfnamefont {Y.}~\bibnamefont {Liu}}, \bibinfo
  {author} {\bibfnamefont {L.~T.}\ \bibnamefont {Varghese}}, \bibinfo {author}
  {\bibfnamefont {A.~J.}\ \bibnamefont {Metcalf}}, \bibinfo {author}
  {\bibfnamefont {X.}~\bibnamefont {Xue}}, \bibinfo {author} {\bibfnamefont
  {P.-H.}\ \bibnamefont {Wang}}, \bibinfo {author} {\bibfnamefont
  {K.}~\bibnamefont {Han}}, \bibinfo {author} {\bibfnamefont {J.~A.}\
  \bibnamefont {Jaramillo-Villegas}}, \bibinfo {author} {\bibfnamefont {A.~A.}\
  \bibnamefont {Noman}}, \bibinfo {author} {\bibfnamefont {C.}~\bibnamefont
  {Wang}}, \bibinfo {author} {\bibfnamefont {S.}~\bibnamefont {Kim}}, \bibinfo
  {author} {\bibfnamefont {M.}~\bibnamefont {Teng}}, \bibinfo {author}
  {\bibfnamefont {Y.~J.}\ \bibnamefont {Lee}}, \bibinfo {author} {\bibfnamefont
  {B.}~\bibnamefont {Niu}}, \bibinfo {author} {\bibfnamefont {L.}~\bibnamefont
  {Fan}}, \bibinfo {author} {\bibfnamefont {J.}~\bibnamefont {Wang}}, \bibinfo
  {author} {\bibfnamefont {D.~E.}\ \bibnamefont {Leaird}}, \bibinfo {author}
  {\bibfnamefont {A.~M.}\ \bibnamefont {Weiner}}, \ and\ \bibinfo {author}
  {\bibfnamefont {M.}~\bibnamefont {Qi}},\ }\href {\doibase
  10.1364/OPTICA.3.001171} {\bibfield  {journal} {\bibinfo  {journal} {Optica}\
  }\textbf {\bibinfo {volume} {3}},\ \bibinfo {pages} {1171} (\bibinfo {year}
  {2016})}\BibitemShut {NoStop}%
\bibitem [{\citenamefont {Ji}\ \emph {et~al.}(2017)\citenamefont {Ji},
  \citenamefont {Barbosa}, \citenamefont {Roberts}, \citenamefont {Dutt},
  \citenamefont {Cardenas}, \citenamefont {Okawachi}, \citenamefont {Bryant},
  \citenamefont {Gaeta},\ and\ \citenamefont {Lipson}}]{Ji:17}%
  \BibitemOpen
  \bibfield  {author} {\bibinfo {author} {\bibfnamefont {X.}~\bibnamefont
  {Ji}}, \bibinfo {author} {\bibfnamefont {F.~A.~S.}\ \bibnamefont {Barbosa}},
  \bibinfo {author} {\bibfnamefont {S.~P.}\ \bibnamefont {Roberts}}, \bibinfo
  {author} {\bibfnamefont {A.}~\bibnamefont {Dutt}}, \bibinfo {author}
  {\bibfnamefont {J.}~\bibnamefont {Cardenas}}, \bibinfo {author}
  {\bibfnamefont {Y.}~\bibnamefont {Okawachi}}, \bibinfo {author}
  {\bibfnamefont {A.}~\bibnamefont {Bryant}}, \bibinfo {author} {\bibfnamefont
  {A.~L.}\ \bibnamefont {Gaeta}}, \ and\ \bibinfo {author} {\bibfnamefont
  {M.}~\bibnamefont {Lipson}},\ }\href {\doibase 10.1364/OPTICA.4.000619}
  {\bibfield  {journal} {\bibinfo  {journal} {Optica}\ }\textbf {\bibinfo
  {volume} {4}},\ \bibinfo {pages} {619} (\bibinfo {year} {2017})}\BibitemShut
  {NoStop}%
\bibitem [{\citenamefont {Liu}\ \emph {et~al.}(2021)\citenamefont {Liu},
  \citenamefont {Huang}, \citenamefont {Wang}, \citenamefont {He},
  \citenamefont {Raja}, \citenamefont {Liu}, \citenamefont {Engelsen},\ and\
  \citenamefont {Kippenberg}}]{Liu:21}%
  \BibitemOpen
  \bibfield  {author} {\bibinfo {author} {\bibfnamefont {J.}~\bibnamefont
  {Liu}}, \bibinfo {author} {\bibfnamefont {G.}~\bibnamefont {Huang}}, \bibinfo
  {author} {\bibfnamefont {R.~N.}\ \bibnamefont {Wang}}, \bibinfo {author}
  {\bibfnamefont {J.}~\bibnamefont {He}}, \bibinfo {author} {\bibfnamefont
  {A.~S.}\ \bibnamefont {Raja}}, \bibinfo {author} {\bibfnamefont
  {T.}~\bibnamefont {Liu}}, \bibinfo {author} {\bibfnamefont {N.~J.}\
  \bibnamefont {Engelsen}}, \ and\ \bibinfo {author} {\bibfnamefont {T.~J.}\
  \bibnamefont {Kippenberg}},\ }\href {\doibase 10.1038/s41467-021-21973-z}
  {\bibfield  {journal} {\bibinfo  {journal} {Nature Communications}\ }\textbf
  {\bibinfo {volume} {12}},\ \bibinfo {pages} {2236} (\bibinfo {year}
  {2021})}\BibitemShut {NoStop}%
\bibitem [{\citenamefont {Spencer}\ \emph {et~al.}(2014)\citenamefont
  {Spencer}, \citenamefont {Bauters}, \citenamefont {Heck},\ and\ \citenamefont
  {Bowers}}]{Spencer:14}%
  \BibitemOpen
  \bibfield  {author} {\bibinfo {author} {\bibfnamefont {D.~T.}\ \bibnamefont
  {Spencer}}, \bibinfo {author} {\bibfnamefont {J.~F.}\ \bibnamefont
  {Bauters}}, \bibinfo {author} {\bibfnamefont {M.~J.~R.}\ \bibnamefont
  {Heck}}, \ and\ \bibinfo {author} {\bibfnamefont {J.~E.}\ \bibnamefont
  {Bowers}},\ }\href {\doibase 10.1364/OPTICA.1.000153} {\bibfield  {journal}
  {\bibinfo  {journal} {Optica}\ }\textbf {\bibinfo {volume} {1}},\ \bibinfo
  {pages} {153} (\bibinfo {year} {2014})}\BibitemShut {NoStop}%
\bibitem [{\citenamefont {Zhang}\ \emph {et~al.}(2017)\citenamefont {Zhang},
  \citenamefont {Wang}, \citenamefont {Cheng}, \citenamefont {Shams-Ansari},\
  and\ \citenamefont {Lon\v{c}ar}}]{Zhang:17}%
  \BibitemOpen
  \bibfield  {author} {\bibinfo {author} {\bibfnamefont {M.}~\bibnamefont
  {Zhang}}, \bibinfo {author} {\bibfnamefont {C.}~\bibnamefont {Wang}},
  \bibinfo {author} {\bibfnamefont {R.}~\bibnamefont {Cheng}}, \bibinfo
  {author} {\bibfnamefont {A.}~\bibnamefont {Shams-Ansari}}, \ and\ \bibinfo
  {author} {\bibfnamefont {M.}~\bibnamefont {Lon\v{c}ar}},\ }\href {\doibase
  10.1364/OPTICA.4.001536} {\bibfield  {journal} {\bibinfo  {journal} {Optica}\
  }\textbf {\bibinfo {volume} {4}},\ \bibinfo {pages} {1536} (\bibinfo {year}
  {2017})}\BibitemShut {NoStop}%
\bibitem [{\citenamefont {He}\ \emph {et~al.}(2019)\citenamefont {He},
  \citenamefont {Yang}, \citenamefont {Ling}, \citenamefont {Luo},
  \citenamefont {Liang}, \citenamefont {Li}, \citenamefont {Shen},
  \citenamefont {Wang}, \citenamefont {Vahala},\ and\ \citenamefont
  {Lin}}]{He:19}%
  \BibitemOpen
  \bibfield  {author} {\bibinfo {author} {\bibfnamefont {Y.}~\bibnamefont
  {He}}, \bibinfo {author} {\bibfnamefont {Q.-F.}\ \bibnamefont {Yang}},
  \bibinfo {author} {\bibfnamefont {J.}~\bibnamefont {Ling}}, \bibinfo {author}
  {\bibfnamefont {R.}~\bibnamefont {Luo}}, \bibinfo {author} {\bibfnamefont
  {H.}~\bibnamefont {Liang}}, \bibinfo {author} {\bibfnamefont
  {M.}~\bibnamefont {Li}}, \bibinfo {author} {\bibfnamefont {B.}~\bibnamefont
  {Shen}}, \bibinfo {author} {\bibfnamefont {H.}~\bibnamefont {Wang}}, \bibinfo
  {author} {\bibfnamefont {K.}~\bibnamefont {Vahala}}, \ and\ \bibinfo {author}
  {\bibfnamefont {Q.}~\bibnamefont {Lin}},\ }\href {\doibase
  10.1364/OPTICA.6.001138} {\bibfield  {journal} {\bibinfo  {journal} {Optica}\
  }\textbf {\bibinfo {volume} {6}},\ \bibinfo {pages} {1138} (\bibinfo {year}
  {2019})}\BibitemShut {NoStop}%
\bibitem [{\citenamefont {Gao}\ \emph {et~al.}(2022)\citenamefont {Gao},
  \citenamefont {Yao}, \citenamefont {Guan}, \citenamefont {Deng},
  \citenamefont {Lin}, \citenamefont {Wang}, \citenamefont {Qiao},
  \citenamefont {Fang},\ and\ \citenamefont {Cheng}}]{GaoR:22}%
  \BibitemOpen
  \bibfield  {author} {\bibinfo {author} {\bibfnamefont {R.}~\bibnamefont
  {Gao}}, \bibinfo {author} {\bibfnamefont {N.}~\bibnamefont {Yao}}, \bibinfo
  {author} {\bibfnamefont {J.}~\bibnamefont {Guan}}, \bibinfo {author}
  {\bibfnamefont {L.}~\bibnamefont {Deng}}, \bibinfo {author} {\bibfnamefont
  {J.}~\bibnamefont {Lin}}, \bibinfo {author} {\bibfnamefont {M.}~\bibnamefont
  {Wang}}, \bibinfo {author} {\bibfnamefont {L.}~\bibnamefont {Qiao}}, \bibinfo
  {author} {\bibfnamefont {W.}~\bibnamefont {Fang}}, \ and\ \bibinfo {author}
  {\bibfnamefont {Y.}~\bibnamefont {Cheng}},\ }\href
  {https://opg.optica.org/col/abstract.cfm?URI=col-20-1-011902} {\bibfield
  {journal} {\bibinfo  {journal} {Chin. Opt. Lett.}\ }\textbf {\bibinfo
  {volume} {20}},\ \bibinfo {pages} {011902} (\bibinfo {year}
  {2022})}\BibitemShut {NoStop}%
\bibitem [{\citenamefont {Guidry}\ \emph {et~al.}(2020)\citenamefont {Guidry},
  \citenamefont {Yang}, \citenamefont {Lukin}, \citenamefont {Markosyan},
  \citenamefont {Yang}, \citenamefont {Fejer},\ and\ \citenamefont
  {Vu\v{c}kovi\'{c}}}]{Guidry:20}%
  \BibitemOpen
  \bibfield  {author} {\bibinfo {author} {\bibfnamefont {M.~A.}\ \bibnamefont
  {Guidry}}, \bibinfo {author} {\bibfnamefont {K.~Y.}\ \bibnamefont {Yang}},
  \bibinfo {author} {\bibfnamefont {D.~M.}\ \bibnamefont {Lukin}}, \bibinfo
  {author} {\bibfnamefont {A.}~\bibnamefont {Markosyan}}, \bibinfo {author}
  {\bibfnamefont {J.}~\bibnamefont {Yang}}, \bibinfo {author} {\bibfnamefont
  {M.~M.}\ \bibnamefont {Fejer}}, \ and\ \bibinfo {author} {\bibfnamefont
  {J.}~\bibnamefont {Vu\v{c}kovi\'{c}}},\ }\href {\doibase
  10.1364/OPTICA.394138} {\bibfield  {journal} {\bibinfo  {journal} {Optica}\
  }\textbf {\bibinfo {volume} {7}},\ \bibinfo {pages} {1139} (\bibinfo {year}
  {2020})}\BibitemShut {NoStop}%
\bibitem [{\citenamefont {Wang}\ \emph {et~al.}(2022)\citenamefont {Wang},
  \citenamefont {Li}, \citenamefont {Yi}, \citenamefont {Fang}, \citenamefont
  {Zhou}, \citenamefont {Wang}, \citenamefont {Niu}, \citenamefont {Chen},
  \citenamefont {Zhang}, \citenamefont {Cheng}, \citenamefont {Liu},
  \citenamefont {Dong},\ and\ \citenamefont {Ou}}]{Wang:22}%
  \BibitemOpen
  \bibfield  {author} {\bibinfo {author} {\bibfnamefont {C.}~\bibnamefont
  {Wang}}, \bibinfo {author} {\bibfnamefont {J.}~\bibnamefont {Li}}, \bibinfo
  {author} {\bibfnamefont {A.}~\bibnamefont {Yi}}, \bibinfo {author}
  {\bibfnamefont {Z.}~\bibnamefont {Fang}}, \bibinfo {author} {\bibfnamefont
  {L.}~\bibnamefont {Zhou}}, \bibinfo {author} {\bibfnamefont {Z.}~\bibnamefont
  {Wang}}, \bibinfo {author} {\bibfnamefont {R.}~\bibnamefont {Niu}}, \bibinfo
  {author} {\bibfnamefont {Y.}~\bibnamefont {Chen}}, \bibinfo {author}
  {\bibfnamefont {J.}~\bibnamefont {Zhang}}, \bibinfo {author} {\bibfnamefont
  {Y.}~\bibnamefont {Cheng}}, \bibinfo {author} {\bibfnamefont
  {J.}~\bibnamefont {Liu}}, \bibinfo {author} {\bibfnamefont {C.-H.}\
  \bibnamefont {Dong}}, \ and\ \bibinfo {author} {\bibfnamefont
  {X.}~\bibnamefont {Ou}},\ }\href {\doibase 10.1038/s41377-022-01042-w}
  {\bibfield  {journal} {\bibinfo  {journal} {Light: Science \& Applications}\
  }\textbf {\bibinfo {volume} {11}},\ \bibinfo {pages} {341} (\bibinfo {year}
  {2022})}\BibitemShut {NoStop}%
\bibitem [{\citenamefont {Cai}\ \emph {et~al.}(2022)\citenamefont {Cai},
  \citenamefont {Li}, \citenamefont {Wang},\ and\ \citenamefont {Li}}]{Cai:22}%
  \BibitemOpen
  \bibfield  {author} {\bibinfo {author} {\bibfnamefont {L.}~\bibnamefont
  {Cai}}, \bibinfo {author} {\bibfnamefont {J.}~\bibnamefont {Li}}, \bibinfo
  {author} {\bibfnamefont {R.}~\bibnamefont {Wang}}, \ and\ \bibinfo {author}
  {\bibfnamefont {Q.}~\bibnamefont {Li}},\ }\href {\doibase 10.1364/PRJ.449267}
  {\bibfield  {journal} {\bibinfo  {journal} {Photon. Res.}\ }\textbf {\bibinfo
  {volume} {10}},\ \bibinfo {pages} {870} (\bibinfo {year} {2022})}\BibitemShut
  {NoStop}%
\bibitem [{\citenamefont {Kippenberg}\ \emph {et~al.}(2018)\citenamefont
  {Kippenberg}, \citenamefont {Gaeta}, \citenamefont {Lipson},\ and\
  \citenamefont {Gorodetsky}}]{Kippenberg:18}%
  \BibitemOpen
  \bibfield  {author} {\bibinfo {author} {\bibfnamefont {T.~J.}\ \bibnamefont
  {Kippenberg}}, \bibinfo {author} {\bibfnamefont {A.~L.}\ \bibnamefont
  {Gaeta}}, \bibinfo {author} {\bibfnamefont {M.}~\bibnamefont {Lipson}}, \
  and\ \bibinfo {author} {\bibfnamefont {M.~L.}\ \bibnamefont {Gorodetsky}},\
  }\href {\doibase 10.1126/science.aan8083} {\bibfield  {journal} {\bibinfo
  {journal} {Science}\ }\textbf {\bibinfo {volume} {361}},\ \bibinfo {pages}
  {eaan8083} (\bibinfo {year} {2018})}\BibitemShut {NoStop}%
\bibitem [{\citenamefont {Gaeta}\ \emph {et~al.}(2019)\citenamefont {Gaeta},
  \citenamefont {Lipson},\ and\ \citenamefont {Kippenberg}}]{Gaeta:19}%
  \BibitemOpen
  \bibfield  {author} {\bibinfo {author} {\bibfnamefont {A.~L.}\ \bibnamefont
  {Gaeta}}, \bibinfo {author} {\bibfnamefont {M.}~\bibnamefont {Lipson}}, \
  and\ \bibinfo {author} {\bibfnamefont {T.~J.}\ \bibnamefont {Kippenberg}},\
  }\href {\doibase 10.1038/s41566-019-0358-x} {\bibfield  {journal} {\bibinfo
  {journal} {Nature Photonics}\ }\textbf {\bibinfo {volume} {13}},\ \bibinfo
  {pages} {158} (\bibinfo {year} {2019})}\BibitemShut {NoStop}%
\bibitem [{\citenamefont {Zhang}\ \emph {et~al.}(2019)\citenamefont {Zhang},
  \citenamefont {Buscaino}, \citenamefont {Wang}, \citenamefont {Shams-Ansari},
  \citenamefont {Reimer}, \citenamefont {Zhu}, \citenamefont {Kahn},\ and\
  \citenamefont {Lon{\v c}ar}}]{Zhang:19}%
  \BibitemOpen
  \bibfield  {author} {\bibinfo {author} {\bibfnamefont {M.}~\bibnamefont
  {Zhang}}, \bibinfo {author} {\bibfnamefont {B.}~\bibnamefont {Buscaino}},
  \bibinfo {author} {\bibfnamefont {C.}~\bibnamefont {Wang}}, \bibinfo {author}
  {\bibfnamefont {A.}~\bibnamefont {Shams-Ansari}}, \bibinfo {author}
  {\bibfnamefont {C.}~\bibnamefont {Reimer}}, \bibinfo {author} {\bibfnamefont
  {R.}~\bibnamefont {Zhu}}, \bibinfo {author} {\bibfnamefont {J.~M.}\
  \bibnamefont {Kahn}}, \ and\ \bibinfo {author} {\bibfnamefont
  {M.}~\bibnamefont {Lon{\v c}ar}},\ }\href {\doibase
  10.1038/s41586-019-1008-7} {\bibfield  {journal} {\bibinfo  {journal}
  {Nature}\ }\textbf {\bibinfo {volume} {568}},\ \bibinfo {pages} {373}
  (\bibinfo {year} {2019})}\BibitemShut {NoStop}%
\bibitem [{\citenamefont {Siddharth}\ \emph {et~al.}(2022)\citenamefont
  {Siddharth}, \citenamefont {Wunderer}, \citenamefont {Lihachev},
  \citenamefont {Voloshin}, \citenamefont {Haller}, \citenamefont {Wang},
  \citenamefont {Teepe}, \citenamefont {Yang}, \citenamefont {Liu},
  \citenamefont {Riemensberger}, \citenamefont {Grandjean}, \citenamefont
  {Johnson},\ and\ \citenamefont {Kippenberg}}]{Siddharth:22}%
  \BibitemOpen
  \bibfield  {author} {\bibinfo {author} {\bibfnamefont {A.}~\bibnamefont
  {Siddharth}}, \bibinfo {author} {\bibfnamefont {T.}~\bibnamefont {Wunderer}},
  \bibinfo {author} {\bibfnamefont {G.}~\bibnamefont {Lihachev}}, \bibinfo
  {author} {\bibfnamefont {A.~S.}\ \bibnamefont {Voloshin}}, \bibinfo {author}
  {\bibfnamefont {C.}~\bibnamefont {Haller}}, \bibinfo {author} {\bibfnamefont
  {R.~N.}\ \bibnamefont {Wang}}, \bibinfo {author} {\bibfnamefont
  {M.}~\bibnamefont {Teepe}}, \bibinfo {author} {\bibfnamefont
  {Z.}~\bibnamefont {Yang}}, \bibinfo {author} {\bibfnamefont {J.}~\bibnamefont
  {Liu}}, \bibinfo {author} {\bibfnamefont {J.}~\bibnamefont {Riemensberger}},
  \bibinfo {author} {\bibfnamefont {N.}~\bibnamefont {Grandjean}}, \bibinfo
  {author} {\bibfnamefont {N.}~\bibnamefont {Johnson}}, \ and\ \bibinfo
  {author} {\bibfnamefont {T.~J.}\ \bibnamefont {Kippenberg}},\ }\href
  {\doibase 10.1063/5.0081660} {\bibfield  {journal} {\bibinfo  {journal} {APL
  Photonics}\ }\textbf {\bibinfo {volume} {7}},\ \bibinfo {pages} {046108}
  (\bibinfo {year} {2022})}\BibitemShut {NoStop}%
\bibitem [{\citenamefont {Corato-Zanarella}\ \emph {et~al.}(2023)\citenamefont
  {Corato-Zanarella}, \citenamefont {Gil-Molina}, \citenamefont {Ji},
  \citenamefont {Shin}, \citenamefont {Mohanty},\ and\ \citenamefont
  {Lipson}}]{Corato-Zanarella:23}%
  \BibitemOpen
  \bibfield  {author} {\bibinfo {author} {\bibfnamefont {M.}~\bibnamefont
  {Corato-Zanarella}}, \bibinfo {author} {\bibfnamefont {A.}~\bibnamefont
  {Gil-Molina}}, \bibinfo {author} {\bibfnamefont {X.}~\bibnamefont {Ji}},
  \bibinfo {author} {\bibfnamefont {M.~C.}\ \bibnamefont {Shin}}, \bibinfo
  {author} {\bibfnamefont {A.}~\bibnamefont {Mohanty}}, \ and\ \bibinfo
  {author} {\bibfnamefont {M.}~\bibnamefont {Lipson}},\ }\href {\doibase
  10.1038/s41566-022-01120-w} {\bibfield  {journal} {\bibinfo  {journal}
  {Nature Photonics}\ }\textbf {\bibinfo {volume} {17}},\ \bibinfo {pages}
  {157} (\bibinfo {year} {2023})}\BibitemShut {NoStop}%
\bibitem [{\citenamefont {Ling}\ \emph {et~al.}(2023)\citenamefont {Ling},
  \citenamefont {Staffa}, \citenamefont {Wang}, \citenamefont {Shen},
  \citenamefont {Chang}, \citenamefont {Javid}, \citenamefont {Wu},
  \citenamefont {Yuan}, \citenamefont {Lopez-Rios}, \citenamefont {Li},
  \citenamefont {He}, \citenamefont {Li}, \citenamefont {Bowers}, \citenamefont
  {Vahala},\ and\ \citenamefont {Lin}}]{Ling:23}%
  \BibitemOpen
  \bibfield  {author} {\bibinfo {author} {\bibfnamefont {J.}~\bibnamefont
  {Ling}}, \bibinfo {author} {\bibfnamefont {J.}~\bibnamefont {Staffa}},
  \bibinfo {author} {\bibfnamefont {H.}~\bibnamefont {Wang}}, \bibinfo {author}
  {\bibfnamefont {B.}~\bibnamefont {Shen}}, \bibinfo {author} {\bibfnamefont
  {L.}~\bibnamefont {Chang}}, \bibinfo {author} {\bibfnamefont {U.~A.}\
  \bibnamefont {Javid}}, \bibinfo {author} {\bibfnamefont {L.}~\bibnamefont
  {Wu}}, \bibinfo {author} {\bibfnamefont {Z.}~\bibnamefont {Yuan}}, \bibinfo
  {author} {\bibfnamefont {R.}~\bibnamefont {Lopez-Rios}}, \bibinfo {author}
  {\bibfnamefont {M.}~\bibnamefont {Li}}, \bibinfo {author} {\bibfnamefont
  {Y.}~\bibnamefont {He}}, \bibinfo {author} {\bibfnamefont {B.}~\bibnamefont
  {Li}}, \bibinfo {author} {\bibfnamefont {J.~E.}\ \bibnamefont {Bowers}},
  \bibinfo {author} {\bibfnamefont {K.~J.}\ \bibnamefont {Vahala}}, \ and\
  \bibinfo {author} {\bibfnamefont {Q.}~\bibnamefont {Lin}},\ }\href {\doibase
  https://doi.org/10.1002/lpor.202200663} {\bibfield  {journal} {\bibinfo
  {journal} {Laser \& Photonics Reviews}\ }\textbf {\bibinfo {volume} {17}},\
  \bibinfo {pages} {2200663} (\bibinfo {year} {2023})}\BibitemShut {NoStop}%
\bibitem [{\citenamefont {Isichenko}\ \emph {et~al.}(2024)\citenamefont
  {Isichenko}, \citenamefont {Hunter}, \citenamefont {Bose}, \citenamefont
  {Chauhan}, \citenamefont {Song}, \citenamefont {Liu}, \citenamefont
  {Harrington},\ and\ \citenamefont {Blumenthal}}]{Isichenko:24}%
  \BibitemOpen
  \bibfield  {author} {\bibinfo {author} {\bibfnamefont {A.}~\bibnamefont
  {Isichenko}}, \bibinfo {author} {\bibfnamefont {A.~S.}\ \bibnamefont
  {Hunter}}, \bibinfo {author} {\bibfnamefont {D.}~\bibnamefont {Bose}},
  \bibinfo {author} {\bibfnamefont {N.}~\bibnamefont {Chauhan}}, \bibinfo
  {author} {\bibfnamefont {M.}~\bibnamefont {Song}}, \bibinfo {author}
  {\bibfnamefont {K.}~\bibnamefont {Liu}}, \bibinfo {author} {\bibfnamefont
  {M.~W.}\ \bibnamefont {Harrington}}, \ and\ \bibinfo {author} {\bibfnamefont
  {D.~J.}\ \bibnamefont {Blumenthal}},\ }\href {\doibase
  10.1038/s41598-024-76699-x} {\bibfield  {journal} {\bibinfo  {journal}
  {Scientific Reports}\ }\textbf {\bibinfo {volume} {14}},\ \bibinfo {pages}
  {27015} (\bibinfo {year} {2024})}\BibitemShut {NoStop}%
\bibitem [{\citenamefont {Nejadriahi}\ \emph {et~al.}(2024)\citenamefont
  {Nejadriahi}, \citenamefont {Kittlaus}, \citenamefont {Bose}, \citenamefont
  {Chauhan}, \citenamefont {Wang}, \citenamefont {Fradet}, \citenamefont
  {Bagheri}, \citenamefont {Isichenko}, \citenamefont {Heim}, \citenamefont
  {Forouhar},\ and\ \citenamefont {Blumenthal}}]{Nejadriahi:24}%
  \BibitemOpen
  \bibfield  {author} {\bibinfo {author} {\bibfnamefont {H.}~\bibnamefont
  {Nejadriahi}}, \bibinfo {author} {\bibfnamefont {E.}~\bibnamefont
  {Kittlaus}}, \bibinfo {author} {\bibfnamefont {D.}~\bibnamefont {Bose}},
  \bibinfo {author} {\bibfnamefont {N.}~\bibnamefont {Chauhan}}, \bibinfo
  {author} {\bibfnamefont {J.}~\bibnamefont {Wang}}, \bibinfo {author}
  {\bibfnamefont {M.}~\bibnamefont {Fradet}}, \bibinfo {author} {\bibfnamefont
  {M.}~\bibnamefont {Bagheri}}, \bibinfo {author} {\bibfnamefont
  {A.}~\bibnamefont {Isichenko}}, \bibinfo {author} {\bibfnamefont
  {D.}~\bibnamefont {Heim}}, \bibinfo {author} {\bibfnamefont {S.}~\bibnamefont
  {Forouhar}}, \ and\ \bibinfo {author} {\bibfnamefont {D.~J.}\ \bibnamefont
  {Blumenthal}},\ }\href {\doibase 10.1364/OL.543307} {\bibfield  {journal}
  {\bibinfo  {journal} {Opt. Lett.}\ }\textbf {\bibinfo {volume} {49}},\
  \bibinfo {pages} {7254} (\bibinfo {year} {2024})}\BibitemShut {NoStop}%
\bibitem [{\citenamefont {Blumenthal}\ \emph {et~al.}(2024)\citenamefont
  {Blumenthal}, \citenamefont {Isichenko},\ and\ \citenamefont
  {Chauhan}}]{Blumenthal:24}%
  \BibitemOpen
  \bibfield  {author} {\bibinfo {author} {\bibfnamefont {D.~J.}\ \bibnamefont
  {Blumenthal}}, \bibinfo {author} {\bibfnamefont {A.}~\bibnamefont
  {Isichenko}}, \ and\ \bibinfo {author} {\bibfnamefont {N.}~\bibnamefont
  {Chauhan}},\ }\href {\doibase 10.1364/OPTICAQ.532260} {\bibfield  {journal}
  {\bibinfo  {journal} {Optica Quantum}\ }\textbf {\bibinfo {volume} {2}},\
  \bibinfo {pages} {444} (\bibinfo {year} {2024})}\BibitemShut {NoStop}%
\bibitem [{\citenamefont {Isichenko}\ \emph {et~al.}(2023)\citenamefont
  {Isichenko}, \citenamefont {Chauhan}, \citenamefont {Bose}, \citenamefont
  {Wang}, \citenamefont {Kunz},\ and\ \citenamefont
  {Blumenthal}}]{Isichenko:23}%
  \BibitemOpen
  \bibfield  {author} {\bibinfo {author} {\bibfnamefont {A.}~\bibnamefont
  {Isichenko}}, \bibinfo {author} {\bibfnamefont {N.}~\bibnamefont {Chauhan}},
  \bibinfo {author} {\bibfnamefont {D.}~\bibnamefont {Bose}}, \bibinfo {author}
  {\bibfnamefont {J.}~\bibnamefont {Wang}}, \bibinfo {author} {\bibfnamefont
  {P.~D.}\ \bibnamefont {Kunz}}, \ and\ \bibinfo {author} {\bibfnamefont
  {D.~J.}\ \bibnamefont {Blumenthal}},\ }\href {\doibase
  10.1038/s41467-023-38818-6} {\bibfield  {journal} {\bibinfo  {journal}
  {Nature Communications}\ }\textbf {\bibinfo {volume} {14}},\ \bibinfo {pages}
  {3080} (\bibinfo {year} {2023})}\BibitemShut {NoStop}%
\bibitem [{\citenamefont {Lee}\ \emph {et~al.}(2022)\citenamefont {Lee},
  \citenamefont {Ding}, \citenamefont {Christensen}, \citenamefont {Rosenthal},
  \citenamefont {Ison}, \citenamefont {Gillund}, \citenamefont {Bossert},
  \citenamefont {Fuerschbach}, \citenamefont {Kindel}, \citenamefont
  {Finnegan}, \citenamefont {Wendt}, \citenamefont {Gehl}, \citenamefont
  {Kodigala}, \citenamefont {McGuinness}, \citenamefont {Walker}, \citenamefont
  {Kemme}, \citenamefont {Lentine}, \citenamefont {Biedermann},\ and\
  \citenamefont {Schwindt}}]{Lee:22}%
  \BibitemOpen
  \bibfield  {author} {\bibinfo {author} {\bibfnamefont {J.}~\bibnamefont
  {Lee}}, \bibinfo {author} {\bibfnamefont {R.}~\bibnamefont {Ding}}, \bibinfo
  {author} {\bibfnamefont {J.}~\bibnamefont {Christensen}}, \bibinfo {author}
  {\bibfnamefont {R.~R.}\ \bibnamefont {Rosenthal}}, \bibinfo {author}
  {\bibfnamefont {A.}~\bibnamefont {Ison}}, \bibinfo {author} {\bibfnamefont
  {D.~P.}\ \bibnamefont {Gillund}}, \bibinfo {author} {\bibfnamefont
  {D.}~\bibnamefont {Bossert}}, \bibinfo {author} {\bibfnamefont {K.~H.}\
  \bibnamefont {Fuerschbach}}, \bibinfo {author} {\bibfnamefont
  {W.}~\bibnamefont {Kindel}}, \bibinfo {author} {\bibfnamefont {P.~S.}\
  \bibnamefont {Finnegan}}, \bibinfo {author} {\bibfnamefont {J.~R.}\
  \bibnamefont {Wendt}}, \bibinfo {author} {\bibfnamefont {M.}~\bibnamefont
  {Gehl}}, \bibinfo {author} {\bibfnamefont {A.}~\bibnamefont {Kodigala}},
  \bibinfo {author} {\bibfnamefont {H.}~\bibnamefont {McGuinness}}, \bibinfo
  {author} {\bibfnamefont {C.~A.}\ \bibnamefont {Walker}}, \bibinfo {author}
  {\bibfnamefont {S.~A.}\ \bibnamefont {Kemme}}, \bibinfo {author}
  {\bibfnamefont {A.}~\bibnamefont {Lentine}}, \bibinfo {author} {\bibfnamefont
  {G.}~\bibnamefont {Biedermann}}, \ and\ \bibinfo {author} {\bibfnamefont
  {P.~D.~D.}\ \bibnamefont {Schwindt}},\ }\href {\doibase
  10.1038/s41467-022-31410-4} {\bibfield  {journal} {\bibinfo  {journal}
  {Nature Communications}\ }\textbf {\bibinfo {volume} {13}},\ \bibinfo {pages}
  {5131} (\bibinfo {year} {2022})}\BibitemShut {NoStop}%
\bibitem [{\citenamefont {Ropp}\ \emph {et~al.}(2023)\citenamefont {Ropp},
  \citenamefont {Zhu}, \citenamefont {Yulaev}, \citenamefont {Westly},
  \citenamefont {Simelgor}, \citenamefont {Rakholia}, \citenamefont {Lunden},
  \citenamefont {Sheredy}, \citenamefont {Boyd}, \citenamefont {Papp},
  \citenamefont {Agrawal},\ and\ \citenamefont {Aksyuk}}]{Ropp:23}%
  \BibitemOpen
  \bibfield  {author} {\bibinfo {author} {\bibfnamefont {C.}~\bibnamefont
  {Ropp}}, \bibinfo {author} {\bibfnamefont {W.}~\bibnamefont {Zhu}}, \bibinfo
  {author} {\bibfnamefont {A.}~\bibnamefont {Yulaev}}, \bibinfo {author}
  {\bibfnamefont {D.}~\bibnamefont {Westly}}, \bibinfo {author} {\bibfnamefont
  {G.}~\bibnamefont {Simelgor}}, \bibinfo {author} {\bibfnamefont
  {A.}~\bibnamefont {Rakholia}}, \bibinfo {author} {\bibfnamefont
  {W.}~\bibnamefont {Lunden}}, \bibinfo {author} {\bibfnamefont
  {D.}~\bibnamefont {Sheredy}}, \bibinfo {author} {\bibfnamefont {M.~M.}\
  \bibnamefont {Boyd}}, \bibinfo {author} {\bibfnamefont {S.}~\bibnamefont
  {Papp}}, \bibinfo {author} {\bibfnamefont {A.}~\bibnamefont {Agrawal}}, \
  and\ \bibinfo {author} {\bibfnamefont {V.}~\bibnamefont {Aksyuk}},\ }\href
  {\doibase 10.1038/s41377-023-01081-x} {\bibfield  {journal} {\bibinfo
  {journal} {Light: Science \& Applications}\ }\textbf {\bibinfo {volume}
  {12}},\ \bibinfo {pages} {83} (\bibinfo {year} {2023})}\BibitemShut {NoStop}%
\bibitem [{\citenamefont {Kitching}(2018)}]{Kitching:18}%
  \BibitemOpen
  \bibfield  {author} {\bibinfo {author} {\bibfnamefont {J.}~\bibnamefont
  {Kitching}},\ }\href {\doibase 10.1063/1.5026238} {\bibfield  {journal}
  {\bibinfo  {journal} {Applied Physics Reviews}\ }\textbf {\bibinfo {volume}
  {5}},\ \bibinfo {pages} {031302} (\bibinfo {year} {2018})}\BibitemShut
  {NoStop}%
\bibitem [{\citenamefont {Newman}\ \emph {et~al.}(2019)\citenamefont {Newman},
  \citenamefont {Maurice}, \citenamefont {Drake}, \citenamefont {Stone},
  \citenamefont {Briles}, \citenamefont {Spencer}, \citenamefont {Fredrick},
  \citenamefont {Li}, \citenamefont {Westly}, \citenamefont {Ilic},
  \citenamefont {Shen}, \citenamefont {Suh}, \citenamefont {Yang},
  \citenamefont {Johnson}, \citenamefont {Johnson}, \citenamefont {Hollberg},
  \citenamefont {Vahala}, \citenamefont {Srinivasan}, \citenamefont {Diddams},
  \citenamefont {Kitching}, \citenamefont {Papp},\ and\ \citenamefont
  {Hummon}}]{Newman:19}%
  \BibitemOpen
  \bibfield  {author} {\bibinfo {author} {\bibfnamefont {Z.~L.}\ \bibnamefont
  {Newman}}, \bibinfo {author} {\bibfnamefont {V.}~\bibnamefont {Maurice}},
  \bibinfo {author} {\bibfnamefont {T.}~\bibnamefont {Drake}}, \bibinfo
  {author} {\bibfnamefont {J.~R.}\ \bibnamefont {Stone}}, \bibinfo {author}
  {\bibfnamefont {T.~C.}\ \bibnamefont {Briles}}, \bibinfo {author}
  {\bibfnamefont {D.~T.}\ \bibnamefont {Spencer}}, \bibinfo {author}
  {\bibfnamefont {C.}~\bibnamefont {Fredrick}}, \bibinfo {author}
  {\bibfnamefont {Q.}~\bibnamefont {Li}}, \bibinfo {author} {\bibfnamefont
  {D.}~\bibnamefont {Westly}}, \bibinfo {author} {\bibfnamefont {B.~R.}\
  \bibnamefont {Ilic}}, \bibinfo {author} {\bibfnamefont {B.}~\bibnamefont
  {Shen}}, \bibinfo {author} {\bibfnamefont {M.-G.}\ \bibnamefont {Suh}},
  \bibinfo {author} {\bibfnamefont {K.~Y.}\ \bibnamefont {Yang}}, \bibinfo
  {author} {\bibfnamefont {C.}~\bibnamefont {Johnson}}, \bibinfo {author}
  {\bibfnamefont {D.~M.~S.}\ \bibnamefont {Johnson}}, \bibinfo {author}
  {\bibfnamefont {L.}~\bibnamefont {Hollberg}}, \bibinfo {author}
  {\bibfnamefont {K.~J.}\ \bibnamefont {Vahala}}, \bibinfo {author}
  {\bibfnamefont {K.}~\bibnamefont {Srinivasan}}, \bibinfo {author}
  {\bibfnamefont {S.~A.}\ \bibnamefont {Diddams}}, \bibinfo {author}
  {\bibfnamefont {J.}~\bibnamefont {Kitching}}, \bibinfo {author}
  {\bibfnamefont {S.~B.}\ \bibnamefont {Papp}}, \ and\ \bibinfo {author}
  {\bibfnamefont {M.~T.}\ \bibnamefont {Hummon}},\ }\href {\doibase
  10.1364/OPTICA.6.000680} {\bibfield  {journal} {\bibinfo  {journal} {Optica}\
  }\textbf {\bibinfo {volume} {6}},\ \bibinfo {pages} {680} (\bibinfo {year}
  {2019})}\BibitemShut {NoStop}%
\bibitem [{\citenamefont {Will}(2014)}]{Will:14}%
  \BibitemOpen
  \bibfield  {author} {\bibinfo {author} {\bibfnamefont {C.~M.}\ \bibnamefont
  {Will}},\ }\href {\doibase 10.12942/lrr-2014-4} {\bibfield  {journal}
  {\bibinfo  {journal} {Living Reviews in Relativity}\ }\textbf {\bibinfo
  {volume} {17}},\ \bibinfo {pages} {4} (\bibinfo {year} {2014})}\BibitemShut
  {NoStop}%
\bibitem [{\citenamefont {Flechtner}\ \emph {et~al.}(2021)\citenamefont
  {Flechtner}, \citenamefont {Reigber}, \citenamefont {Rummel},\ and\
  \citenamefont {Balmino}}]{Flechtner:21}%
  \BibitemOpen
  \bibfield  {author} {\bibinfo {author} {\bibfnamefont {F.}~\bibnamefont
  {Flechtner}}, \bibinfo {author} {\bibfnamefont {C.}~\bibnamefont {Reigber}},
  \bibinfo {author} {\bibfnamefont {R.}~\bibnamefont {Rummel}}, \ and\ \bibinfo
  {author} {\bibfnamefont {G.}~\bibnamefont {Balmino}},\ }\href {\doibase
  10.1007/s10712-021-09658-0} {\bibfield  {journal} {\bibinfo  {journal}
  {Surveys in Geophysics}\ }\textbf {\bibinfo {volume} {42}},\ \bibinfo {pages}
  {1029} (\bibinfo {year} {2021})}\BibitemShut {NoStop}%
\bibitem [{\citenamefont {Lauf}\ \emph {et~al.}(2003)\citenamefont {Lauf},
  \citenamefont {Calhoun}, \citenamefont {Kuhnle}, \citenamefont {Sydnor},\
  and\ \citenamefont {Tjoelker}}]{Lauf:03}%
  \BibitemOpen
  \bibfield  {author} {\bibinfo {author} {\bibfnamefont {J.}~\bibnamefont
  {Lauf}}, \bibinfo {author} {\bibfnamefont {M.}~\bibnamefont {Calhoun}},
  \bibinfo {author} {\bibfnamefont {P.}~\bibnamefont {Kuhnle}}, \bibinfo
  {author} {\bibfnamefont {R.}~\bibnamefont {Sydnor}}, \ and\ \bibinfo {author}
  {\bibfnamefont {R.}~\bibnamefont {Tjoelker}},\ }in\ \href@noop {} {\emph
  {\bibinfo {booktitle} {Proceedings of the 35th Annual Precise Time and Time
  Interval Systems and Applications Meeting}}}\ (\bibinfo {year} {2003})\ pp.\
  \bibinfo {pages} {371--382}\BibitemShut {NoStop}%
\bibitem [{\citenamefont {Lechner}\ and\ \citenamefont
  {Baumann}(2000)}]{Lechner:00}%
  \BibitemOpen
  \bibfield  {author} {\bibinfo {author} {\bibfnamefont {W.}~\bibnamefont
  {Lechner}}\ and\ \bibinfo {author} {\bibfnamefont {S.}~\bibnamefont
  {Baumann}},\ }\href {\doibase https://doi.org/10.1016/S0168-1699(99)00056-3}
  {\bibfield  {journal} {\bibinfo  {journal} {Computers and Electronics in
  Agriculture}\ }\textbf {\bibinfo {volume} {25}},\ \bibinfo {pages} {67}
  (\bibinfo {year} {2000})}\BibitemShut {NoStop}%
\bibitem [{\citenamefont {Grewal}(2011)}]{Grewal:11}%
  \BibitemOpen
  \bibfield  {author} {\bibinfo {author} {\bibfnamefont {M.~S.}\ \bibnamefont
  {Grewal}},\ }\href {\doibase https://doi.org/10.1002/wics.158} {\bibfield
  {journal} {\bibinfo  {journal} {WIREs Computational Statistics}\ }\textbf
  {\bibinfo {volume} {3}},\ \bibinfo {pages} {383} (\bibinfo {year}
  {2011})}\BibitemShut {NoStop}%
\bibitem [{\citenamefont {Brasch}\ \emph {et~al.}(2014)\citenamefont {Brasch},
  \citenamefont {Chen}, \citenamefont {Schiller},\ and\ \citenamefont
  {Kippenberg}}]{Brasch:14}%
  \BibitemOpen
  \bibfield  {author} {\bibinfo {author} {\bibfnamefont {V.}~\bibnamefont
  {Brasch}}, \bibinfo {author} {\bibfnamefont {Q.-F.}\ \bibnamefont {Chen}},
  \bibinfo {author} {\bibfnamefont {S.}~\bibnamefont {Schiller}}, \ and\
  \bibinfo {author} {\bibfnamefont {T.~J.}\ \bibnamefont {Kippenberg}},\ }\href
  {\doibase 10.1364/OE.22.030786} {\bibfield  {journal} {\bibinfo  {journal}
  {Opt. Express}\ }\textbf {\bibinfo {volume} {22}},\ \bibinfo {pages} {30786}
  (\bibinfo {year} {2014})}\BibitemShut {NoStop}%
\bibitem [{\citenamefont {Du}\ \emph {et~al.}(2020)\citenamefont {Du},
  \citenamefont {Michon}, \citenamefont {Li}, \citenamefont {Kita},
  \citenamefont {Ma}, \citenamefont {Zuo}, \citenamefont {Yu}, \citenamefont
  {Gu}, \citenamefont {Agarwal}, \citenamefont {Li},\ and\ \citenamefont
  {Hu}}]{Du:20}%
  \BibitemOpen
  \bibfield  {author} {\bibinfo {author} {\bibfnamefont {Q.}~\bibnamefont
  {Du}}, \bibinfo {author} {\bibfnamefont {J.}~\bibnamefont {Michon}}, \bibinfo
  {author} {\bibfnamefont {B.}~\bibnamefont {Li}}, \bibinfo {author}
  {\bibfnamefont {D.}~\bibnamefont {Kita}}, \bibinfo {author} {\bibfnamefont
  {D.}~\bibnamefont {Ma}}, \bibinfo {author} {\bibfnamefont {H.}~\bibnamefont
  {Zuo}}, \bibinfo {author} {\bibfnamefont {S.}~\bibnamefont {Yu}}, \bibinfo
  {author} {\bibfnamefont {T.}~\bibnamefont {Gu}}, \bibinfo {author}
  {\bibfnamefont {A.}~\bibnamefont {Agarwal}}, \bibinfo {author} {\bibfnamefont
  {M.}~\bibnamefont {Li}}, \ and\ \bibinfo {author} {\bibfnamefont
  {J.}~\bibnamefont {Hu}},\ }\href {\doibase 10.1364/PRJ.379019} {\bibfield
  {journal} {\bibinfo  {journal} {Photon. Res.}\ }\textbf {\bibinfo {volume}
  {8}},\ \bibinfo {pages} {186} (\bibinfo {year} {2020})}\BibitemShut {NoStop}%
\bibitem [{\citenamefont {Ma}\ \emph {et~al.}(2016)\citenamefont {Ma},
  \citenamefont {Han}, \citenamefont {Du}, \citenamefont {Hu}, \citenamefont
  {Kimerling}, \citenamefont {Agarwal},\ and\ \citenamefont {Tan}}]{Ma:16}%
  \BibitemOpen
  \bibfield  {author} {\bibinfo {author} {\bibfnamefont {D.}~\bibnamefont
  {Ma}}, \bibinfo {author} {\bibfnamefont {Z.}~\bibnamefont {Han}}, \bibinfo
  {author} {\bibfnamefont {Q.}~\bibnamefont {Du}}, \bibinfo {author}
  {\bibfnamefont {J.}~\bibnamefont {Hu}}, \bibinfo {author} {\bibfnamefont
  {L.}~\bibnamefont {Kimerling}}, \bibinfo {author} {\bibfnamefont
  {A.}~\bibnamefont {Agarwal}}, \ and\ \bibinfo {author} {\bibfnamefont
  {D.~T.~H.}\ \bibnamefont {Tan}},\ }in\ \href {\doibase
  10.1109/ICSENS.2016.7808700} {\emph {\bibinfo {booktitle} {2016 IEEE
  SENSORS}}}\ (\bibinfo {year} {2016})\ pp.\ \bibinfo {pages}
  {1--3}\BibitemShut {NoStop}%
\bibitem [{\citenamefont {Moss}\ \emph {et~al.}(2013)\citenamefont {Moss},
  \citenamefont {Morandotti}, \citenamefont {Gaeta},\ and\ \citenamefont
  {Lipson}}]{Moss:13}%
  \BibitemOpen
  \bibfield  {author} {\bibinfo {author} {\bibfnamefont {D.~J.}\ \bibnamefont
  {Moss}}, \bibinfo {author} {\bibfnamefont {R.}~\bibnamefont {Morandotti}},
  \bibinfo {author} {\bibfnamefont {A.~L.}\ \bibnamefont {Gaeta}}, \ and\
  \bibinfo {author} {\bibfnamefont {M.}~\bibnamefont {Lipson}},\ }\href
  {https://doi.org/10.1038/nphoton.2013.183} {\bibfield  {journal} {\bibinfo
  {journal} {Nature Photonics}\ }\textbf {\bibinfo {volume} {7}},\ \bibinfo
  {pages} {597} (\bibinfo {year} {2013})}\BibitemShut {NoStop}%
\bibitem [{\citenamefont {Xiang}\ \emph {et~al.}(2022)\citenamefont {Xiang},
  \citenamefont {Jin},\ and\ \citenamefont {Bowers}}]{Xiang:22a}%
  \BibitemOpen
  \bibfield  {author} {\bibinfo {author} {\bibfnamefont {C.}~\bibnamefont
  {Xiang}}, \bibinfo {author} {\bibfnamefont {W.}~\bibnamefont {Jin}}, \ and\
  \bibinfo {author} {\bibfnamefont {J.~E.}\ \bibnamefont {Bowers}},\ }\href
  {\doibase 10.1364/PRJ.452936} {\bibfield  {journal} {\bibinfo  {journal}
  {Photon. Res.}\ }\textbf {\bibinfo {volume} {10}},\ \bibinfo {pages} {A82}
  (\bibinfo {year} {2022})}\BibitemShut {NoStop}%
\bibitem [{\citenamefont {Zhu}\ \emph {et~al.}(2021)\citenamefont {Zhu},
  \citenamefont {Shao}, \citenamefont {Yu}, \citenamefont {Cheng},
  \citenamefont {Desiatov}, \citenamefont {Xin}, \citenamefont {Hu},
  \citenamefont {Holzgrafe}, \citenamefont {Ghosh}, \citenamefont
  {Shams-Ansari}, \citenamefont {Puma}, \citenamefont {Sinclair}, \citenamefont
  {Reimer}, \citenamefont {Zhang},\ and\ \citenamefont {Lon\v{c}ar}}]{Zhu:21}%
  \BibitemOpen
  \bibfield  {author} {\bibinfo {author} {\bibfnamefont {D.}~\bibnamefont
  {Zhu}}, \bibinfo {author} {\bibfnamefont {L.}~\bibnamefont {Shao}}, \bibinfo
  {author} {\bibfnamefont {M.}~\bibnamefont {Yu}}, \bibinfo {author}
  {\bibfnamefont {R.}~\bibnamefont {Cheng}}, \bibinfo {author} {\bibfnamefont
  {B.}~\bibnamefont {Desiatov}}, \bibinfo {author} {\bibfnamefont {C.~J.}\
  \bibnamefont {Xin}}, \bibinfo {author} {\bibfnamefont {Y.}~\bibnamefont
  {Hu}}, \bibinfo {author} {\bibfnamefont {J.}~\bibnamefont {Holzgrafe}},
  \bibinfo {author} {\bibfnamefont {S.}~\bibnamefont {Ghosh}}, \bibinfo
  {author} {\bibfnamefont {A.}~\bibnamefont {Shams-Ansari}}, \bibinfo {author}
  {\bibfnamefont {E.}~\bibnamefont {Puma}}, \bibinfo {author} {\bibfnamefont
  {N.}~\bibnamefont {Sinclair}}, \bibinfo {author} {\bibfnamefont
  {C.}~\bibnamefont {Reimer}}, \bibinfo {author} {\bibfnamefont
  {M.}~\bibnamefont {Zhang}}, \ and\ \bibinfo {author} {\bibfnamefont
  {M.}~\bibnamefont {Lon\v{c}ar}},\ }\href {\doibase 10.1364/AOP.411024}
  {\bibfield  {journal} {\bibinfo  {journal} {Adv. Opt. Photon.}\ }\textbf
  {\bibinfo {volume} {13}},\ \bibinfo {pages} {242} (\bibinfo {year}
  {2021})}\BibitemShut {NoStop}%
\bibitem [{\citenamefont {Boes}\ \emph {et~al.}()\citenamefont {Boes},
  \citenamefont {Chang}, \citenamefont {Langrock}, \citenamefont {Yu},
  \citenamefont {Zhang}, \citenamefont {Lin}, \citenamefont {Lon{\v c}ar},
  \citenamefont {Fejer}, \citenamefont {Bowers},\ and\ \citenamefont
  {Mitchell}}]{Boes:23}%
  \BibitemOpen
  \bibfield  {author} {\bibinfo {author} {\bibfnamefont {A.}~\bibnamefont
  {Boes}}, \bibinfo {author} {\bibfnamefont {L.}~\bibnamefont {Chang}},
  \bibinfo {author} {\bibfnamefont {C.}~\bibnamefont {Langrock}}, \bibinfo
  {author} {\bibfnamefont {M.}~\bibnamefont {Yu}}, \bibinfo {author}
  {\bibfnamefont {M.}~\bibnamefont {Zhang}}, \bibinfo {author} {\bibfnamefont
  {Q.}~\bibnamefont {Lin}}, \bibinfo {author} {\bibfnamefont {M.}~\bibnamefont
  {Lon{\v c}ar}}, \bibinfo {author} {\bibfnamefont {M.}~\bibnamefont {Fejer}},
  \bibinfo {author} {\bibfnamefont {J.}~\bibnamefont {Bowers}}, \ and\ \bibinfo
  {author} {\bibfnamefont {A.}~\bibnamefont {Mitchell}},\ }\href {\doibase
  10.1126/science.abj4396} {\bibfield  {journal} {\bibinfo  {journal}
  {Science}\ }\textbf {\bibinfo {volume} {379}},\ \bibinfo {pages}
  {eabj4396}}\BibitemShut {NoStop}%
\bibitem [{\citenamefont {Lukin}\ \emph {et~al.}(2020)\citenamefont {Lukin},
  \citenamefont {Dory}, \citenamefont {Guidry}, \citenamefont {Yang},
  \citenamefont {Mishra}, \citenamefont {Trivedi}, \citenamefont {Radulaski},
  \citenamefont {Sun}, \citenamefont {Vercruysse}, \citenamefont {Ahn},\ and\
  \citenamefont {Vu{\v c}kovi{\'c}}}]{Lukin:20}%
  \BibitemOpen
  \bibfield  {author} {\bibinfo {author} {\bibfnamefont {D.~M.}\ \bibnamefont
  {Lukin}}, \bibinfo {author} {\bibfnamefont {C.}~\bibnamefont {Dory}},
  \bibinfo {author} {\bibfnamefont {M.~A.}\ \bibnamefont {Guidry}}, \bibinfo
  {author} {\bibfnamefont {K.~Y.}\ \bibnamefont {Yang}}, \bibinfo {author}
  {\bibfnamefont {S.~D.}\ \bibnamefont {Mishra}}, \bibinfo {author}
  {\bibfnamefont {R.}~\bibnamefont {Trivedi}}, \bibinfo {author} {\bibfnamefont
  {M.}~\bibnamefont {Radulaski}}, \bibinfo {author} {\bibfnamefont
  {S.}~\bibnamefont {Sun}}, \bibinfo {author} {\bibfnamefont {D.}~\bibnamefont
  {Vercruysse}}, \bibinfo {author} {\bibfnamefont {G.~H.}\ \bibnamefont {Ahn}},
  \ and\ \bibinfo {author} {\bibfnamefont {J.}~\bibnamefont {Vu{\v
  c}kovi{\'c}}},\ }\href {\doibase 10.1038/s41566-019-0556-6} {\bibfield
  {journal} {\bibinfo  {journal} {Nature Photonics}\ }\textbf {\bibinfo
  {volume} {14}},\ \bibinfo {pages} {330} (\bibinfo {year} {2020})}\BibitemShut
  {NoStop}%
\bibitem [{\citenamefont {Yi}\ \emph {et~al.}(2022)\citenamefont {Yi},
  \citenamefont {Wang}, \citenamefont {Zhou}, \citenamefont {Zhu},
  \citenamefont {Zhang}, \citenamefont {You}, \citenamefont {Zhang},\ and\
  \citenamefont {Ou}}]{Yi:22}%
  \BibitemOpen
  \bibfield  {author} {\bibinfo {author} {\bibfnamefont {A.}~\bibnamefont
  {Yi}}, \bibinfo {author} {\bibfnamefont {C.}~\bibnamefont {Wang}}, \bibinfo
  {author} {\bibfnamefont {L.}~\bibnamefont {Zhou}}, \bibinfo {author}
  {\bibfnamefont {Y.}~\bibnamefont {Zhu}}, \bibinfo {author} {\bibfnamefont
  {S.}~\bibnamefont {Zhang}}, \bibinfo {author} {\bibfnamefont
  {T.}~\bibnamefont {You}}, \bibinfo {author} {\bibfnamefont {J.}~\bibnamefont
  {Zhang}}, \ and\ \bibinfo {author} {\bibfnamefont {X.}~\bibnamefont {Ou}},\
  }\href {\doibase 10.1063/5.0079649} {\bibfield  {journal} {\bibinfo
  {journal} {Applied Physics Reviews}\ }\textbf {\bibinfo {volume} {9}},\
  \bibinfo {pages} {031302} (\bibinfo {year} {2022})}\BibitemShut {NoStop}%
\bibitem [{\citenamefont {Lu}\ \emph {et~al.}(2022)\citenamefont {Lu},
  \citenamefont {Cao}, \citenamefont {Peng},\ and\ \citenamefont
  {Pan}}]{LuCY:22}%
  \BibitemOpen
  \bibfield  {author} {\bibinfo {author} {\bibfnamefont {C.-Y.}\ \bibnamefont
  {Lu}}, \bibinfo {author} {\bibfnamefont {Y.}~\bibnamefont {Cao}}, \bibinfo
  {author} {\bibfnamefont {C.-Z.}\ \bibnamefont {Peng}}, \ and\ \bibinfo
  {author} {\bibfnamefont {J.-W.}\ \bibnamefont {Pan}},\ }\href {\doibase
  10.1103/RevModPhys.94.035001} {\bibfield  {journal} {\bibinfo  {journal}
  {Rev. Mod. Phys.}\ }\textbf {\bibinfo {volume} {94}},\ \bibinfo {pages}
  {035001} (\bibinfo {year} {2022})}\BibitemShut {NoStop}%
\bibitem [{\citenamefont {Stassinopoulos}\ and\ \citenamefont
  {Raymond}(1988)}]{Stassinopoulos:88}%
  \BibitemOpen
  \bibfield  {author} {\bibinfo {author} {\bibfnamefont {E.}~\bibnamefont
  {Stassinopoulos}}\ and\ \bibinfo {author} {\bibfnamefont {J.}~\bibnamefont
  {Raymond}},\ }\href {\doibase 10.1109/5.90113} {\bibfield  {journal}
  {\bibinfo  {journal} {Proceedings of the IEEE}\ }\textbf {\bibinfo {volume}
  {76}},\ \bibinfo {pages} {1423} (\bibinfo {year} {1988})}\BibitemShut
  {NoStop}%
\bibitem [{\citenamefont {Stassinopoulos}(1980)}]{Stassinopoulos:80}%
  \BibitemOpen
  \bibfield  {author} {\bibinfo {author} {\bibfnamefont {E.~G.}\ \bibnamefont
  {Stassinopoulos}},\ }\href {\doibase 10.2514/3.28028} {\bibfield  {journal}
  {\bibinfo  {journal} {Journal of Spacecraft and Rockets}\ }\textbf {\bibinfo
  {volume} {17}},\ \bibinfo {pages} {145} (\bibinfo {year} {1980})},\ \Eprint
  {http://arxiv.org/abs/https://doi.org/10.2514/3.28028}
  {https://doi.org/10.2514/3.28028} \BibitemShut {NoStop}%
\bibitem [{\citenamefont {Curto}(2020)}]{Curto:20}%
  \BibitemOpen
  \bibfield  {author} {\bibinfo {author} {\bibfnamefont {J.~J.}\ \bibnamefont
  {Curto}},\ }\href@noop {} {\bibfield  {journal} {\bibinfo  {journal} {Journal
  of Space Weather and Space Climate}\ }\textbf {\bibinfo {volume} {10}},\
  \bibinfo {pages} {27} (\bibinfo {year} {2020})}\BibitemShut {NoStop}%
\bibitem [{\citenamefont {Leka}\ and\ \citenamefont {Barnes}(2018)}]{Leka:18}%
  \BibitemOpen
  \bibfield  {author} {\bibinfo {author} {\bibfnamefont {K.}~\bibnamefont
  {Leka}}\ and\ \bibinfo {author} {\bibfnamefont {G.}~\bibnamefont {Barnes}},\
  }in\ \href {\doibase https://doi.org/10.1016/B978-0-12-812700-1.00003-0}
  {\emph {\bibinfo {booktitle} {Extreme Events in Geospace}}},\ \bibinfo
  {editor} {edited by\ \bibinfo {editor} {\bibfnamefont {N.}~\bibnamefont
  {Buzulukova}}}\ (\bibinfo  {publisher} {Elsevier},\ \bibinfo {year} {2018})\
  pp.\ \bibinfo {pages} {65--98}\BibitemShut {NoStop}%
\bibitem [{\citenamefont {Bhat}\ \emph {et~al.}(2005)\citenamefont {Bhat},
  \citenamefont {Upadhyaya},\ and\ \citenamefont {Kulkarni}}]{Bhat:05}%
  \BibitemOpen
  \bibfield  {author} {\bibinfo {author} {\bibfnamefont {B.}~\bibnamefont
  {Bhat}}, \bibinfo {author} {\bibfnamefont {N.}~\bibnamefont {Upadhyaya}}, \
  and\ \bibinfo {author} {\bibfnamefont {R.}~\bibnamefont {Kulkarni}},\ }\href
  {\doibase 10.1109/TNS.2005.846881} {\bibfield  {journal} {\bibinfo  {journal}
  {IEEE Transactions on Nuclear Science}\ }\textbf {\bibinfo {volume} {52}},\
  \bibinfo {pages} {530} (\bibinfo {year} {2005})}\BibitemShut {NoStop}%
\bibitem [{\citenamefont {McMillan}(1945)}]{McMillan:45}%
  \BibitemOpen
  \bibfield  {author} {\bibinfo {author} {\bibfnamefont {E.~M.}\ \bibnamefont
  {McMillan}},\ }\href {\doibase 10.1103/PhysRev.68.143} {\bibfield  {journal}
  {\bibinfo  {journal} {Phys. Rev.}\ }\textbf {\bibinfo {volume} {68}},\
  \bibinfo {pages} {143} (\bibinfo {year} {1945})}\BibitemShut {NoStop}%
\bibitem [{\citenamefont {Wilson}(1977)}]{Wilson:77}%
  \BibitemOpen
  \bibfield  {author} {\bibinfo {author} {\bibfnamefont {E.~J.~N.}\
  \bibnamefont {Wilson}},\ }\href
  {http://inis.iaea.org/search/search.aspx?orig_q=RN:09395643} {\emph {\bibinfo
  {title} {Proton synchrotron accelerator theory}}},\ \bibinfo {type} {Tech.
  Rep.}\ (\bibinfo {address} {European Organization for Nuclear Research
  (CERN)},\ \bibinfo {year} {Mar 1977})\ \bibinfo {note}
  {cERN--77-07}\BibitemShut {NoStop}%
\bibitem [{\citenamefont {Bauters}\ \emph {et~al.}(2011)\citenamefont
  {Bauters}, \citenamefont {Heck}, \citenamefont {John}, \citenamefont
  {Barton}, \citenamefont {Bruinink}, \citenamefont {Leinse}, \citenamefont
  {Heideman}, \citenamefont {Blumenthal},\ and\ \citenamefont
  {Bowers}}]{Bauters:11}%
  \BibitemOpen
  \bibfield  {author} {\bibinfo {author} {\bibfnamefont {J.~F.}\ \bibnamefont
  {Bauters}}, \bibinfo {author} {\bibfnamefont {M.~J.~R.}\ \bibnamefont
  {Heck}}, \bibinfo {author} {\bibfnamefont {D.~D.}\ \bibnamefont {John}},
  \bibinfo {author} {\bibfnamefont {J.~S.}\ \bibnamefont {Barton}}, \bibinfo
  {author} {\bibfnamefont {C.~M.}\ \bibnamefont {Bruinink}}, \bibinfo {author}
  {\bibfnamefont {A.}~\bibnamefont {Leinse}}, \bibinfo {author} {\bibfnamefont
  {R.~G.}\ \bibnamefont {Heideman}}, \bibinfo {author} {\bibfnamefont {D.~J.}\
  \bibnamefont {Blumenthal}}, \ and\ \bibinfo {author} {\bibfnamefont {J.~E.}\
  \bibnamefont {Bowers}},\ }\href {\doibase 10.1364/OE.19.024090} {\bibfield
  {journal} {\bibinfo  {journal} {Opt. Express}\ }\textbf {\bibinfo {volume}
  {19}},\ \bibinfo {pages} {24090} (\bibinfo {year} {2011})}\BibitemShut
  {NoStop}%
\bibitem [{\citenamefont {Soller}\ \emph {et~al.}(2005)\citenamefont {Soller},
  \citenamefont {Gifford}, \citenamefont {Wolfe},\ and\ \citenamefont
  {Froggatt}}]{Soller:05}%
  \BibitemOpen
  \bibfield  {author} {\bibinfo {author} {\bibfnamefont {B.~J.}\ \bibnamefont
  {Soller}}, \bibinfo {author} {\bibfnamefont {D.~K.}\ \bibnamefont {Gifford}},
  \bibinfo {author} {\bibfnamefont {M.~S.}\ \bibnamefont {Wolfe}}, \ and\
  \bibinfo {author} {\bibfnamefont {M.~E.}\ \bibnamefont {Froggatt}},\ }\href
  {\doibase 10.1364/OPEX.13.000666} {\bibfield  {journal} {\bibinfo  {journal}
  {Opt. Express}\ }\textbf {\bibinfo {volume} {13}},\ \bibinfo {pages} {666}
  (\bibinfo {year} {2005})}\BibitemShut {NoStop}%
\bibitem [{\citenamefont {Shi}\ \emph {et~al.}(2024)\citenamefont {Shi},
  \citenamefont {Luo}, \citenamefont {Sun}, \citenamefont {Hu}, \citenamefont
  {Long}, \citenamefont {Bai}, \citenamefont {Wang},\ and\ \citenamefont
  {Liu}}]{Shi:24}%
  \BibitemOpen
  \bibfield  {author} {\bibinfo {author} {\bibfnamefont {B.}~\bibnamefont
  {Shi}}, \bibinfo {author} {\bibfnamefont {Y.-H.}\ \bibnamefont {Luo}},
  \bibinfo {author} {\bibfnamefont {W.}~\bibnamefont {Sun}}, \bibinfo {author}
  {\bibfnamefont {Y.}~\bibnamefont {Hu}}, \bibinfo {author} {\bibfnamefont
  {J.}~\bibnamefont {Long}}, \bibinfo {author} {\bibfnamefont {X.}~\bibnamefont
  {Bai}}, \bibinfo {author} {\bibfnamefont {A.}~\bibnamefont {Wang}}, \ and\
  \bibinfo {author} {\bibfnamefont {J.}~\bibnamefont {Liu}},\ }\href {\doibase
  10.1364/PRJ.510795} {\bibfield  {journal} {\bibinfo  {journal} {Photon.
  Res.}\ }\textbf {\bibinfo {volume} {12}},\ \bibinfo {pages} {663} (\bibinfo
  {year} {2024})}\BibitemShut {NoStop}%
\bibitem [{\citenamefont {Luo}\ \emph {et~al.}(2024)\citenamefont {Luo},
  \citenamefont {Shi}, \citenamefont {Sun}, \citenamefont {Chen}, \citenamefont
  {Huang}, \citenamefont {Wang}, \citenamefont {Long}, \citenamefont {Shen},
  \citenamefont {Ye}, \citenamefont {Guo},\ and\ \citenamefont {Liu}}]{Luo:24}%
  \BibitemOpen
  \bibfield  {author} {\bibinfo {author} {\bibfnamefont {Y.-H.}\ \bibnamefont
  {Luo}}, \bibinfo {author} {\bibfnamefont {B.}~\bibnamefont {Shi}}, \bibinfo
  {author} {\bibfnamefont {W.}~\bibnamefont {Sun}}, \bibinfo {author}
  {\bibfnamefont {R.}~\bibnamefont {Chen}}, \bibinfo {author} {\bibfnamefont
  {S.}~\bibnamefont {Huang}}, \bibinfo {author} {\bibfnamefont
  {Z.}~\bibnamefont {Wang}}, \bibinfo {author} {\bibfnamefont {J.}~\bibnamefont
  {Long}}, \bibinfo {author} {\bibfnamefont {C.}~\bibnamefont {Shen}}, \bibinfo
  {author} {\bibfnamefont {Z.}~\bibnamefont {Ye}}, \bibinfo {author}
  {\bibfnamefont {H.}~\bibnamefont {Guo}}, \ and\ \bibinfo {author}
  {\bibfnamefont {J.}~\bibnamefont {Liu}},\ }\href
  {https://doi.org/10.1038/s41377-024-01435-z} {\bibfield  {journal} {\bibinfo
  {journal} {Light: Science \& Applications}\ }\textbf {\bibinfo {volume} {13}}
  (\bibinfo {year} {2024})}\BibitemShut {NoStop}%
\bibitem [{\citenamefont {Liu}\ \emph {et~al.}(2018)\citenamefont {Liu},
  \citenamefont {Raja}, \citenamefont {Pfeiffer}, \citenamefont {Herkommer},
  \citenamefont {Guo}, \citenamefont {Zervas}, \citenamefont {Geiselmann},\
  and\ \citenamefont {Kippenberg}}]{Liu:18}%
  \BibitemOpen
  \bibfield  {author} {\bibinfo {author} {\bibfnamefont {J.}~\bibnamefont
  {Liu}}, \bibinfo {author} {\bibfnamefont {A.~S.}\ \bibnamefont {Raja}},
  \bibinfo {author} {\bibfnamefont {M.~H.~P.}\ \bibnamefont {Pfeiffer}},
  \bibinfo {author} {\bibfnamefont {C.}~\bibnamefont {Herkommer}}, \bibinfo
  {author} {\bibfnamefont {H.}~\bibnamefont {Guo}}, \bibinfo {author}
  {\bibfnamefont {M.}~\bibnamefont {Zervas}}, \bibinfo {author} {\bibfnamefont
  {M.}~\bibnamefont {Geiselmann}}, \ and\ \bibinfo {author} {\bibfnamefont
  {T.~J.}\ \bibnamefont {Kippenberg}},\ }\href {\doibase 10.1364/OL.43.003200}
  {\bibfield  {journal} {\bibinfo  {journal} {Opt. Lett.}\ }\textbf {\bibinfo
  {volume} {43}},\ \bibinfo {pages} {3200} (\bibinfo {year}
  {2018})}\BibitemShut {NoStop}%
\bibitem [{\citenamefont {Pfeiffer}\ \emph {et~al.}(2017)\citenamefont
  {Pfeiffer}, \citenamefont {Herkommer}, \citenamefont {Liu}, \citenamefont
  {Guo}, \citenamefont {Karpov}, \citenamefont {Lucas}, \citenamefont
  {Zervas},\ and\ \citenamefont {Kippenberg}}]{Pfeiffer:17}%
  \BibitemOpen
  \bibfield  {author} {\bibinfo {author} {\bibfnamefont {M.~H.~P.}\
  \bibnamefont {Pfeiffer}}, \bibinfo {author} {\bibfnamefont {C.}~\bibnamefont
  {Herkommer}}, \bibinfo {author} {\bibfnamefont {J.}~\bibnamefont {Liu}},
  \bibinfo {author} {\bibfnamefont {H.}~\bibnamefont {Guo}}, \bibinfo {author}
  {\bibfnamefont {M.}~\bibnamefont {Karpov}}, \bibinfo {author} {\bibfnamefont
  {E.}~\bibnamefont {Lucas}}, \bibinfo {author} {\bibfnamefont
  {M.}~\bibnamefont {Zervas}}, \ and\ \bibinfo {author} {\bibfnamefont {T.~J.}\
  \bibnamefont {Kippenberg}},\ }\href {\doibase 10.1364/OPTICA.4.000684}
  {\bibfield  {journal} {\bibinfo  {journal} {Optica}\ }\textbf {\bibinfo
  {volume} {4}},\ \bibinfo {pages} {684} (\bibinfo {year} {2017})}\BibitemShut
  {NoStop}%
\bibitem [{\citenamefont {Li}\ \emph {et~al.}(2013)\citenamefont {Li},
  \citenamefont {Eftekhar}, \citenamefont {Xia},\ and\ \citenamefont
  {Adibi}}]{Li:13}%
  \BibitemOpen
  \bibfield  {author} {\bibinfo {author} {\bibfnamefont {Q.}~\bibnamefont
  {Li}}, \bibinfo {author} {\bibfnamefont {A.~A.}\ \bibnamefont {Eftekhar}},
  \bibinfo {author} {\bibfnamefont {Z.}~\bibnamefont {Xia}}, \ and\ \bibinfo
  {author} {\bibfnamefont {A.}~\bibnamefont {Adibi}},\ }\href {\doibase
  10.1103/PhysRevA.88.033816} {\bibfield  {journal} {\bibinfo  {journal} {Phys.
  Rev. A}\ }\textbf {\bibinfo {volume} {88}},\ \bibinfo {pages} {033816}
  (\bibinfo {year} {2013})}\BibitemShut {NoStop}%
\end{thebibliography}%


\begin{thebibliography}{12}%
\makeatletter
\providecommand \@ifxundefined [1]{%
 \@ifx{#1\undefined}
}%
\providecommand \@ifnum [1]{%
 \ifnum #1\expandafter \@firstoftwo
 \else \expandafter \@secondoftwo
 \fi
}%
\providecommand \@ifx [1]{%
 \ifx #1\expandafter \@firstoftwo
 \else \expandafter \@secondoftwo
 \fi
}%
\providecommand \natexlab [1]{#1}%
\providecommand \enquote  [1]{``#1''}%
\providecommand \bibnamefont  [1]{#1}%
\providecommand \bibfnamefont [1]{#1}%
\providecommand \citenamefont [1]{#1}%
\providecommand \href@noop [0]{\@secondoftwo}%
\providecommand \href [0]{\begingroup \@sanitize@url \@href}%
\providecommand \@href[1]{\@@startlink{#1}\@@href}%
\providecommand \@@href[1]{\endgroup#1\@@endlink}%
\providecommand \@sanitize@url [0]{\catcode `\\12\catcode `\$12\catcode
  `\&12\catcode `\#12\catcode `\^12\catcode `\_12\catcode `\%12\relax}%
\providecommand \@@startlink[1]{}%
\providecommand \@@endlink[0]{}%
\providecommand \url  [0]{\begingroup\@sanitize@url \@url }%
\providecommand \@url [1]{\endgroup\@href {#1}{\urlprefix }}%
\providecommand \urlprefix  [0]{URL }%
\providecommand \Eprint [0]{\href }%
\providecommand \doibase [0]{http://dx.doi.org/}%
\providecommand \selectlanguage [0]{\@gobble}%
\providecommand \bibinfo  [0]{\@secondoftwo}%
\providecommand \bibfield  [0]{\@secondoftwo}%
\providecommand \translation [1]{[#1]}%
\providecommand \BibitemOpen [0]{}%
\providecommand \bibitemStop [0]{}%
\providecommand \bibitemNoStop [0]{.\EOS\space}%
\providecommand \EOS [0]{\spacefactor3000\relax}%
\providecommand \BibitemShut  [1]{\csname bibitem#1\endcsname}%
\let\auto@bib@innerbib\@empty
\bibitem [{\citenamefont {Ye}\ \emph {et~al.}(2023)\citenamefont {Ye},
  \citenamefont {Jia}, \citenamefont {Huang}, \citenamefont {Shen},
  \citenamefont {Long}, \citenamefont {Shi}, \citenamefont {Luo}, \citenamefont
  {Gao}, \citenamefont {Sun}, \citenamefont {Guo}, \citenamefont {He},\ and\
  \citenamefont {Liu}}]{Ye:23}%
  \BibitemOpen
  \bibfield  {author} {\bibinfo {author} {\bibfnamefont {Z.}~\bibnamefont
  {Ye}}, \bibinfo {author} {\bibfnamefont {H.}~\bibnamefont {Jia}}, \bibinfo
  {author} {\bibfnamefont {Z.}~\bibnamefont {Huang}}, \bibinfo {author}
  {\bibfnamefont {C.}~\bibnamefont {Shen}}, \bibinfo {author} {\bibfnamefont
  {J.}~\bibnamefont {Long}}, \bibinfo {author} {\bibfnamefont {B.}~\bibnamefont
  {Shi}}, \bibinfo {author} {\bibfnamefont {Y.-H.}\ \bibnamefont {Luo}},
  \bibinfo {author} {\bibfnamefont {L.}~\bibnamefont {Gao}}, \bibinfo {author}
  {\bibfnamefont {W.}~\bibnamefont {Sun}}, \bibinfo {author} {\bibfnamefont
  {H.}~\bibnamefont {Guo}}, \bibinfo {author} {\bibfnamefont {J.}~\bibnamefont
  {He}}, \ and\ \bibinfo {author} {\bibfnamefont {J.}~\bibnamefont {Liu}},\
  }\href {\doibase 10.1364/PRJ.486379} {\bibfield  {journal} {\bibinfo
  {journal} {Photon. Res.}\ }\textbf {\bibinfo {volume} {11}},\ \bibinfo
  {pages} {558} (\bibinfo {year} {2023})}\BibitemShut {NoStop}%
\bibitem [{\citenamefont {Sun}\ \emph {et~al.}(2024)\citenamefont {Sun},
  \citenamefont {Chen}, \citenamefont {Li}, \citenamefont {Shen}, \citenamefont
  {Long}, \citenamefont {Zheng}, \citenamefont {Wang}, \citenamefont {Chen},
  \citenamefont {Zhang}, \citenamefont {Shi}, \citenamefont {Li}, \citenamefont
  {Gao}, \citenamefont {Luo}, \citenamefont {Chen},\ and\ \citenamefont
  {Liu.}}]{Sun:24}%
  \BibitemOpen
  \bibfield  {author} {\bibinfo {author} {\bibfnamefont {W.}~\bibnamefont
  {Sun}}, \bibinfo {author} {\bibfnamefont {Z.}~\bibnamefont {Chen}}, \bibinfo
  {author} {\bibfnamefont {L.}~\bibnamefont {Li}}, \bibinfo {author}
  {\bibfnamefont {C.}~\bibnamefont {Shen}}, \bibinfo {author} {\bibfnamefont
  {J.}~\bibnamefont {Long}}, \bibinfo {author} {\bibfnamefont {H.}~\bibnamefont
  {Zheng}}, \bibinfo {author} {\bibfnamefont {L.}~\bibnamefont {Wang}},
  \bibinfo {author} {\bibfnamefont {Q.}~\bibnamefont {Chen}}, \bibinfo {author}
  {\bibfnamefont {Z.}~\bibnamefont {Zhang}}, \bibinfo {author} {\bibfnamefont
  {B.}~\bibnamefont {Shi}}, \bibinfo {author} {\bibfnamefont {S.}~\bibnamefont
  {Li}}, \bibinfo {author} {\bibfnamefont {L.}~\bibnamefont {Gao}}, \bibinfo
  {author} {\bibfnamefont {Y.-H.}\ \bibnamefont {Luo}}, \bibinfo {author}
  {\bibfnamefont {B.}~\bibnamefont {Chen}}, \ and\ \bibinfo {author}
  {\bibfnamefont {J.}~\bibnamefont {Liu.}},\ }\href@noop {} {\bibfield
  {journal} {\bibinfo  {journal} {arXiv}\ }\textbf {\bibinfo {volume}
  {2403.02828}} (\bibinfo {year} {2024})}\BibitemShut {NoStop}%
\bibitem [{\citenamefont {Wang}\ \emph {et~al.}(2022)\citenamefont {Wang},
  \citenamefont {Li}, \citenamefont {Yi}, \citenamefont {Fang}, \citenamefont
  {Zhou}, \citenamefont {Wang}, \citenamefont {Niu}, \citenamefont {Chen},
  \citenamefont {Zhang}, \citenamefont {Cheng}, \citenamefont {Liu},
  \citenamefont {Dong},\ and\ \citenamefont {Ou}}]{Wang:22}%
  \BibitemOpen
  \bibfield  {author} {\bibinfo {author} {\bibfnamefont {C.}~\bibnamefont
  {Wang}}, \bibinfo {author} {\bibfnamefont {J.}~\bibnamefont {Li}}, \bibinfo
  {author} {\bibfnamefont {A.}~\bibnamefont {Yi}}, \bibinfo {author}
  {\bibfnamefont {Z.}~\bibnamefont {Fang}}, \bibinfo {author} {\bibfnamefont
  {L.}~\bibnamefont {Zhou}}, \bibinfo {author} {\bibfnamefont {Z.}~\bibnamefont
  {Wang}}, \bibinfo {author} {\bibfnamefont {R.}~\bibnamefont {Niu}}, \bibinfo
  {author} {\bibfnamefont {Y.}~\bibnamefont {Chen}}, \bibinfo {author}
  {\bibfnamefont {J.}~\bibnamefont {Zhang}}, \bibinfo {author} {\bibfnamefont
  {Y.}~\bibnamefont {Cheng}}, \bibinfo {author} {\bibfnamefont
  {J.}~\bibnamefont {Liu}}, \bibinfo {author} {\bibfnamefont {C.-H.}\
  \bibnamefont {Dong}}, \ and\ \bibinfo {author} {\bibfnamefont
  {X.}~\bibnamefont {Ou}},\ }\href {\doibase 10.1038/s41377-022-01042-w}
  {\bibfield  {journal} {\bibinfo  {journal} {Light: Science \& Applications}\
  }\textbf {\bibinfo {volume} {11}},\ \bibinfo {pages} {341} (\bibinfo {year}
  {2022})}\BibitemShut {NoStop}%
\bibitem [{\citenamefont {Yi}\ \emph {et~al.}(2022)\citenamefont {Yi},
  \citenamefont {Wang}, \citenamefont {Zhou}, \citenamefont {Zhu},
  \citenamefont {Zhang}, \citenamefont {You}, \citenamefont {Zhang},\ and\
  \citenamefont {Ou}}]{Yi:22}%
  \BibitemOpen
  \bibfield  {author} {\bibinfo {author} {\bibfnamefont {A.}~\bibnamefont
  {Yi}}, \bibinfo {author} {\bibfnamefont {C.}~\bibnamefont {Wang}}, \bibinfo
  {author} {\bibfnamefont {L.}~\bibnamefont {Zhou}}, \bibinfo {author}
  {\bibfnamefont {Y.}~\bibnamefont {Zhu}}, \bibinfo {author} {\bibfnamefont
  {S.}~\bibnamefont {Zhang}}, \bibinfo {author} {\bibfnamefont
  {T.}~\bibnamefont {You}}, \bibinfo {author} {\bibfnamefont {J.}~\bibnamefont
  {Zhang}}, \ and\ \bibinfo {author} {\bibfnamefont {X.}~\bibnamefont {Ou}},\
  }\href {\doibase 10.1063/5.0079649} {\bibfield  {journal} {\bibinfo
  {journal} {Applied Physics Reviews}\ }\textbf {\bibinfo {volume} {9}},\
  \bibinfo {pages} {031302} (\bibinfo {year} {2022})}\BibitemShut {NoStop}%
\bibitem [{\citenamefont {Wang}\ \emph {et~al.}(2018)\citenamefont {Wang},
  \citenamefont {Zhang}, \citenamefont {Chen}, \citenamefont {Bertrand},
  \citenamefont {Shams-Ansari}, \citenamefont {Chandrasekhar}, \citenamefont
  {Winzer},\ and\ \citenamefont {Lon{\v c}ar}}]{WangC:18}%
  \BibitemOpen
  \bibfield  {author} {\bibinfo {author} {\bibfnamefont {C.}~\bibnamefont
  {Wang}}, \bibinfo {author} {\bibfnamefont {M.}~\bibnamefont {Zhang}},
  \bibinfo {author} {\bibfnamefont {X.}~\bibnamefont {Chen}}, \bibinfo {author}
  {\bibfnamefont {M.}~\bibnamefont {Bertrand}}, \bibinfo {author}
  {\bibfnamefont {A.}~\bibnamefont {Shams-Ansari}}, \bibinfo {author}
  {\bibfnamefont {S.}~\bibnamefont {Chandrasekhar}}, \bibinfo {author}
  {\bibfnamefont {P.}~\bibnamefont {Winzer}}, \ and\ \bibinfo {author}
  {\bibfnamefont {M.}~\bibnamefont {Lon{\v c}ar}},\ }\href {\doibase
  10.1038/s41586-018-0551-y} {\bibfield  {journal} {\bibinfo  {journal}
  {Nature}\ }\textbf {\bibinfo {volume} {562}},\ \bibinfo {pages} {101}
  (\bibinfo {year} {2018})}\BibitemShut {NoStop}%
\bibitem [{\citenamefont {Wan}\ \emph {et~al.}(2024)\citenamefont {Wan},
  \citenamefont {Wang}, \citenamefont {Ma}, \citenamefont {Wang}, \citenamefont
  {Niu}, \citenamefont {He}, \citenamefont {Guo}, \citenamefont {Bo},
  \citenamefont {Liu},\ and\ \citenamefont {Dong}}]{Wan:24}%
  \BibitemOpen
  \bibfield  {author} {\bibinfo {author} {\bibfnamefont {S.}~\bibnamefont
  {Wan}}, \bibinfo {author} {\bibfnamefont {P.-Y.}\ \bibnamefont {Wang}},
  \bibinfo {author} {\bibfnamefont {R.}~\bibnamefont {Ma}}, \bibinfo {author}
  {\bibfnamefont {Z.-Y.}\ \bibnamefont {Wang}}, \bibinfo {author}
  {\bibfnamefont {R.}~\bibnamefont {Niu}}, \bibinfo {author} {\bibfnamefont
  {D.-Y.}\ \bibnamefont {He}}, \bibinfo {author} {\bibfnamefont {G.-C.}\
  \bibnamefont {Guo}}, \bibinfo {author} {\bibfnamefont {F.}~\bibnamefont
  {Bo}}, \bibinfo {author} {\bibfnamefont {J.}~\bibnamefont {Liu}}, \ and\
  \bibinfo {author} {\bibfnamefont {C.-H.}\ \bibnamefont {Dong}},\ }\href
  {\doibase https://doi.org/10.1002/lpor.202300627} {\bibfield  {journal}
  {\bibinfo  {journal} {Laser \& Photonics Reviews}\ }\textbf {\bibinfo
  {volume} {18}},\ \bibinfo {pages} {2300627} (\bibinfo {year}
  {2024})}\BibitemShut {NoStop}%
\bibitem [{\citenamefont {Li}\ \emph {et~al.}(2013)\citenamefont {Li},
  \citenamefont {Eftekhar}, \citenamefont {Xia},\ and\ \citenamefont
  {Adibi}}]{Li:13}%
  \BibitemOpen
  \bibfield  {author} {\bibinfo {author} {\bibfnamefont {Q.}~\bibnamefont
  {Li}}, \bibinfo {author} {\bibfnamefont {A.~A.}\ \bibnamefont {Eftekhar}},
  \bibinfo {author} {\bibfnamefont {Z.}~\bibnamefont {Xia}}, \ and\ \bibinfo
  {author} {\bibfnamefont {A.}~\bibnamefont {Adibi}},\ }\href {\doibase
  10.1103/PhysRevA.88.033816} {\bibfield  {journal} {\bibinfo  {journal} {Phys.
  Rev. A}\ }\textbf {\bibinfo {volume} {88}},\ \bibinfo {pages} {033816}
  (\bibinfo {year} {2013})}\BibitemShut {NoStop}%
\bibitem [{\citenamefont {Sergo}\ \emph {et~al.}(1996)\citenamefont {Sergo},
  \citenamefont {Pezzotti}, \citenamefont {Katagiri}, \citenamefont {Muraki},\
  and\ \citenamefont {Nishida}}]{Sergo:96}%
  \BibitemOpen
  \bibfield  {author} {\bibinfo {author} {\bibfnamefont {V.}~\bibnamefont
  {Sergo}}, \bibinfo {author} {\bibfnamefont {G.}~\bibnamefont {Pezzotti}},
  \bibinfo {author} {\bibfnamefont {G.}~\bibnamefont {Katagiri}}, \bibinfo
  {author} {\bibfnamefont {N.}~\bibnamefont {Muraki}}, \ and\ \bibinfo {author}
  {\bibfnamefont {T.}~\bibnamefont {Nishida}},\ }\href {\doibase
  https://doi.org/10.1111/j.1151-2916.1996.tb07944.x} {\bibfield  {journal}
  {\bibinfo  {journal} {Journal of the American Ceramic Society}\ }\textbf
  {\bibinfo {volume} {79}},\ \bibinfo {pages} {781} (\bibinfo {year}
  {1996})}\BibitemShut {NoStop}%
\bibitem [{\citenamefont {Yang}\ \emph {et~al.}(1987)\citenamefont {Yang},
  \citenamefont {Lan}, \citenamefont {Li},\ and\ \citenamefont
  {Wang}}]{Yang:87}%
  \BibitemOpen
  \bibfield  {author} {\bibinfo {author} {\bibfnamefont {X.}~\bibnamefont
  {Yang}}, \bibinfo {author} {\bibfnamefont {G.}~\bibnamefont {Lan}}, \bibinfo
  {author} {\bibfnamefont {B.}~\bibnamefont {Li}}, \ and\ \bibinfo {author}
  {\bibfnamefont {H.}~\bibnamefont {Wang}},\ }\href {\doibase
  https://doi.org/10.1002/pssb.2221420130} {\bibfield  {journal} {\bibinfo
  {journal} {physica status solidi (b)}\ }\textbf {\bibinfo {volume} {142}},\
  \bibinfo {pages} {287} (\bibinfo {year} {1987})}\BibitemShut {NoStop}%
\bibitem [{\citenamefont {Wu}\ \emph {et~al.}(1995)\citenamefont {Wu},
  \citenamefont {Zhang}, \citenamefont {Yan},\ and\ \citenamefont
  {Feng}}]{Wu:95}%
  \BibitemOpen
  \bibfield  {author} {\bibinfo {author} {\bibfnamefont {X.}~\bibnamefont
  {Wu}}, \bibinfo {author} {\bibfnamefont {M.}~\bibnamefont {Zhang}}, \bibinfo
  {author} {\bibfnamefont {F.}~\bibnamefont {Yan}}, \ and\ \bibinfo {author}
  {\bibfnamefont {D.}~\bibnamefont {Feng}},\ }\href {\doibase 10.1063/1.360166}
  {\bibfield  {journal} {\bibinfo  {journal} {Journal of Applied Physics}\
  }\textbf {\bibinfo {volume} {78}},\ \bibinfo {pages} {1953} (\bibinfo {year}
  {1995})}\BibitemShut {NoStop}%
\bibitem [{\citenamefont {Han}\ \emph {et~al.}(2011)\citenamefont {Han},
  \citenamefont {Han}, \citenamefont {Wang},\ and\ \citenamefont
  {Li}}]{Han:11}%
  \BibitemOpen
  \bibfield  {author} {\bibinfo {author} {\bibfnamefont {R.}~\bibnamefont
  {Han}}, \bibinfo {author} {\bibfnamefont {B.}~\bibnamefont {Han}}, \bibinfo
  {author} {\bibfnamefont {D.~H.}\ \bibnamefont {Wang}}, \ and\ \bibinfo
  {author} {\bibfnamefont {C.}~\bibnamefont {Li}},\ }\href {\doibase
  10.1063/1.3609009} {\bibfield  {journal} {\bibinfo  {journal} {Applied
  Physics Letters}\ }\textbf {\bibinfo {volume} {99}},\ \bibinfo {pages}
  {011912} (\bibinfo {year} {2011})}\BibitemShut {NoStop}%
\bibitem [{\citenamefont {Sun}\ \emph {et~al.}(2013)\citenamefont {Sun},
  \citenamefont {Lien}, \citenamefont {Qiu}, \citenamefont {Wang},
  \citenamefont {Mei}, \citenamefont {Liu},\ and\ \citenamefont
  {Feng}}]{Sun:13}%
  \BibitemOpen
  \bibfield  {author} {\bibinfo {author} {\bibfnamefont {H.~Y.}\ \bibnamefont
  {Sun}}, \bibinfo {author} {\bibfnamefont {S.-C.}\ \bibnamefont {Lien}},
  \bibinfo {author} {\bibfnamefont {Z.~R.}\ \bibnamefont {Qiu}}, \bibinfo
  {author} {\bibfnamefont {H.~C.}\ \bibnamefont {Wang}}, \bibinfo {author}
  {\bibfnamefont {T.}~\bibnamefont {Mei}}, \bibinfo {author} {\bibfnamefont
  {C.~W.}\ \bibnamefont {Liu}}, \ and\ \bibinfo {author} {\bibfnamefont
  {Z.~C.}\ \bibnamefont {Feng}},\ }\href {\doibase 10.1364/OE.21.026475}
  {\bibfield  {journal} {\bibinfo  {journal} {Opt. Express}\ }\textbf {\bibinfo
  {volume} {21}},\ \bibinfo {pages} {26475} (\bibinfo {year}
  {2013})}\BibitemShut {NoStop}%
\end{thebibliography}%
\end{document}


\title{Supplementary Information for: Space compatibility of emerging, wide-bandgap, ultralow-loss integrated photonics}

\author{Yue Hu}
\affiliation{International Quantum Academy, Shenzhen 518048, China}
\affiliation{Shenzhen Institute for Quantum Science and Engineering, Southern University of Science and Technology, Shenzhen 518055, China}

\author{Xue Bai}
\email[]{baixue@iqasz.cn}
\affiliation{International Quantum Academy, Shenzhen 518048, China}

\author{Baoqi Shi}
\affiliation{International Quantum Academy, Shenzhen 518048, China}

\author{Jiahao Sun}
\affiliation{Shenzhen Institute for Quantum Science and Engineering, Southern University of Science and Technology, Shenzhen 518055, China}
\affiliation{International Quantum Academy, Shenzhen 518048, China}

\author{Yafei Ding} 
\affiliation{International Quantum Academy, Shenzhen 518048, China}
\affiliation{Shenzhen Institute for Quantum Science and Engineering, Southern University of Science and Technology, Shenzhen 518055, China}

\author{Zhenyuan Shang}
\affiliation{International Quantum Academy, Shenzhen 518048, China}
\affiliation{Shenzhen Institute for Quantum Science and Engineering, Southern University of Science and Technology, Shenzhen 518055, China}

 \author{Hanke Feng}
\affiliation{Department of Electrical Engineering \& State Key Laboratory of Terahertz and Millimeter Waves, City University of Hong Kong, Kowloon, China}

\author{Liping Zhou}
\affiliation{State Key Laboratory of Materials for Integrated Circuits, Shanghai Institute of Microsystem and Information Technology, Chinese Academy of Sciences, Shanghai 200050, China}

\author{Bingcheng Yang}
\affiliation{State Key Laboratory of Materials for Integrated Circuits, Shanghai Institute of Microsystem and Information Technology, Chinese Academy of Sciences, Shanghai 200050, China}
\affiliation{The Center of Materials Science and Optoelectronics Engineering, University of Chinese Academy of Sciences, Beijing, China}

\author{Shuting Kang}
\affiliation{MOE Key Laboratory of Weak-Light Nonlinear Photonics, TEDA Applied Physics Institute and School of Physics, Nankai University, Tianjin 300457, China.}

\author{Yuan Chen}
\affiliation{International Quantum Academy, Shenzhen 518048, China}
\affiliation{Shenzhen Institute for Quantum Science and Engineering, Southern University of Science and Technology, Shenzhen 518055, China}

\author{Shuyi Li}
\affiliation{International Quantum Academy, Shenzhen 518048, China}

\author{Jinbao Long}
\affiliation{International Quantum Academy, Shenzhen 518048, China}

\author{Chen Shen}
\affiliation{International Quantum Academy, Shenzhen 518048, China}
\affiliation{Qaleido Photonics, 518048 Shenzhen, China}

\author{Fang Bo}
\email[]{bofang@nankai.edu.cn}
\affiliation{MOE Key Laboratory of Weak-Light Nonlinear Photonics, TEDA Applied Physics Institute and School of Physics, Nankai University, Tianjin 300457, China.}

\author{Xin Ou}
\email[]{ouxin@mail.sim.ac.cn}
\affiliation{State Key Laboratory of Materials for Integrated Circuits, Shanghai Institute of Microsystem and Information Technology, Chinese Academy of Sciences, Shanghai 200050, China}

\author{Cheng Wang}
\email[]{cwang257@cityu.edu.hk}
\affiliation{Department of Electrical Engineering \& State Key Laboratory of Terahertz and Millimeter Waves, City University of Hong Kong, Kowloon, China}

\author{Junqiu Liu}
\email[]{liujq@iqasz.cn}
\affiliation{International Quantum Academy, Shenzhen 518048, China}
\affiliation{Hefei National Laboratory, University of Science and Technology of China, Hefei 230088, China}
\maketitle

\section{Sample fabrication}
\vspace{0.2cm}

The fabrication processes of PICs based on Si$_3$N$_4$, 4H-SiC, and LiNbO$_3$ are described respectively in the following:

\vspace{0.2cm}

\noindent\textbf{Silicon nitride}: 
The Si$_3$N$_4$ PIC is fabricated using a deep-ultraviolet (DUV) subtractive process \cite{Ye:23, Sun:24}. 
The process flow is presented in Supplementary Fig. \ref{Fig:S1}a. 
First,a 6-inch (150-mm-diameter) wafer with wet thermal SiO$_2$ bottom cladding is thoroughly cleaned. 
Then, 300-nm-thick Si$_3$N$_4$ film is deposited on the wafer via low-pressure chemical vapor deposition (LPCVD), followed by deposition of SiO$_2$ film used as an etch hardmask.
Afterwards, DUV stepper photolithography based on KrF source (248 nm emission) and with 110 nm resolution is used to create waveguide pattern on the photoresist mask.
The pattern is subsequently transferred from the photoresist mask to the SiO$_2$ hardmask, and then into the Si$_3$N$_4$ layer to form waveguides, via ICP dry etching.  
After resist removal, the etched substrate is annealed to eliminate hydrogen content in Si$_3$N$_4$, whose existence causes optical absorption loss.
Finally, top SiO$_2$ cladding is deposited via LPCVD on the wafer, which also undergoes high-temperature annealing.
The wafer is diced into chips for following characterization and experiment. 

\vspace{0.2cm}

\noindent\textbf{Silicon carbide}: 
The 4H-SiC PIC is fabricated using a standard subtractive process \cite{Wang:22, Yi:22}. 
The process flow is presented in Supplementary Fig. \ref{Fig:S1}b. 
First, a 4H-SiC substrate is wafer-bonded on a silicon substrate with 2800-nm-thick SiO$_2$ cladding. 
Then the 4H-SiC substrate is mechanically grinded and polished to 410 nm thickness, forming a 4H-SiC-on-insulator (4H-SiCOI) substrate.  
The wafer is divided into multiple dies.
One die is taken and rinsed with acetone, alcohol, SC-1 solution and deionized water in sequence.
Then, 800-nm-thick HSQ resist is spin-coated on the die, and electron-beam lithography (EBL) is used to expose the PIC pattern. 
The pattern is etched into the 4H-SiC layer by ICP-RIE dry etching with SF$_6$/O$_2$ etchants. 
The EBL resist is removed using hydrofluoric acid solution.
Afterwards, 3-$\mu$m-thick top SiO$_2$ cladding is deposited via plasma-enhanced chemical vapor deposition (PECVD).
Finally, the die is cleaved to create smooth facet for edge-coupling of light with fibers.

\vspace{0.2cm}

\noindent\textbf{Lithium niobate}: 
The LiNbO$_3$ PIC is fabricated using a standard subtractive process \cite{WangC:18, Wan:24}. 
The process flow is presented in Supplementary Fig. \ref{Fig:S1}c. 
We use a commercially available, 4-inch, x-cut, lithium-niobate-on-insulator (LNOI) substrate (from NANOLN), which has a 500-nm-thick LiNbO$_3$ layer on 2-$\mu$m-thick SiO$_2$ bottom cladding. 
First, SiO$_2$ is deposited via PECVD as an etch hardmask for following LiNbO$_3$ etch. 
The PIC pattern is exposed using an ASML UV stepper lithography system (Nanosystem Fabrication Facility at the Hong Kong University of Science and Technology) with field size of $1.5\times1.5$ cm$^2$ and 500 nm resolution. 
The pattern is subsequently transferred from the photoresist mask to the SiO$_2$ hardmask using a standard fluorine-based dry etching process, and then into the LiNbO$_3$ layer using an optimized Ar$^+$-based ICP-RIE process.
The LiNbO$_3$  etch depth is approximately 250 nm, leaving a 250-nm-thick slab.
After removing the residual SiO$_2$ mask, top SiO$_2$ cladding is deposited via PECVD.
Finally,the wafer is diced into chips for following characterization and experiment.

\begin{figure*}[h!]
\renewcommand{\figurename}{Supplementary Figure}
\centering
\includegraphics{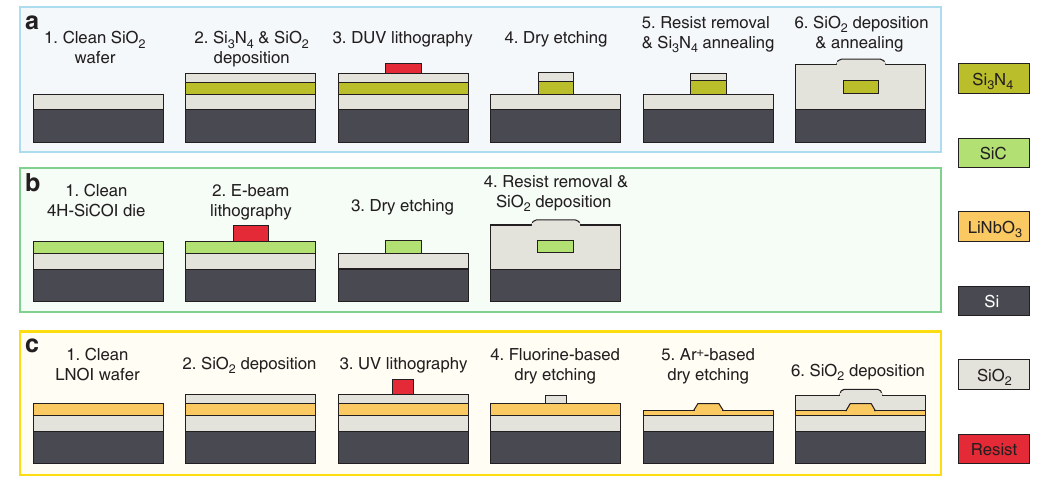}
\caption{
Fabrication process flow of Si$_3$N$_4$ (\textbf{a}), 4H-SiC (\textbf{b}) and LiNbO$_3$ (\textbf{c}) photonic chips.
}
\label{Fig:S1}
\end{figure*}

\clearpage

\section{Characterization of resonance linewidth}
\vspace{0.2cm}
To analyze the changes in optical loss of our chips before and after irradiation, we characterized the resonance linewidths of integrated microresonators based on Si$_3$N$_4$, 4H-SiC, and LiNbO$_3$.
For each resonance, we fit the profile using the formula \cite{Li:13}:
\begin{equation}
    T(\Delta\omega) = \left|1 - \frac{\kappa_\mathrm{ex}\left[i\Delta\omega+(\kappa_0+\kappa_\mathrm{ex})/2\right]}{\left[i\Delta\omega+(\kappa_0+\kappa_\mathrm{ex})/2\right]^2+\kappa_\mathrm{c}^2/4} \right|^2, 
    \label{eqn:S1}
\end{equation}
where $T$ represents the transmission, $\Delta\omega/2\pi$ is the laser detuning relative to the resonance, $\kappa_0/2\pi$ is the resonance's intrinsic loss, $\kappa_\mathrm{ex}/2\pi$ is the external coupling strength of the resonance to the bus waveguide, and $\kappa_\mathrm{c}/2\pi$ the complex coupling coefficient between the clockwise and counter-clockwise modes, in the presence of mode split.
Notably, $\kappa_\mathrm{c}/2\pi=0$ corresponds to resonances that do not exhibit visible mode splitting.

The fitting results of $\kappa_0$ and $\kappa_\mathrm{ex}$ for each resonance, both before and after irradiation, are presented in Supplementary Figs.\ref{Fig:S4}--\ref{Fig:S6} for microresonators based on Si$_3$N$_4$, 4H-SiC, and LiNbO$_3$, respectively.
Based on the fitting results, we conclude that: 
1. For Si$_3$N$_4$ and 4H-SiC microresonators, neither proton nor $\gamma$-ray irradiation induces any noticeable changes in $\kappa_0$ and $\kappa_\mathrm{ex}$.
2. For LiNbO$_3$ microresonators with SiO$_2$ top cladding, proton irradiation does not affect $\kappa_0$ and $\kappa_\mathrm{ex}$.
However, $\gamma$-ray irradiation marginally increases $\kappa_0$, indicating an increase in optical loss, while $\kappa_\mathrm{ex}$ remains unaffected.

\begin{figure}[h]
\renewcommand{\figurename}{Supplementary Figure}
\centering
\includegraphics{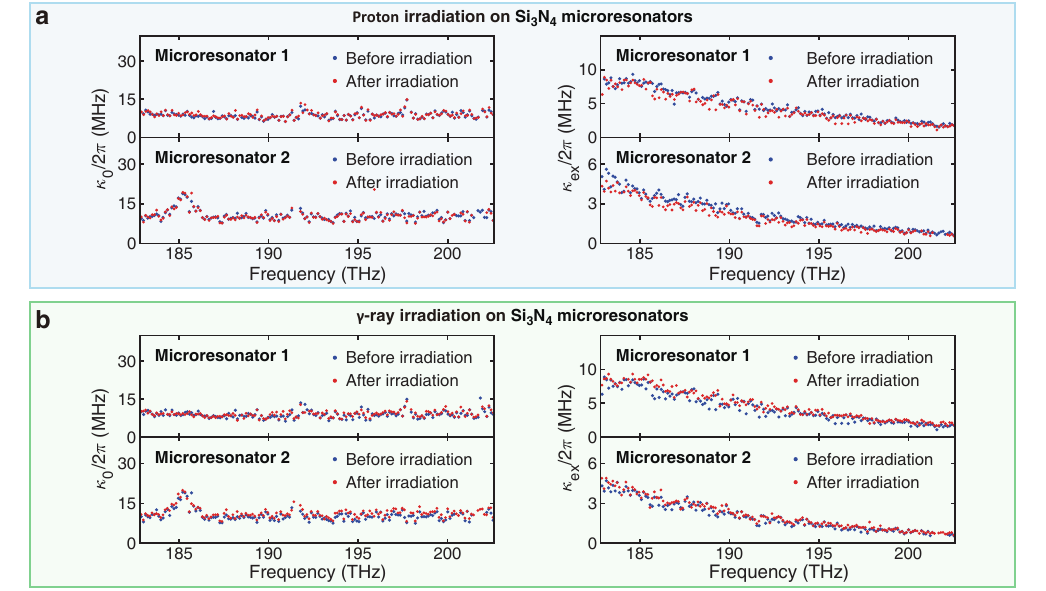}
\caption{
\textbf{Loss change due to irradiation for Si$_3$N$_4$ microresonators.}
Comparison of $\kappa_0/2\pi$ and $\kappa_\text{ex}/2\pi$ before and after proton (\textbf{a}) or $\gamma$-ray (\textbf{b}) irradiation.
Data of two microresonators are presented. 
Left column: $\kappa_0/2\pi$; Right column: $\kappa_\text{ex}/2\pi$.}
\label{Fig:S4}
\end{figure}

\begin{figure}[h]
\renewcommand{\figurename}{Supplementary Figure}
\centering
\includegraphics{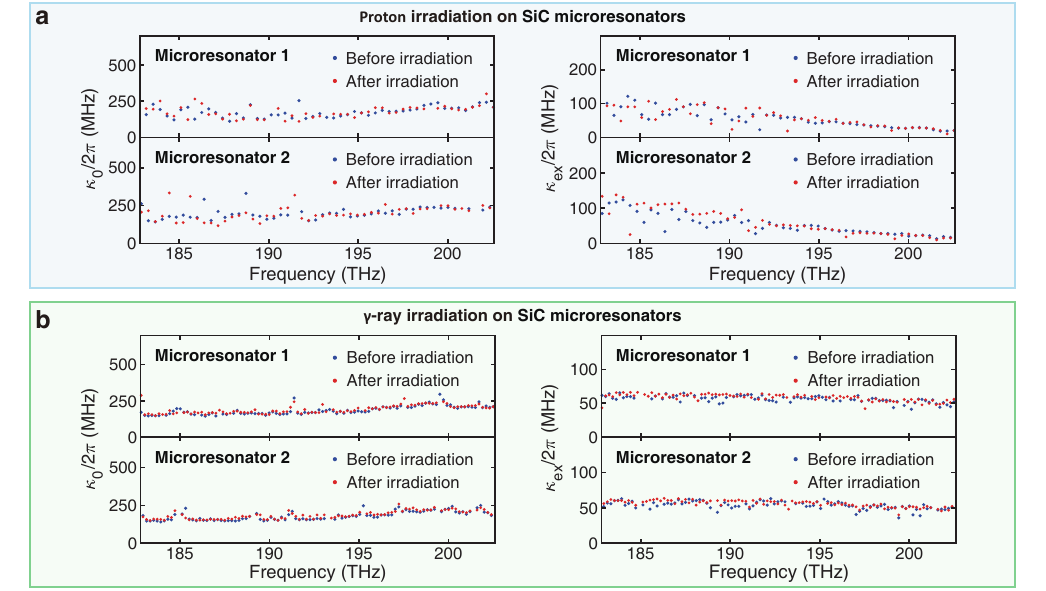}
\caption{
\textbf{Loss change due to irradiation for 4H-SiC microresonators.}
Comparison of $\kappa_0/2\pi$ and $\kappa_\text{ex}/2\pi$ before and after proton (\textbf{a}) or $\gamma$-ray (\textbf{b}) irradiation.
Data of two microresonators are presented. 
Left column: $\kappa_0/2\pi$; Right column: $\kappa_\text{ex}/2\pi$.}
\label{Fig:S5}
\end{figure}

\begin{figure}[h]
\renewcommand{\figurename}{Supplementary Figure}
\centering
\includegraphics{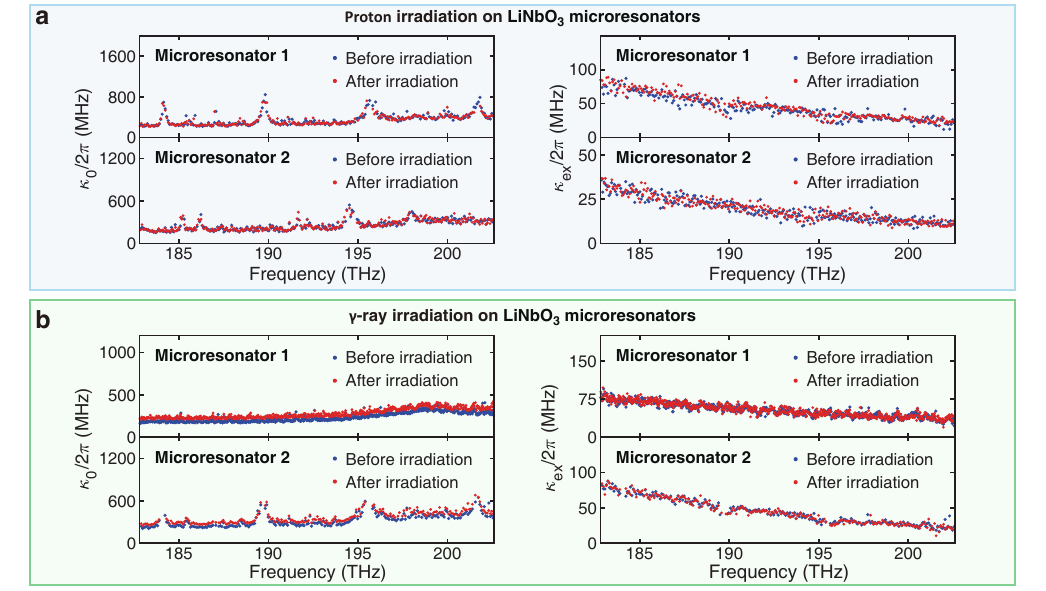}
\caption{
\textbf{Loss change due to irradiation for LiNbO$_3$ microresonators.} 
Comparison of $\kappa_0/2\pi$ and $\kappa_\text{ex}/2\pi$ before and after proton (\textbf{a}) or $\gamma$-ray (\textbf{b}) irradiation.
Data of two microresonators are presented. 
Left column: $\kappa_0/2\pi$; Right column: $\kappa_\text{ex}/2\pi$.}
\label{Fig:S6}
\end{figure}

\clearpage

\section{Characterization of uncladded LiNbO$_3$ chips}
\vspace{0.2cm}

\begin{figure}[h]
\renewcommand{\figurename}{Supplementary Figure}
\centering
\includegraphics{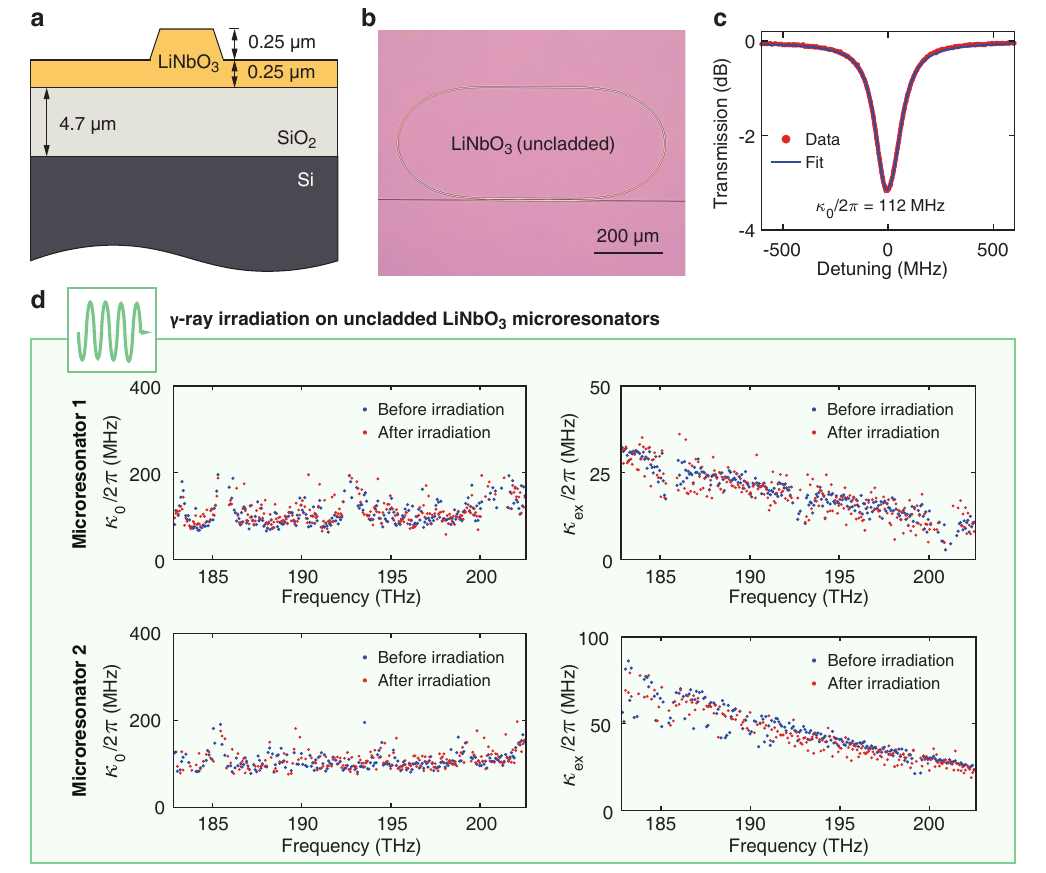}
\caption{
\textbf{Characterization of uncladded LiNbO$_3$ microresonator.} 
\textbf{a.} 
Cross-section of the uncladded LiNbO$_3$ waveguide.
\textbf{b.} 
Optical microscope image of an uncladded LiNbO$_3$ microresonator. 
\textbf{c.}
A representative resonance with Lorentzian fit. 
Extracted $\kappa_0/2\pi=112$ MHz is marked.
\textbf{d.} 
Comparison of $\kappa_0/2\pi$ and $\kappa_\text{ex}/2\pi$ of two uncladded LiNbO$_3$ microresonators before and after $\gamma$-ray irradiation.
}
\label{Fig:S7}
\end{figure}

The results in Supplementary Fig.\ref{Fig:S6} b suggest that LiNbO$_3$ microresonators fully cladded with SiO$_2$ exhibit a noticeable increase in $\kappa_0/2\pi$ after $\gamma$-ray irradiation.
To investigate whether the loss increase is caused by the radiation impact on the LiNbO$_3$ waveguides or on the SiO$_2$ top cladding, we conduct another $\gamma$-ray irradiation experiment of the same dose on an uncladded LiNbO$_3$ chip.
Supplementary Fig. \ref{Fig:S7}a shows the cross-section geometry of the uncladded LiNbO$_3$ waveguide used here. 
Supplementary Fig. \ref{Fig:S7}b shows the optical microscope image of an uncladded LiNbO$_3$ microresonator. 
Supplementary Fig.\ref{Fig:S7}c displays a representative microresonator resonance with Lorentzian fit.
Supplementary Fig.\ref{Fig:S7}d presents the $\kappa_0/2\pi$ and $\kappa_\text{ex}/2\pi$ of two uncladded LiNbO$_3$ microresonators before and after $\gamma$-ray irradiation, indicating that $\kappa_0/2\pi$ remains unchanged.
The results suggest that the loss increase is primarily due to the radiation impact on the SiO$_2$ top cladding.

\clearpage

\section{Material characterization}
\vspace{0.2cm}

In addition to the optical characterization of microresonators, material characterization of thin films has been conducted to further investigate radiation effects on material defects, lattice structure, and surface topography. This section presents the analysis results obtained from Raman spectroscopy, X-ray diffraction (XRD), and atomic force microscopy (AFM), for thin films made of Si$_3$N$_4$, 4H-SiC, and LiNbO$_3$, both before and after exposure to 1.2 Mrad $\gamma$-ray irradiation. 
The study included bare thin films of Si$_3$N$_4$, 4H-SiC, and LiNbO$_3$, as well as films cladded with SiO$_2$.

\vspace{0.2cm}

\noindent\textbf{Raman spectra}: 
The Raman spectra of Si$_3$N$_4$, 4H-SiC and LiNbO$_3$ thin films is performed using a confocal Raman spectrometer (Oxford-WITec:alpha300R) before and after exposure to 1.2 Mrad $\gamma$-ray irradiation.
To facilitate direct comparison, the spectra are normalized and overlapped, as shown in Supplementary Fig. \ref{Fig:S8}.
Supplementary Fig. \ref{Fig:S8}a presents the Raman spectra of the Si$_3$N$_4$ film, showing a characteristic peak within the 900--1100 cm$^{-1}$ range, consistent with the results reported in Ref. \cite{Sergo:96}.
Notably, the Si$_3$N$_4$ film in this study is amorphous and deposited using LPCVD, different from that used in Ref. \cite{Sergo:96} which accounts for the minor variations observed in the Raman spectra.
Supplementary Fig. \ref{Fig:S8}(b, c) presents the Raman spectra of the 4H-SiC and LiNbO$_3$ films, respectively, which are consistent with Refs.~\cite{Yang:87, Wu:95, Han:11, Sun:13}.
No significant peak shifts are observed in the Raman spectra before and after $\gamma$-ray irradiation, indicating that no detectable changes in stress, defects, or crystal structure occurred in these films.

\begin{figure}[H]
\renewcommand{\figurename}{Supplementary Figure}
\centering
\includegraphics{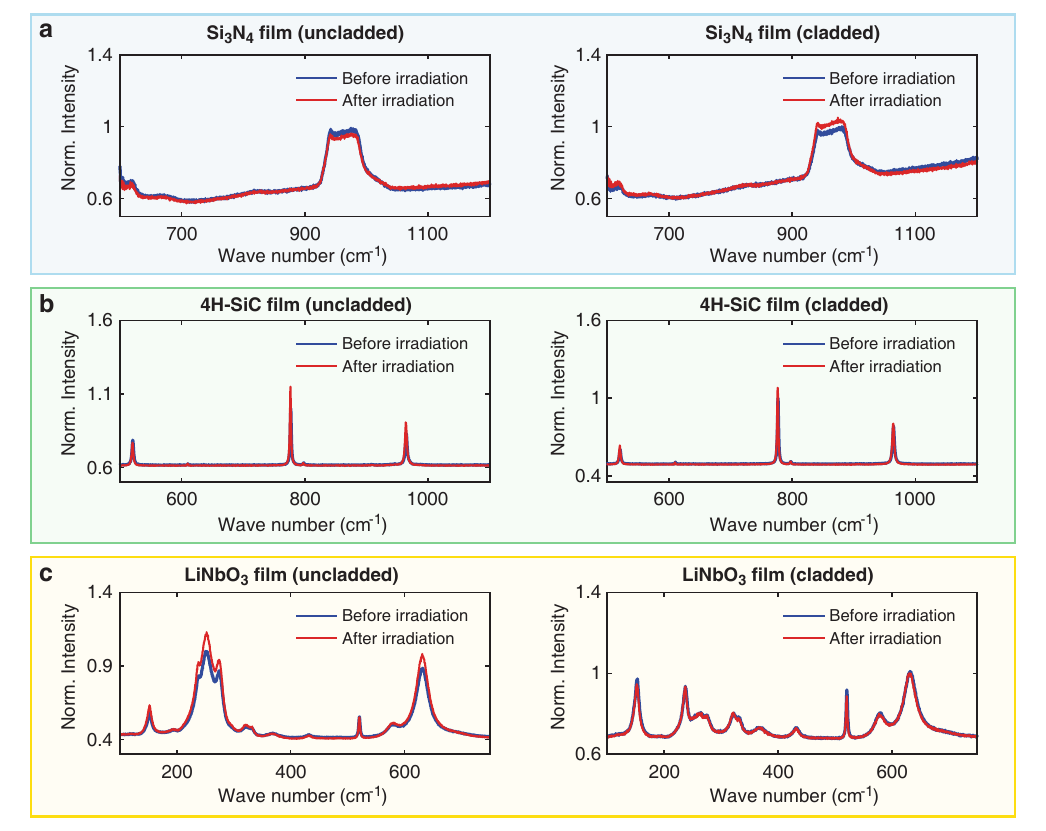}
\caption{
Raman spectra of the Si$_3$N$_4$ (\textbf{a}), 4H-SiC (\textbf{b}) and LiNbO$_3$ (\textbf{c}) before and after 1.2 Mrad $\gamma$-ray irradiation.
Left column: uncladded films; 
Right column: SiO$_2$ cladded films.
No peak shift has been observed within the detection accuracy for these films.
}
\label{Fig:S8}
\end{figure}

\clearpage
\noindent\textbf{X-ray diffraction (XRD)}: 
To further investigate the radiation effects on lattice structure, we measure the XRD spectra of LiNbO$_3$ and 4H-SiC films using an X-ray diffractometer (TD-3500) before and after 1.2 Mrad $\gamma$-ray irradiation.
The pre- and post-irradiation XRD spectra, shown in Supplementary Fig. \ref{Fig:S9}, indicate that the diffraction patterns remain unchanged,. 
Thus we conclude that the crystal lattices have not been altered.

\begin{figure}[h!]
\renewcommand{\figurename}{Supplementary Figure}
\centering
\includegraphics{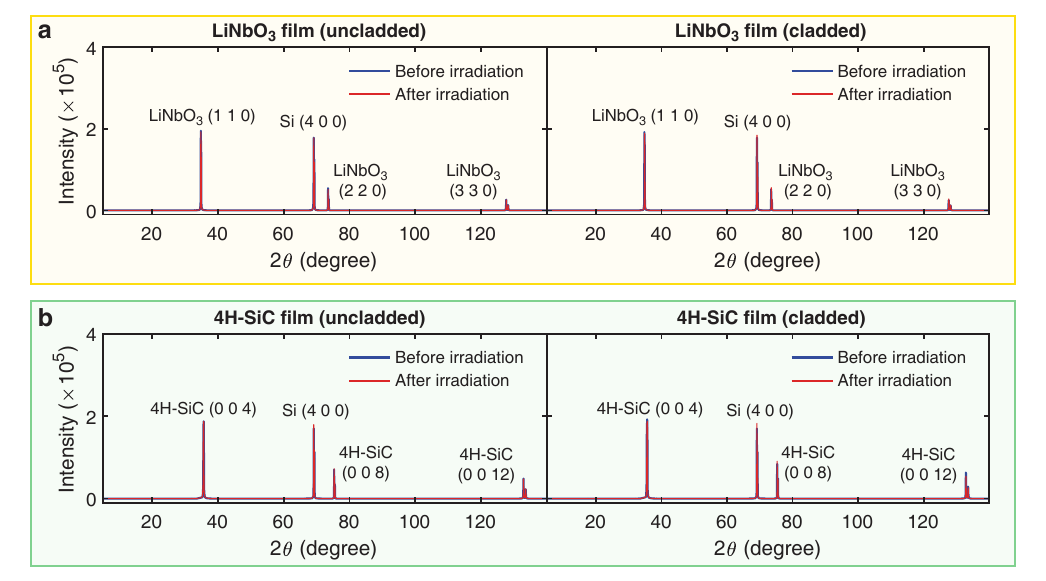}
\caption{
XRD spectra of LiNbO$_3$ (\textbf{a}) and 4H-SiC (\textbf{b}) films before and after 1.2 Mrad $\gamma$-ray irradiation.
Left column: uncladded films; 
Right column: SiO$_2$ cladded films.
The diffraction patterns remain unchanged after irradiation.
}
\label{Fig:S9}
\end{figure}

\begin{figure}[H]
\renewcommand{\figurename}{Supplementary Figure}
\centering
\includegraphics{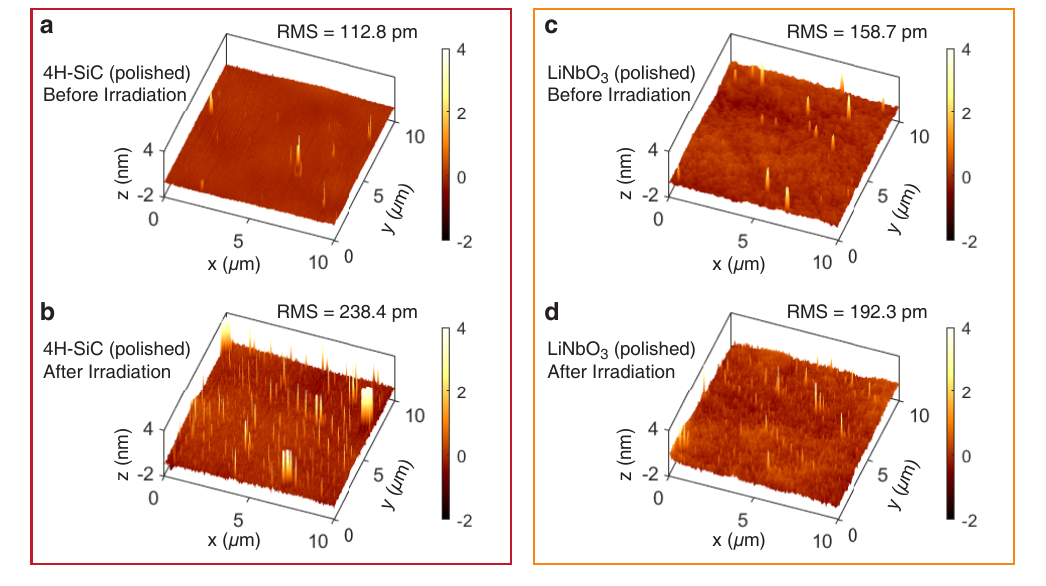}
\caption{
Surface topography data of a polished 4H-SiC film before (\textbf{a}) and after (\textbf{b}) $\gamma$-ray irradiation, and a polished LiNbO$_3$ film before (\textbf{c}) and after (\textbf{d}) $\gamma$-ray irradiation. 
}
\label{Fig:S10}
\end{figure}

\clearpage
\noindent\textbf{Atomic force microscopy (AFM)}:
To investigate the irradiation effects on surface topography, we measure the surface topography of polished 4H-SiC and LiNbO$_3$ films using an AFM (Oxford-MFP-3D) before and after exposure to 1.2 Mrad $\gamma$-ray irradiation.
Supplementary Fig.\ref{Fig:S10}(a, b) and (c, d) show the surface topography of 4H-SiC and LiNbO$_3$, respectively, and reveal minor changes in surface topography.
The average height deviation for each surface is calculated, and the surface roughness of polished 4H-SiC increased from 112.8 pm to 238.4 pm, while that of polished LiNbO$_3$ increased from 158.7 pm to 192.3 pm. 

\section*{Supplementary References}
\bigskip
\bibliographystyle{apsrev4-1}
\bibliography{bibliography}